\documentclass[submission, Phys]{SciPost}

\hypersetup{
    colorlinks,
    linkcolor={red!50!black},
    citecolor={blue!50!black},
    urlcolor={blue!80!black}
}

\usepackage[bitstream-charter]{mathdesign}
\DeclareSymbolFont{usualmathcal}{OMS}{cmsy}{m}{n}
\DeclareSymbolFontAlphabet{\mathcal}{usualmathcal}

\usepackage{bbm}
\usepackage{multirow}
\usepackage{color}
\usepackage{hyperref}
\usepackage{braket}
\usepackage[justification=justified]{caption}
\usepackage{subcaption}

\graphicspath{
{./}
{./FIGS/}
}
\begin{document}

\begin{center}{\Large \textbf{Strong Hilbert space fragmentation via emergent quantum drums in two dimensions
      \\
}}\end{center}
\begin{center}
Anwesha Chattopadhyay\textsuperscript{1,2}, Bhaskar
Mukherjee\textsuperscript{3}, K. Sengupta\textsuperscript{1}, Arnab
Sen\textsuperscript{1*}
\end{center}

\begin{center}
{\bf 1} School of Physical Sciences, Indian Association for the
Cultivation of Science, Jadavpur, Kolkata 700032, India.
\\
{\bf 2} Department of Physics, School of Mathematical Sciences,
Ramakrishna Mission Vivekananda Educational and Research Institute,
Belur, Howrah 711202, India.
\\
{\bf 3} Department of Physics and Astronomy, University College
London, Gower Street, London WC1E 6BT, United Kingdom.
\\
${}^\star$ {\small \sf tpars@iacs.res.in}
\end{center}

\begin{center}
\today
\end{center}


\section*{Abstract}
{\bf
  We introduce a disorder-free model of $S=1/2$ spins on the square lattice
  in a constrained Hilbert space where two up-spins are not allowed
  simultaneously on any two neighboring sites of the lattice.
  The interactions are given by
  ring-exchange terms on elementary
  plaquettes that conserve both the total magnetization as well as
  dipole moment.
  We show that this model provides a tractable example of
  strong Hilbert space fragmentation in two dimensions with 
    typical
  initial states evading thermalization with respect to the full Hilbert space.
  Given any product state,
  the system can be decomposed into
  disjoint spatial regions made of edge and/or vertex sharing
  plaquettes that we dub as ``quantum drums''.
  These quantum drums come in
  many shapes and sizes
  and specifying the plaquettes that belong to a drum fixes its spectrum.
  The spectra of some small drums is calculated analytically.
  We study two bigger quasi-one-dimensional
  drums numerically, dubbed ``wire'' and a ``junction of two wires''
  respectively. We find that these possess
  a chaotic spectrum but also support distinct families of quantum
  many-body scars that cause periodic revivals from different initial states.
  The wire is shown to be equivalent to the
  one-dimensional PXP chain with open boundaries,
  a paradigmatic model for quantum many-body
  scarring; while the junction of two wires represents a
  distinct constrained model.

}

\vspace{10pt}
\noindent\rule{\textwidth}{1pt}
\tableofcontents\thispagestyle{fancy}
\noindent\rule{\textwidth}{1pt}
\vspace{10pt}

\section{Introduction}
\label{int}
A generic isolated quantum system with many degrees of freedom is expected to
``self-thermalize'' as it evolves unitarily under the dynamics of its own
Hamiltonian~\cite{Rigol2008}.
This implies that pure states obtained from the time evolution of
different initial states that share the same energy density cannot be
distinguished from each other at late times using only local probes.
A microscopic justification for this self-thermalization is provided by the
eigenstate thermalization hypothesis (ETH)~\cite{Deutsch1991, Srednicki1994,
Alessio2016, Dymarsky2018}
that posits that high-energy eigenstates of such systems appear locally thermal
with the temperature being set by the energy density of the eigenstate.

Rapid progress in producing and manipulating well-isolated quantum simulators
such as ultracold gases~\cite{Bloch2005, Langen2015},
trapped ions~\cite{Bohnet2016}, Rydberg atom arrays~\cite{Bluvstein2021} and
superconducting qubits~\cite{Kjaergaard2020} has made it possible to study
thermalization and its violations in such platforms.
In particular, the experimental observation of late-time coherent oscillations
from certain simple high-energy initial states in a kinetically-constrained
chain of $51$ Rydberg atoms~\cite{Bernien2017} generated great interest
in understanding thermalization in interacting theories
with constrained Hilbert spaces. The revivals reported in
Ref.~\cite{Bernien2017} were shown to arise due to
the large overlap of some simple initial states with a small set of
nonthermal high-energy eigenstates, dubbed quantum many-body scars (QMBS) in
Refs.~\cite{Turner2018a,Turner2018b}, in
an otherwise non-integrable PXP model~\cite{Sachdev2002, Lesanovsky2012}
that served as the minimal model for the experiment.

Subsequent theoretical
studies have shown a plethora of interesting non-ergodic behavior in
various models with constrained Hilbert spaces, including Hamiltonian
formulations of lattice gauge theories~\cite{Kogut1975, Kogut1979, Horn1981,
  Orland1989, Chandrasekharan1997} that may be realizable on quantum
simulators~\cite{Mil2020, Yang2020, Brower2020}.
These include different varieties of QMBS~\cite{
  Choi2019, Ho2019, Lin2019, Iadecola2019, Shiraishi2019, Chattopadhyay2020,
  Mukherjee2020a, Mukherjee2021spin1, Pai2019, Mukherjee2020b, Mukherjee2020c,
  Mizuta2020, Michailidis2020, Lin2020,
  Surace2020, Banerjee2021, Biswas2022, wildeboer2021, Zhao2020, Zhao2021, 
  Mukherjee2021periodically, Desaules2022a, Desaules2022b, Hudomal2022},
disorder-free localization~\cite{Smith2017, Karpov2021, Chakraborty2022,
Brenes2018} as well as a richer ergodicity-breaking
paradigm dubbed Hilbert space fragmentation~\cite{Sala2020, Khemani2020}.
Such forms of ETH-violation
are distinct from
the breakdown of ETH due to
many-body localization~\cite{Pal2010, Nandkishore2015, Abanin2019} where
strong disorder plays a crucial role.

Systems with Hilbert space fragmentation~\cite{Moudgalyabookchapter,
  Pai2019b, Yang2020b,
  Rakovszky2020, Feldmeier2020, McClarty2020, Herviou2021, Kyungmin2021,
  Roy2020, Mukherjee2021fragment,
  Pozsgay2021, Langlett2021, Hahn2021, Khudorozhkov2021} often
feature multiple conservation laws~\cite{Sala2020, Khemani2020}
which severely restrict the mobility of excitations.
In such cases, the Hilbert space can split
into exponentially many dynamically disconnected
{\it fragments}. These fragments cannot be distinguished by any
obvious global symmetries of the Hamiltonian~\cite{Sala2020, Khemani2020}.
Such fragments can either be finite or
infinite-dimensional in size in the
thermodynamic limit and can show vastly different dynamical properties,
such as integrability~\cite{Yang2020b, Pozsgay2021}, disorder-free
localization~\cite{Smith2017, Karpov2021, Chakraborty2022, Brenes2018,
  Pai2019b, McClarty2020} or QMBS~\cite{Langlett2021, Mukherjee2021fragment}
though large
  fragments are expected to typically satisfy a Krylov-restricted version of ETH
  ~\cite{Moudgalyabookchapter}.
Both {\it weak} and {\it strong fragmentation}
is known to exist in one-dimensional (1D) models~\cite{Sala2020, Khemani2020},
with the two cases distinguished
by whether the fraction of eigenstates violating the ETH are a set of
measure zero or not in the thermodynamic limit.
Weakly fragmented systems are similar to systems with QMBS since
both situations lead to weak ergodicity breaking where typical initial states
still thermalize~\cite{Moudgalyareview}.
However, strongly fragmented systems present a distinct form of ergodicity
breaking that is different from systems with QMBS.

In Ref.~\cite{Sala2020}, 1D spin models with both global charge and dipole
conservation laws were considered and it was argued that such
dipole-conserving models should exhibit Hilbert space
fragmentation in any dimension~\cite{Sala2020, Khemani2020}
(for examples of fragmentation without global dipole
conservation, see Refs.~\cite{Mukherjee2021fragment, Yang2020b, McClarty2020,
  Hahn2021, Rakovszky2020, Kyungmin2021, Langlett2021,Mukherjee2021spin1}).
One of the tell-tale signs of fragmentation in such models is an
exponential number of completely inert states that form one-
  dimensional
fragments on their own. While examples of both weak and strong fragmentation
are known in one dimension, it is not clear whether global dipole
conservation alone is sufficient to lead to strong fragmentation in higher
dimensions. This extra conservation ensures that
Hilbert space fragments of different sizes can be constructed by embedding
suitable ``active''
regions into ``inert'' backgrounds and surrounding the ``active'' regions
by ``shielding'' regions; the shielding region, however, turns out to be of the
same size or bigger than the active region it isolates~\cite{Khemani2020}.
This makes it difficult to
construct explicit examples of strong fragmentation in two or higher
dimensions.

In this paper, we will construct a model that shows strong Hilbert
space fragmentation in two dimensions by considering
$S=1/2$ spins (equivalently,
hard-core bosons) on the square
lattice with ring-exchange terms on elementary plaquettes that are
consistent with total magnetization (equivalently, boson number) conservation
as well as global dipole moment conservation. The important
additional ingredient in the model is the presence of a
kinematic constraint that
no two nearest neighbor sites
can have two up-spins (bosons) simultaneously.
Similar models with ring-exchange and other competing
terms, but without the additional hard-core constraints,
are known to have interesting low-energy phases and
transitions~\cite{Paramekanti2002, Sandvik2002, Melko2004}.
High-energy properties of the unconstrained model with only the
ring-exchange terms
were studied recently in Ref.~\cite{Khudorozhkov2021} where it was
realized that such terms imply subsystem symmetries
associated with the conservation of magnetization along each column and
row of the square lattice. This leads to global dipole conservation and
consequently Hilbert space fragmentation. However, the precise
nature of the fragmentation (weak or strong) could not be established for this
unconstrained model in Ref.~\cite{Khudorozhkov2021}.

\begin{figure}[!htb]
  \begin{center}
    \includegraphics[width=0.45\linewidth]{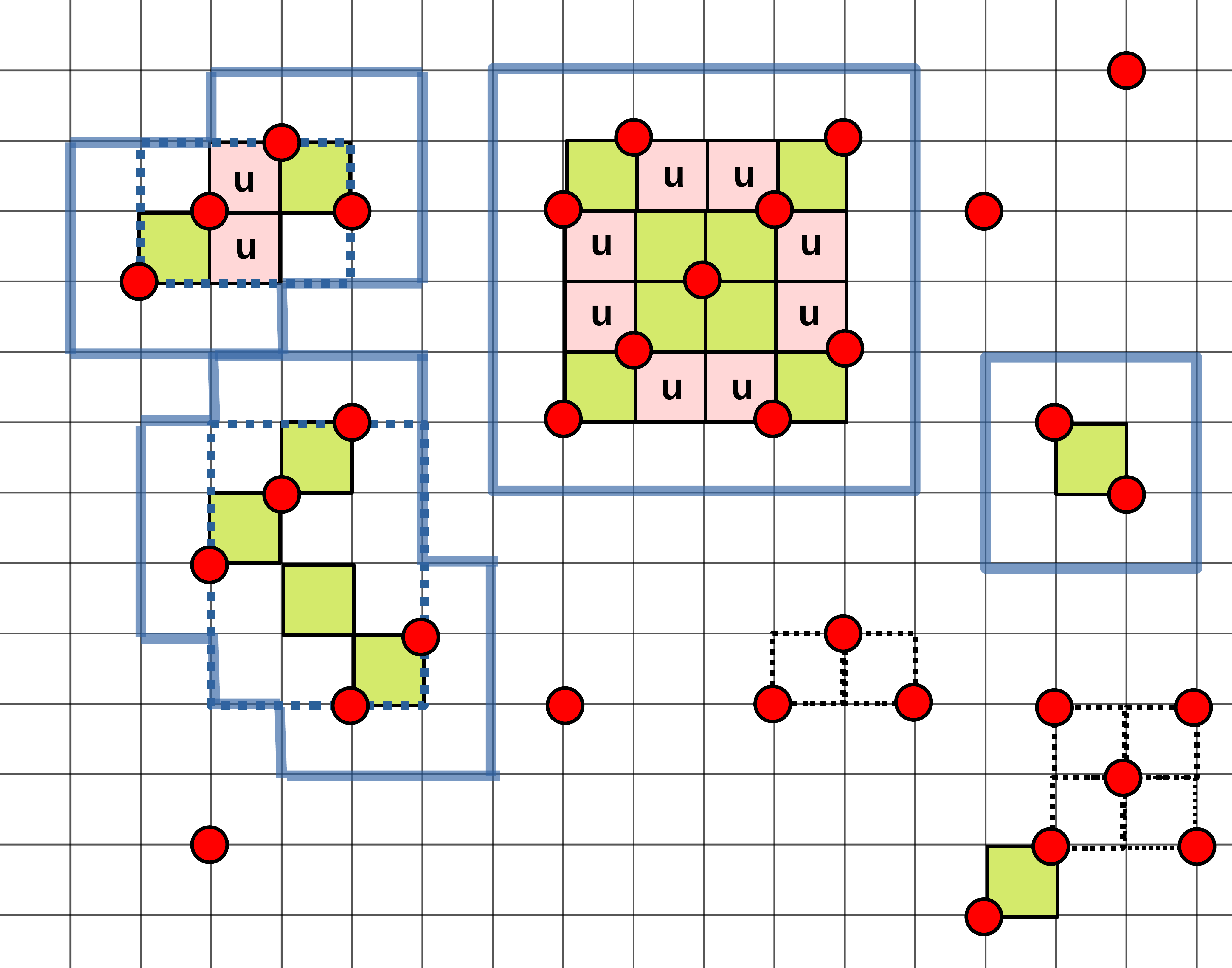}
    \caption{An initial state on the square lattice where the
      up-spins (bosons) are indicated in red while the other sites have
      down-spins (no bosons). The
      five quantum drums that correspond to this initial state are shown
      with their elementary plaquettes shaded. Plaquettes
      that are colored as green can have
      two up-spins (bosons) along both its diagonals
      during quantum evolution as explained in the text.
      The pink plaquettes, also labeled by ``u'',
      can have two up-spins (bosons)
      only along one of its two diagonals.
      Sites that do not
      belong to any of these five drums have inert up/down spins fixed by
      the initial condition.
      Each drum generates a separate fragment in Hilbert space with the
      corresponding fragment size being $3$ for
      the top-left drum, $7$ for the
      bottom-left drum, $24$ for the middle drum and
      $2$ for both the top-right and bottom-right drums.
      The boundary of the shielding region
      for $4$ of the drums are shown using thick blue lines.
    }
    \label{drumsandshields}
\end{center}
\end{figure}

As we will show here, the enforcement of the kinematic constraints
leads to several additional features,
including strong ergodicity breaking and the
emergence of ``quantum drums'', that were
absent in the model considered in Ref.~\cite{Khudorozhkov2021}. 
The quantum drums here can be viewed as the ``active'' regions which can then be
surrounded by ``shielding'' regions of $O(1)$ thickness (in lattice units).
Crucially, the thickness of the shielding regions does not grow with the
size of the quantum drums. Each quantum drum is made of
edge and/or vertex
sharing elementary plaquettes and specifying the plaquettes that make a drum
uniquely fixes its spectrum, thus justifying this particular nomenclature.
We refer the reader to Fig.~\ref{drumsandshields} for
  an example of quantum drums and their corresponding shielding regions
  that emerge from a particular initial state.

All the Hilbert space fragments of this model that are not 
  one-dimensional,
i.e., that do not correspond to
inert Fock states,
can be generated from a combination of appropriate quantum drums embedded
in an otherwise inert background (which may itself shrink to zero for certain
drums) (see Fig.~\ref{drumsandshields}).
Thus, the Hilbert space can be decomposed as a direct sum
  over dynamically disconnected sectors that are completely labeled by
  quantum drums and any remaining inert spins that do not belong to a
quantum drum.
These quantum drums come in a variety of shapes and
sizes and can be made of a finite
number or an arbitrarily large number of plaquettes in the
thermodynamic limit.
Crucially, the nature of the
    Hilbert space fragments
    generated from large drums that form the largest Krylov subspaces and
    are, therefore, relevant for typical initial states allows for a proof
of lack of thermalization by identifying either (a) an extensive number of
    single spin correlators or (b) an extensive number of next-nearest neighbor
    two-spin correlators whose expectation values stay
pinned to their initial (non-thermal) values.
To the best of our knowledge, this interacting theory provides the
first example of {\em strong} Hilbert space fragmentation in two dimensions.

The rest of the paper is arranged as follows. In Sec.~\ref{model}, we introduce
the model and summarize some of its important properties. In Sec.~\ref{qdrums},
we discuss the quantum drums that emerge in this model in more detail.
The classical
construction of the drums, given an initial state, is explained in
Sec.~\ref{drumsIS}. The construction of the shielding regions of drums
and closest approach of two drums such that these can still be considered
independent of each other is explained in Sec.~\ref{drumsShields}.
In Sec.~\ref{drumsR}, a recursive
procedure to generate bigger quantum drums starting from the most elementary
one-plaquette drum is discussed. We introduce some particular drums, dubbed
wires and junctions of wires and some other
quasi-one dimensional (1D) and two-dimensional (2D) drums
in Sec.~\ref{largedrums}. In Sec.~\ref{modED}, we give numerical
evidence that the energy
eigenvalues and their associated degeneracies from exact diagonalization (ED)
on small systems can be completely understood in terms of the spectra of the
quantum drums. We construct a large class of
eigenstates with integer eigenvalues (including zero modes) from the
packing of the simplest one-plaquette quantum drums in a
macroscopic system size in Sec.~\ref{intstates}.
  A wire decomposition
of drums is introduced in
Sec.~\ref{wiresdecom}, with Sec.~\ref{fragdim} showing how to calculate
fragment dimension and Sec.~\ref{wirestodrum} showing how to construct
entire drums
from wire-decomposed reference states for some small drums.
Evidence for
strong Hilbert space fragmentation in this model is presented in
Sec.~\ref{strongfrag}. The numerical evidence from ED is presented in
Sec.~\ref{numfrag}.  In Sec.~\ref{twowires}, we derive the scaling of the
dimension of the Hilbert space fragments for the quantum drums
composed of two long parallel wires to show the
  utility of the wire decomposition in obtaining the fragment
  size scaling for macroscopic drums. The wire decomposition
allows us to derive the
  scaling of the dimension of the Hilbert space fragments associated with
  large 2D drums and determine
  which kinds of drums dominate statistically given a
  certain density of up-spins (bosons) and identify
    the Hilbert space fragment with the largest dimension
  in Sec.~\ref{nonthermal}. We prove that typical initial
  states that belong to
    these large fragments (Krylov subspaces)
    {\it do not} thermalize
    with respect to the full Hilbert space in Sec.~\ref{nonthermal} by
  identifying either an extensive number of single-spin correlators or
  two-spin correlators that stay pinned to their initial non-thermal values.
The analytical study for the spectra of certain small
quantum drums is given in Sec.~\ref{astud}. A tree structure to represent
the action of $H$ in the Fock space of a drum is explained in Sec.~\ref{nstree}.
The spectra of small wires is calculated in Sec.~\ref{astwire} while the
spectra of other small quantum drums that can be viewed as building blocks of
more
complicated wire junctions is calculated in Sec.~\ref{astjunc}.
The spectra of two different
classes of bigger quasi-1D quantum drums, a wire and a particular junction of
two wires, are addressed
numerically using ED in Sec.~\ref{nstud}. Both these
large quantum drums can be interpreted as effective
quasi-1D models with a spectrum that is symmetric around
zero energy. A tree generating algorithm is described and the equivalence
of the wire to the 1D PXP model on an open chain
is shown in Sec.~\ref{treelargedrums}.
The Hilbert space dimensions for both the drums are calculated analytically and
level statistics are computed numerically in Sec.~\ref{nslev}.
The Hilbert space structure of the junction of two wires
turns out to be completely different from that of the wire as discussed in
both Sec.~\ref{treelargedrums} and Sec.~\ref{nslev}. One of these
fragments is shown to have a macroscopically large
number of exact zero modes while the other fragment has no zero modes in
Sec.~\ref{nsind}. Both fragments
satisfy Krylov-restricted ETH but also support distinct families of QMBS
that result in periodic revivals from different simple initial states as
discussed in Sec.~\ref{nsscars}. Our numerical results for the
wire show that open PXP chains of length $3n+1$, where $n$ is an integer,
lead to enhanced fidelity revivals for the period-$3$ ordered initial
$|\mathbb{Z}_3\rangle$ state without adding any optimal perturbations to
the bare Hamiltonian; a feature which may have experimental consequence for
Rydberg chains. The junction of two wires also shows QMBS and
simple initial states from which clear revivals in fidelity are observed.
Finally, we summarize our main results and
conclude in Sec.~\ref{diss}.

\section{Model and its properties}
\label{model}

The Hamiltonian of the model is given by
\begin{eqnarray}
H &=& J \sum_{j_x,j_y} \left( \sigma^+_{j_x,j_y}
\sigma^+_{j_x+1,j_y+1} \sigma^-_{j_x+1,j_y} \sigma^-_{j_x,j_y+1} +
{\rm h.c.} \right) \label{hamdef}
\end{eqnarray}
where $\sigma^{\alpha}_{j_x,j_y}$ for $\alpha=x,y,z$ represent
spin-half Pauli matrices at sites $(j_x,j_y)$ of a 2D
square lattice,
$\sigma_{j_x,j_y}^{\pm} = (\sigma^x_{j_x,j_y} \pm i
\sigma_{j_x,j_y}^y)/2$, and the lattice spacing has been set to
unity ($a=1$). The Hamiltonian is supplemented by the constraint that two
up-spins can not occupy neighboring sites of the lattice; this is
implemented by the operator relation
\begin{eqnarray}
\left(1+\sigma^z_{j_x,j_y}\right)\left(1+\sigma^z_{j_x \pm 1,j_y}\right)
&=& \left(1+\sigma^z_{j_x,j_y}\right)\left(1+\sigma^z_{j_x,j_y \pm
1}\right) =0  \label{constr}
\end{eqnarray}
For finite $L_x \times L_y$ rectangular
lattices, we will consider open boundary conditions
and the constraint (Eq.~\ref{constr}) is then applied to the
three/two nearest neighbors of $(j_x,j_y)$ for the edge/corner sites.

This system maps exactly to hard-core bosons with the following transformations:
\begin{eqnarray}
2b^\dagger_{j_x,j_y}b_{j_x,j_y} - 1 &=& \sigma^z_{j_x,j_y} \nonumber \\
b^\dagger_{j_x,j_y}&=& \sigma^+_{j_x,j_y}
\label{stob}
\end{eqnarray}
where $b^\dagger_{j_x,j_y}$ is the boson creation operator
and $n_{j_x,j_y}$ $=$ $b^\dagger_{j_x,j_y}b_{j_x,j_y}$ is the boson number operator
at site $(j_x,j_y)$. For the rest of this work, we shall set $J=1$.
The terms in Eq.~\ref{hamdef}
can be viewed as ring-exchange terms on elementary plaquettes which
convert a clockwise arrangement of $\sigma^z$ from being
$(+1,-1,+1,-1)$ to $(-1,+1,-1,+1)$ (equivalently, an arrangement of bosons
from $(1,0,1,0)$ to $(0,1,0,1)$) and vice-versa and annihilate other
arrangements on a plaquatte. It is convenient to define a vacuum state
where all sites of the lattice have down-spins, i.e., no bosons for future
reference.
This model has the following properties:
\begin{itemize}
\item The many-body spectrum of $H$ is symmetric around the energy $E=0$ for
  any finite $L_x \times L_y$ lattice with open boundary conditions (OBC).
  This is
  because the operator defined by
  \begin{eqnarray}
    \mathcal{C} = \prod_{{(j_x,j_y) \in (\mathrm{even},\mathrm{even})}}\sigma^z_{j_x,j_y}
    \label{chiralop}
  \end{eqnarray}
  satisfies $\{H,\mathcal{C}\}=0$
  where $ \prod_{{(j_x,j_y) \in (\mathrm{even},\mathrm{even})}}$ denotes a
  product over all the sites $(j_x,j_y)$ of the lattice such that both
  $j_x$ and $j_y$ are even.
  This implies that any many-body eigenstate of $H$ with an
  energy $E$ and denoted by $|E\rangle$ has a partner $\mathcal{C}|E\rangle$
  that has the energy $-E$.

\item Apart from discrete symmetries like rotations by $\pi/2$ (for
  $L_x=L_y$ lattices) and $\pi$ (for $L_x \neq L_y$ lattices),
  the model has a discrete
  reflection symmetry $\mathcal{R}$ where the axis of reflection can be taken
  to be the diagonal through $(0,0)$ for $L_x=L_y$
  or the perpendicular bisector of the longer side when
  $L_x \neq L_y$. $\mathcal{R}$ commutes with both the
  Hamiltonian $H$ and the ``chirality'' operator $\mathcal{C}$.
  This has the important consequence that the spectrum has exact
  zero modes whose number scales exponentially with the system
  size due to an index theorem shown in Ref.~\cite{Schecter2018index}.
  These zero modes are the only eigenstates of $H$ that also possess a
  definite ``chiral charge'' of $\pm 1$ under the action of $\mathcal{C}$.

\item The model conserves the total magnetization (boson number)
  defined by $S^z_{\rm tot} = \sum_{j_x,j_y} \sigma^z_{j_x,j_y}$. More
  interestingly, it conserves the following dipole moments in the
  $x$ and $y$ directions:
  \begin{eqnarray}
    D_x=\sum_{j_x,j_y} j_x \sigma^z_{j_x,j_y}, \mbox{~~~~~~~~~~}D_y=\sum_{j_x,j_y} j_y \sigma^z_{j_x,j_y}.
    \label{dipoles}
    \end{eqnarray}
  This property follows from the fact that the total magnetization on each
  column and each row of the square lattice is separately conserved under the
  dynamics induced by $H$ (Eq.~\ref{hamdef}) as pointed out earlier in
  Ref.~\cite{Khudorozhkov2021} in a similar model, but without the
  Hilbert space constraints. Models with
  the simultaneous conservation of total charge and dipole moment have been
  shown to have the property of Hilbert space fragmentation~\cite{Sala2020,
  Khemani2020}. This model is
  also fragmented due to the same reasons.

  \item The simultaneous conservation of magnetization on each
    column and each row of the $L_x \times L_y$ lattice
    also implies disorder-free localization
  for a large class of initial states. To see this,
  let us consider the vacuum state on a $L_x \times L_y$ lattice with OBC and
  then create an excitation by flipping a subset of spins to $\sigma^z_j =+1$
  such that the sites labelled by $j$ are contained inside or on the
  boundaries of
  a rectangle of finite extent smaller than the entire lattice.
  The aforementioned conservation property then
  ensures that these $\sigma^z=+1$ spins cannot be transported outside
  this bounding rectangle since
  all the
  rows/columns outside this region have their magnetizations
  pinned to their lowest possible value.

   \item This model has an exponentially large number
  (in system size) of zero modes
  that are simply inert states, i.e., Fock states in the computational basis
  that are annihilated by all the local terms in $H$, a property shared by
  other models that simultaneously conserve total charge and dipole moment.
  However, the constrained nature of the Hilbert space also leads to an
  an exponentially large number of {\it non-trivial} zero modes that emerge
  from Hilbert space fragments of various sizes
  larger than $1$,
  ranging from $3$ to $c^{L_xL_y}$, with
  $c>1$, for $L_x,L_y \gg 1$.

\item This model possesses eigenstates
  with exact non-zero integer eigenvalues when the
    Hamiltonian has the normalization of $J=1$ in
    Eq.~\ref{hamdef}. Their number scales exponentially in $L_xL_y$ for
  integer eigenvalues ranging
  from $\pm 1$ to $\pm O(\sqrt{L_xL_y})$ for $L_x,L_y \gg 1$.

  \end{itemize}

While we focus on the model Hamiltonian in
  Eq.~\ref{hamdef} with $J=1$ for the rest of the paper, it is useful to
  point out that that the fragmentation property stays {\it unchanged} even if
  $J$ is replaced by an arbitrary $J(j_x,j_y)$ and/or additional diagonal
  interactions (in the computational basis) are included.
  These only change the associated eigenvalues and
  eigenvectors but not the contributing Fock states in any of the
  Hilbert space fragments. The modified $H$ with an arbitrary $J(j_x,j_y)$ but
  no
  additional diagonal interactions still anticommutes with
  $\mathcal{C}$ (Eq.~\ref{chiralop})
  which means that the many-body spectrum continues to
  have $E$ to $-E$ symmetry.
  While the trivial zero modes of $H$ in
  Eq.~\ref{hamdef} (i.e., the inert states) persist for an
  arbitrary $J(j_x,j_y)$, the
  number of non-trivial zero modes decreases drastically due to the loss of the
  reflection symmetry $\mathcal{R}$,
  with fragments of sizes that are odd (even) integers contributing
  one (no) zero mode each. The
  presence of additional diagonal interactions destroy the
  $E$ to $-E$ symmetry of the many-body spectrum since
  $\mathcal{C}$ no longer anticommutes with the modified $H$.
  
\subsection{Quantum drums}
\label{qdrums}

Due to the structure of $H$ (Eq.~\ref{hamdef}) and the nature of the
constrained Hilbert space (Eq.~\ref{constr}), elementary plaquettes can have a
maximum of two up-spins (bosons), along any one of the two diagonals, and
these are the
only local configurations that can have any dynamics.
Furthermore, a plaquette with two up-spins (bosons) can influence the number of
possible local configurations in
neighboring two-spin plaquettes even if it can have the two up-spins (bosons)
only along one of the diagonals but not the other due to
kinematic constraints (Eq.~\ref{constr}).
These two facts lead
to the emergence of dynamically disconnected
spatial structures called quantum drums
on a $L_x \times L_y$ lattice with OBC.

To understand the origin of these drums,
let us imagine a {\it classical} Markov process in which the
{\it transition} from one Fock state to another is caused by a
ring-exchange on some elementary plaquette with two up-spins (bosons).
In the presence of the hard-core constraints
specified in Eq.~\ref{constr}, the
configuration space splits into mutually inaccessible
fragments, with all configurations within a fragment being mutually accessible
via some finite sequence of the allowed transitions. Crucially, each
such fragment can be associated with a
unique real-space structure composed of a collection of connected
elementary plaquettes that share
edges and/or vertices. The Hamiltonian $H$ (Eq.~\ref{hamdef}) acts
in the space of mutually accessible configurations of each
such fragment to generate the spectra of these quantum drums. From a dynamical
point of view, the precise nature of the quantum drums
is imprinted in the particular
initial state that the system starts from.

 We will specify two complementary construction
procedures for quantum drums below, one which starts from a given product state
in the computational basis (Sec.~\ref{drumsIS}) and the other where the
drums are constructed
recursively starting from the most elementary
one-plaquette drum (Sec.~\ref{drumsR}). Some of the
important properties of quantum drums, which will be detailed out in the
rest of the paper, are summarized below:
\begin{itemize}
  \item Quantum  drums are constructed of connected elementary plaquettes that
    share edges/vertices. A drum has no site that contains an inert
    up-spin (boson).

\item All many-body eigenstates of $H$ (Eq.~\ref{hamdef}) can be
  expressed in terms of the tensor product of eigenstates of
  appropriate quantum drums and of the remaining inert (up/down) spins,
  if any, on sites that do not belong to any quantum drum.
  This point is illustrated in detail using ED
  results in Sec.~\ref{modED}.

\item  The spectrum of a drum is uniquely fixed once
  the plaquettes that belong to it are specified. The spectrum of any
  quantum drum is symmetric around $E=0$. This follows
  from the above mentioned point and implies the existence of a corresponding
  chiral operator $\mathcal{C}_{\mathrm{drum}}$ for each drum.

\item A class of quasi-1D and 2D
  quantum drums have an internal reflection symmetry
  $\mathcal{R}_{\mathrm{drum}}$ that commutes with both $\mathcal{C}_{\mathrm{drum}}$
  and $H$  
  resulting in an exponential number
  of exact zero modes as the size of the drum is increased.

\item Any quantum drum conserves the total magnetization when only the spins
  (bosons) on the sites that belong to the drum are considered.

\item Any quantum drum satisfies an internal subsystem symmetry of simultaneous
  conservation of magnetizations along
  each column and each row (where the column and row is defined with
  respect to the background $L_x \times L_y$ lattice) of the drum.

  \end{itemize}

\subsubsection{Constructing quantum drums associated with an initial
  product state}
\label{drumsIS}

\begin{figure}[!htb]
  \begin{center}
    \includegraphics[width=0.6\linewidth]{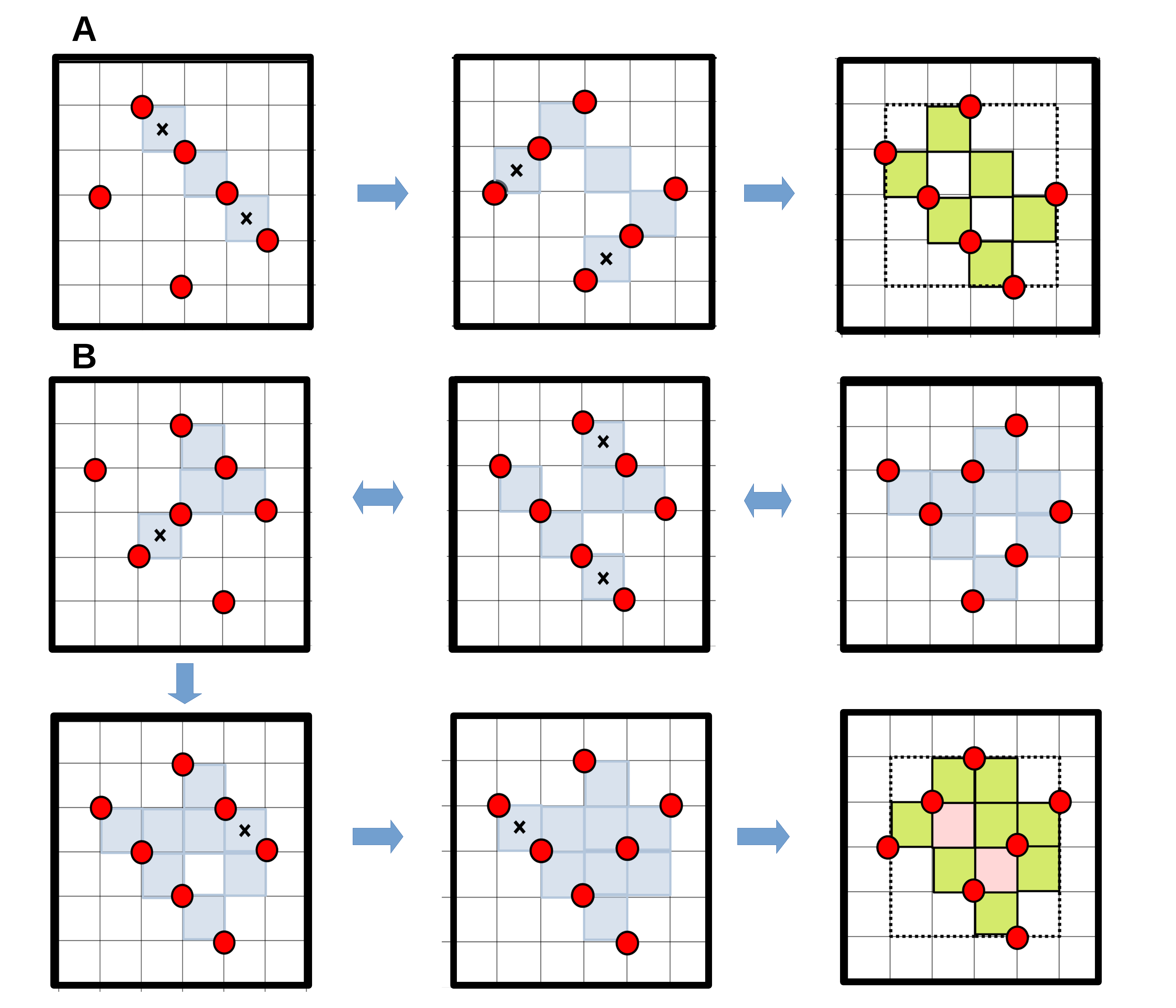}
    \caption{Illustration of the recursive construction of a quantum
      drum given two different initial states (marked by A and B in the figure)
      on a $7 \times 7$ lattice with
      OBC where the filled red dots indicate up-spins.
      The shaded plaquettes form part of a drum with the top-right and bottom-right panels indicating the drums for the initial states marked A and B
      respectively. A cross at
      the center of a plaquette indicates that a ring-exchange move is
      carried out for that plaquette. The green plaquettes in the
      top-right panel and the green and pink plaquettes in the
      bottom-right panel follow the same convention as
    used in Fig.~\ref{drumsandshields}.} \label{drumsfromIS}
\end{center}
\end{figure}

We first give a construction procedure that fixes all the
quantum drums given a classical Fock state on a $L_x \times L_y$ lattice with
OBC. An initial Fock state and its associated drums are
  shown in Fig.~\ref{drumsandshields}. The construction procedure is
  schematically shown in Fig.~\ref{drumsfromIS} for two drums starting
  from different Fock states.
Given the Fock state, firstly all plaquettes with two up-spins (bosons)
are shaded. Ring-exchange moves are attempted on such plaquettes
to see whether any
additional plaquettes with two up-spins (bosons) are generated which are again
shaded. This process is repeated with the newly shaded plaquettes
until no additional shaded plaquettes are generated.
The shaded plaquettes are then subdivided into connected regions
that comprise of elementary plaquettes that share edges and/or vertices.
A final check has to be performed on each of these connected regions
separately to construct the quantum drums. If the mutually accessible
Fock states from a connected region have certain sites where any
up-spin (boson) remains the same in each of the configurations, these
up-spins (bosons) are then labelled as inert and the shaded plaquettes
containing any inert up-spins (bosons) are unshaded. The remaining shaded
plaquettes that are still connected to each other
via an edge or a vertex forms a quantum drum. This last check is necessary
to rule out inert structures made entirely of plaquettes with
two up-spins (bosons) [see Fig.~\ref{drumsandshields} for an example composed of
three up-spins (bosons) on two edge-sharing plaquettes]
and to find spatial structures that can be decomposed into an inert region of
up-spins (bosons) and
a smaller quantum drum [see Fig.~\ref{drumsandshields}, bottom right for an
example of
such a decomposition].


Two simple examples of this construction are given for initial
classical Fock states on a $7 \times 7$ lattice in Fig.~\ref{drumsfromIS}
(top-left and middle-left panels), where
the filled circles indicate up-spins (bosons) while the other sites have
down-spins (no bosons).
Let us first consider the top three panels. The initial state is given in the
top-left panel marked as A and three plaquettes are shaded at this stage.
Implementing ring-exchange moves on two of the shaded plaquettes indicated by
crosses in the top-left panel
generates two more shaded plaquettes as shown in top-middle panel. Implementing
ring-exchange moves on the shaded plaquettes indicated by crosses in that panel
generates another shaded plaquette in the top-right panel and further
ring-exchanges do not generate any additional shaded plaquettes.
The quantum drum generated by this initial state only contains
elementary plaquettes that share vertices.

The initial state in the middle-left panel marked by B
gives four shaded plaquettes. Implementing ring-exchange to this state for
the plaquette indicated by a cross generates two more shaded plaquettes as
shown in the following panel to the right. Carrying out ring-exchange moves on
two more plaquettes as indicated by crosses generates two additional shaded
plaquettes. To generate the other two shaded plaquettes that form the entire
quantum drum, we go back to the initial Fock state shown in the bottom-left
panel and perform two ring exchange moves on the plaquettes indicated by a
cross one after the other as indicated in the bottom panels. The resulting
quantum drum consists of only edge-sharing plaquettes in this case.

Both the quantum drums shown in Fig.~\ref{drumsfromIS} generate Hilbert
space fragments of size $11$ respectively, diagonalizing which
results in the
following eigenvalues:
\begin{eqnarray}
  \left ( \pm \sqrt{\frac{1}{2} \left(9\pm \sqrt{57} \right)}, \pm \sqrt{3}, \pm \sqrt{2}, 0, 0, 0\right) \nonumber \\
  \left(\pm 2\sqrt{2},\pm \sqrt{3}, -1, -1, +1, +1, 0,0,0 \right)
  \label{eigvaluesoftwodrums}
  \end{eqnarray}
where the top (bottom) line in Eq.~\ref{eigvaluesoftwodrums} refers to the
eigenvalues for the quantum drum shown in the top-right (bottom-right) panel of
Fig.~\ref{drumsfromIS}. These two examples already illustrate that drums can
have non-trivial zero modes, nonzero integer-valued eigenstates as well as
eigenstates with irrational eigenvalues. We refer the reader
  to Sec.~\ref{astud} for the explicit construction of the Hilbert space
  fragments associated with some small quantum drums.

\subsubsection{Shielding region and closest approach of drums}
\label{drumsShields}
Each quantum drum is associated with a shielding region of its own such that
two quantum drums can fluctuate independently as long as the boundaries of
their shielding regions do not cross. Given any Fock state
consistent with a single quantum drum composed of a finite number of
elementary plaquettes with the rest of the sites that do not
belong to the drum being $\sigma^z=-1$ (no bosons), the corresponding
shielding region can again be fixed by a classical construction.
For this purpose, let us define 
  \begin{eqnarray}
n_{\Box_j}= \mathrm{max} [2 + (\sigma^z_{j_x,j_y} +
  \sigma^z_{j_x+1,j_y} + \sigma^z_{j_x,j_y+1} + \sigma^z_{j_x+1,j_y+1})/2]
  \label{defshield}
  \end{eqnarray}
  where $n_{\Box_j}$ is computed using {\it all} the Fock states that are
  accessible from the starting Fock state by ring-exchanges on elementary
  plaquettes. By definition (see Sec.~\ref{drumsIS}), all plaquettes that
  belong to a quantum drum have $n_{\Box_j}=2$. For a quantum drum embedded in
  the vacuum state, all other elementary plaquettes must have
  either $n_{\Box_j}=0$ or $1$. Importantly, only those plaquettes that
  lie to the exterior of the drum and directly share
  an edge or a vertex with the plaquettes on the perimeter of the drum
  {\it can} have
  $n_{\Box_j}=1$ while all other exterior plaquettes necessarily have
  $n_{\Box_j}=0$. To identify the subset of exterior plaquettes with
  $n_{\Box_j}=1$ requires constructing all the Fock states accessible to the
  given quantum drum since certain plaquettes with $n_{\Box_j}=2$ may allow
  two up-spins (bosons) only along one diagonal and not the other (e.g., see
  Fig.~\ref{drumsandshields} for such plaquettes that are
  labeled by ``u'' and also
  indicated in pink.) 

\begin{figure}[!htb]
  \begin{center}
    \includegraphics[width=0.65\linewidth]{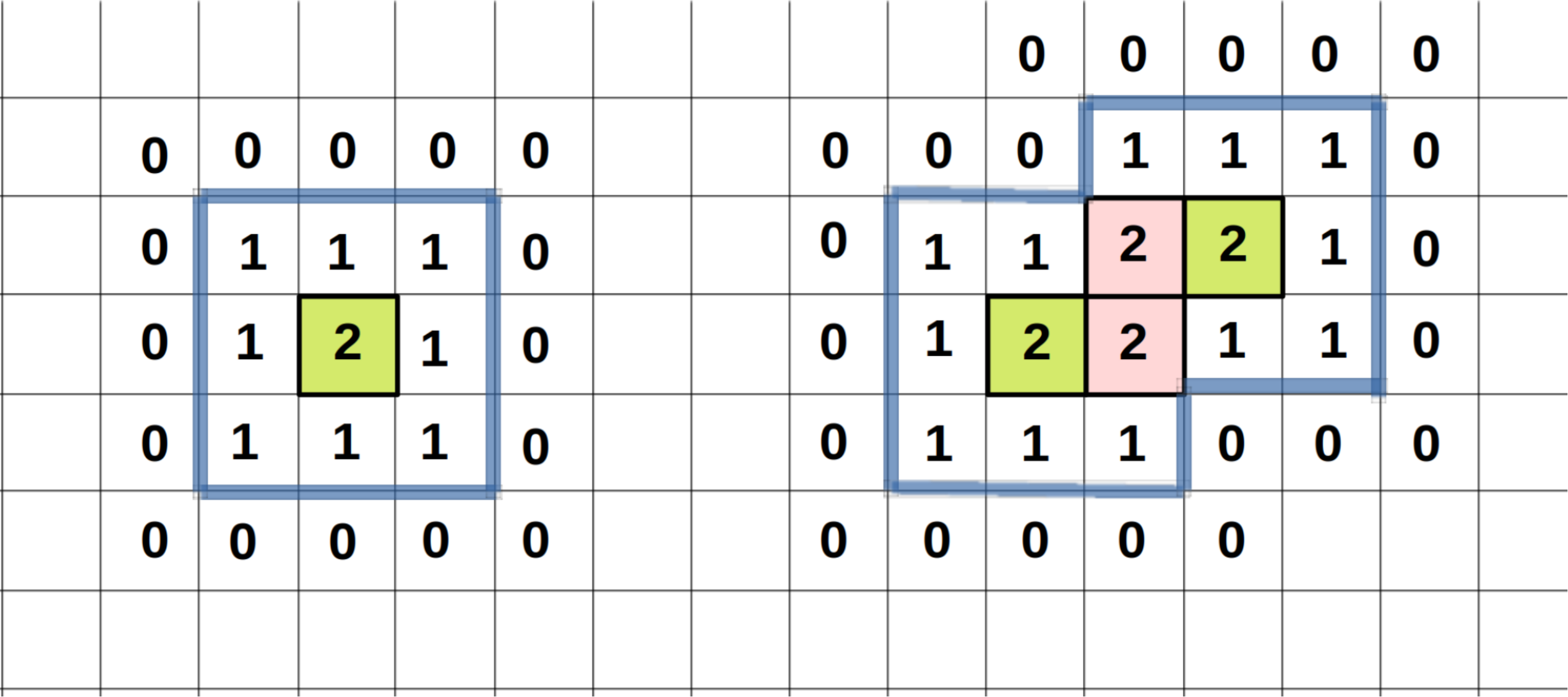}%
    \caption{Two quantum drums are shown in the left and the right panels.
      The integers shown inside each plaquette refers to
      $n_{\Box_j}$ given by Eq.~\ref{defshield} on each
      plaquette of the lattice using all the Fock states generated
      when the quantum drum is embedded in the vacuum state. The green
      (pink) plaquettes in the two quantum drums follow the same convention as
      used in Fig.~\ref{drumsandshields}.
      The boundary of the shielding region of both
      drums is shown using thick blue lines in both
        panels.} \label{shields}
\end{center}
\end{figure}

The shielding
region of a quantum drum composed of a finite number of
  elementary plaquettes
only consists of these exterior plaquettes
with $n_{\Box_j}=1$ that
directly
share edges/vertices with the plaquettes on the perimeter of a quantum drum
irrespective of the size of the drum. Thus, the thickness of the shielding
region {\em does not} scale with the size of the quantum drum and remains
$O(1)$ in lattice units (see Fig.~\ref{drumsandshields} and
Fig.~\ref{shields} for examples).
The boundary of the shielding region is defined as the
  closed curve formed by the edges that are common to the the exterior
  plaquettes with $n_{\Box_j}=1$ and $n_{\Box_j}=0$. The sites belonging to this
boundary do not carry any $\sigma^z=+1$ spins (bosons).

Let us illustrate the construction of the shielding regions
  using two examples.
First consider an elementary one-plaquette quantum drum starting
from the vacuum state and then placing
two $\sigma^z=+1$ spins (bosons) along any one of the diagonals of an elementary
plaquette. Given this Fock state, ring-exchange is possible only on this
elementary plaquette which then generates another Fock state where the
$\sigma_z=+1$ spins (bosons) get transported to the other diagonal of this
plaquette. Considering both these Fock states to compute
$n_{\Box_j}$ (Eq.~\ref{defshield}) on each
plaquette of the lattice, we see that $n_{\Box_j}=2$ for the flippable plaquette
which is surrounded by $n_{\Box_j}=1$ and $n_{\Box_j}=0$ plaquettes, respectively
(Fig.~\ref{shields}, left panel).
The $n_{\Box_j}=2$ plaquette defines the quantum drum while the $n_{\Box_j}=1$
plaquettes along the perimeter of the quantum drum define the
shielding region associated with this drum. The shielding region terminates
at the boundary of these $n_{\Box_j}=1$ and the $n_{\Box_j}=0$ plaquettes
(Fig.~\ref{shields}, left panel). By construction, the sites at the boundary
of the shielding region
cannot have $\sigma^z=+1$ spins (bosons).  A more complicated quantum drum
is shown in Fig.~\ref{shields}, right panel which can be generated from the
vacuum state by, e.g., placing two $\sigma^z=+1$ spins
(bosons) each along parallel diagonals of the left-most and the right-most
plaquette contained in the quantum drum such that the hard-core constraints
are not violated. Performing
all possible ring-exchanges for this quantum drum generates two more Fock
states. Considering these three Fock states to compute
$n_{\Box_j}$ on each
plaquette of the lattice, the four $n_{\Box_j}=2$ plaquettes, which are
all connected to each other by edges for this particular drum,
now define this bigger quantum drum (Fig.~\ref{shields}, right panel)
while the $n_{\Box_j}=1$
plaquettes along the perimeter of the quantum drum define the
shielding region as before (Fig.~\ref{shields}, right panel).
The shielding region is more complicated compared to the
one-plaquette drum and its boundary is again defined by the
boundary of the $n_{\Box_j}=1$ and the $n_{\Box_j}=0$
plaquettes (Fig.~\ref{shields}, right panel). This classical
construction procedure for the shielding region can be carried out for any
arbitrary quantum drum composed of a finite number of elementary plaquettes.

We can now ask
for the closest approach of any two quantum drums embedded
in the vacuum state
such that both the
drums can be viewed to be independent of each other.
Up-spins (bosons) cannot be transported outside the
  closed boundary of the shielding region of a quantum drum. Since the boundary
  sites do not carry any up-spin (boson), two quantum drums stay
  independent of each other as long as the boundaries of their corresponding 
  shielding regions do not intersect; they may at most touch each other.
For example, this is the case in
Fig.~\ref{drumsandshields} which explains
why the different quantum drums can be considered to be independent of
each other. When the boundaries of
  the shielding regions first
cross each other, the corresponding quantum drums
in
their interior have to necessary change 
according to one of the following three possibilities: (i) the two quantum
drums fuse to produce a bigger quantum drum, (ii) a spatial structure is
produced such that it can be decomposed into an inert region of
up-spins (bosons) and a smaller quantum drum, and (iii) a fully inert region of
up-spins (bosons) is formed.

  \subsubsection{Recursive construction of bigger drums from smaller drums}
  
  \label{drumsR}

We now present a complementary drum construction procedure to the one
explained in Sec.~\ref{drumsIS} which does not need the specification of
a product state on the entire $L_x \times L_y$ lattice. Instead, this
recursive construction creates larger
drums starting from smaller ones. We start with a Fock state consistent with
a particular quantum drum composed of certain connected plaquettes, where the
sites that do not belong to the drum are assigned $\sigma^z=-1$ spins
(no bosons). This is equivalent to embedding the quantum drum in the vacuum
state.
By definition, the boundary of the shielding region of such a drum has
$\sigma^z=-1$ (no bosons). A natural way to construct bigger drums is to
choose a subset of the sites that belong to this boundary of the shielding
region and then
replace $\sigma^z=-1$ by $\sigma^z=+1$ at these selected sites.
This generates a new Fock state from which, using the
procedure of identifying a drum from a Fock state explained in
Sec.~\ref{drumsIS},
one gets one of the
following three possibilities: (i) a bigger
quantum drum with no inert up-spins (bosons), (ii) a partially active
structure that can be decomposed into a smaller quantum drum
and a non-zero number of inert up-spins (bosons), and
(iii) a completely frozen structure with all up-spins (bosons) being inert.

\begin{figure}[!htb]
  \begin{center}
    \includegraphics[width=0.9\linewidth]{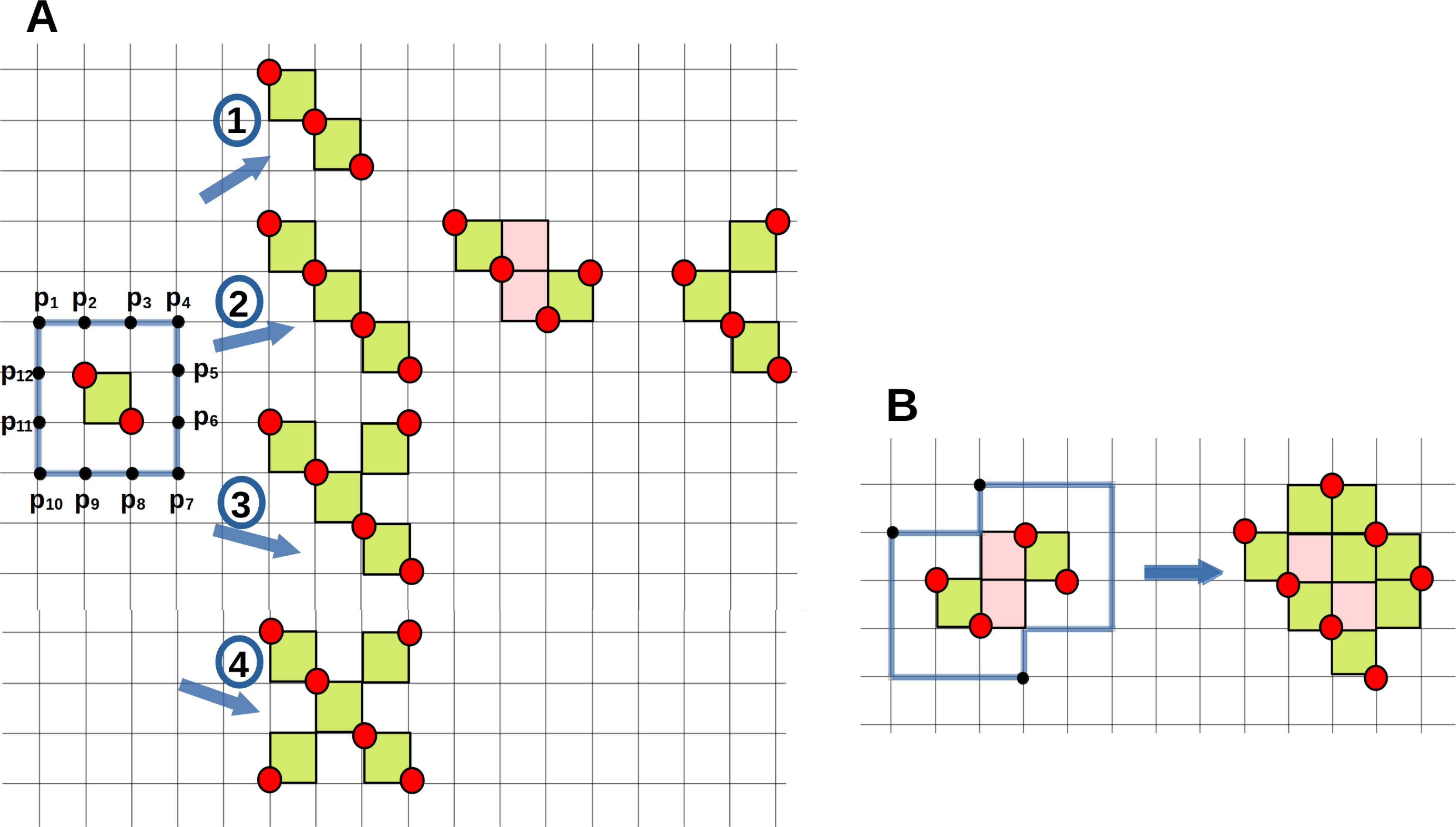}%
    \caption{The left panel shows how adding up-spins (bosons) to the sites
      at the boundary of the shielding region (indicated by thick blue lines
      and also labelled from p$_1$ to p$_{12}$)
      of an elementary
    one-plaquette drum (marked as A) leads to a variety of larger
    quantum drums which are grouped according to the
      addition of \textcircled{$n$}
      up-spins (bosons) where
    $n=1,2,3,4$. The right panel illustrates the same concept for a more
    complicated quantum drum (marked as B) where three up-spins
    (bosons) (indicated by
    black dots) are added to three sites of the boundary of the shielding
    region (indicated by thick blue lines).
    Filled red dots indicate up-spins (bosons). The
      green (pink) plaquettes in all the quantum drums
      follow the same convention as
    used in Fig.~\ref{drumsandshields}.
     } \label{Figrecursion}
\end{center}
\end{figure}

Two examples of this recursive construction to generate bigger drums starting
from a smaller drum are shown in Fig.~\ref{Figrecursion}.
We first start with a Fock state
consistent with an elementary one-plaquette drum in the left panel of
Fig.~\ref{Figrecursion}, marked as A. In this case, the boundary
of the shielding region is a square that
consists of twelve sites, labelled as p$_1$, $\cdots$,
p$_{12}$ in Fig.~\ref{Figrecursion}.
New Fock states, consistent with larger drums, can be created by
adding one/two/three or four up-spins (bosons) in this boundary region as
indicated by the groups labelled by \textcircled{$1$}, \textcircled{$2$},
\textcircled{$3$} and \textcircled{$4$} in Fig.~\ref{Figrecursion}, left panel.
Adding a single up-spin (boson) at p$_1$ or p$_4$
generates Fock states consistent with a
drum composed of two elementary plaquettes that share a vertex as
shown in Fig.~\ref{Figrecursion}, left panel, group labelled by
\textcircled{$1$}.
Adding two up-spins
(bosons) on the boundary of the shielding region in different ways
leads leads to Fock states consistent with three different
drums as shown in Fig.~\ref{Figrecursion}, left panel, group
labelled by \textcircled{$2$}. E.g.,
adding up-spins (bosons) at
p$_1$ and p$_7$ leads to a Fock state consistent with a drum with three
plaquettes that share vertices along
a single diagonal (leftmost drum shown in the group labelled by
\textcircled{$2$} in left panel of Fig.~\ref{Figrecursion}),
at p$_1$ and p$_5$ leads to a Fock state consistent with a
drum with four plaquettes that are connected by edges
(middle drum shown in the group labelled by
\textcircled{$2$} in left panel of Fig.~\ref{Figrecursion}),
and at p$_4$ and p$_7$
leads to a Fock state consistent with a drum with three plaquettes that again
share vertices, but not along a single diagonal
(rightmost drum shown in the group labelled by
\textcircled{$2$} in left panel of Fig.~\ref{Figrecursion})).
Adding three up-spins (bosons)
at, e.g., p$_1$, p$_4$ and p$_7$, leads to a Fock state consistent with
a quantum drum with four plaquettes
that are connected by vertices as shown in the group labelled by
\textcircled{$3$} in left panel of Fig.~\ref{Figrecursion}.
Finally, adding four up-spins (bosons) at
p$_1$, p$_4$, p$_7$, and p$_{10}$ leads to a Fock state consistent with a
quantum drum with five plaquettes connected by
vertices (Fig.~\ref{Figrecursion}, left panel, group labelled by
\textcircled{$4$}). To illustrate
possibility (ii), we can add four up-spins (bosons)
at p$_1$, p$_3$, p$_7$ and p$_{11}$ (Fig.~\ref{Figrecursion}, left panel)
which leads to a Fock state consistent with a single-plaquette
quantum drum containing sites p$_6$, p$_7$ and p$_8$ while the other up-spins
(bosons) become inert. To illustrate possibility (iii), we can add a single
up-spin (boson) at p$_5$ (Fig.~\ref{Figrecursion}, left panel)
to generate a Fock state that has only inert up-spins (bosons).

This recursive procedure can be carried forth for the bigger
quantum drums to produce more complicated quantum drums. An example is shown in
panel marked as B (Fig.~\ref{Figrecursion}, right panel) where three up-spins
(bosons), indicated by filled black dots, are
placed on the boundary of the shielding region of a quantum drum, previously
produced by adding two up-spins (bosons) at
the boundary of the shielding region of the elementary single-plaquette drum,
which leads to a bigger quantum drum with ten elementary plaquettes that are
connected by edges. In principle, this recursive procedure can be used to
generate and enumerate all possible quantum drums until a given stage of
the recursion starting from the most elementary
one-plaquette drum, but we leave this for a possible future investigation.

\subsubsection{Wires, junctions of wires, other quasi-1D and 2D drums}
\label{largedrums}

As is already evident from the examples we have constructed so far,
quantum drums come in several shapes and sizes, from being composed of a
single elementary plaquette (Fig.~\ref{drumsandshields}) to
  a finite number of plaquettes
  (Fig.~\ref{drumsandshields} and Fig.~\ref{drumsfromIS}).
  One can even construct quantum drums with an
  arbitrarily large number of plaquettes in the thermodynamic limit. These
  varieties of drums can be quasi-1D or 2D in nature. We dub the
  {\it simplest} quasi-1D drum as a wire. A wire is
composed of $N_p$ plaquettes that share vertices along a single diagonal and
resemble straight wires (see Fig.~\ref{Figrecursion}, left panel
for three such drums
with $N_p=1,2,3$). Such a wire can be constructed with
any $N_p \ge 1$ that leads to a quasi-1D structure for $N_p \gg 1$.

\begin{figure}[!htb]
  \begin{center}
    \includegraphics[width=0.7\linewidth]{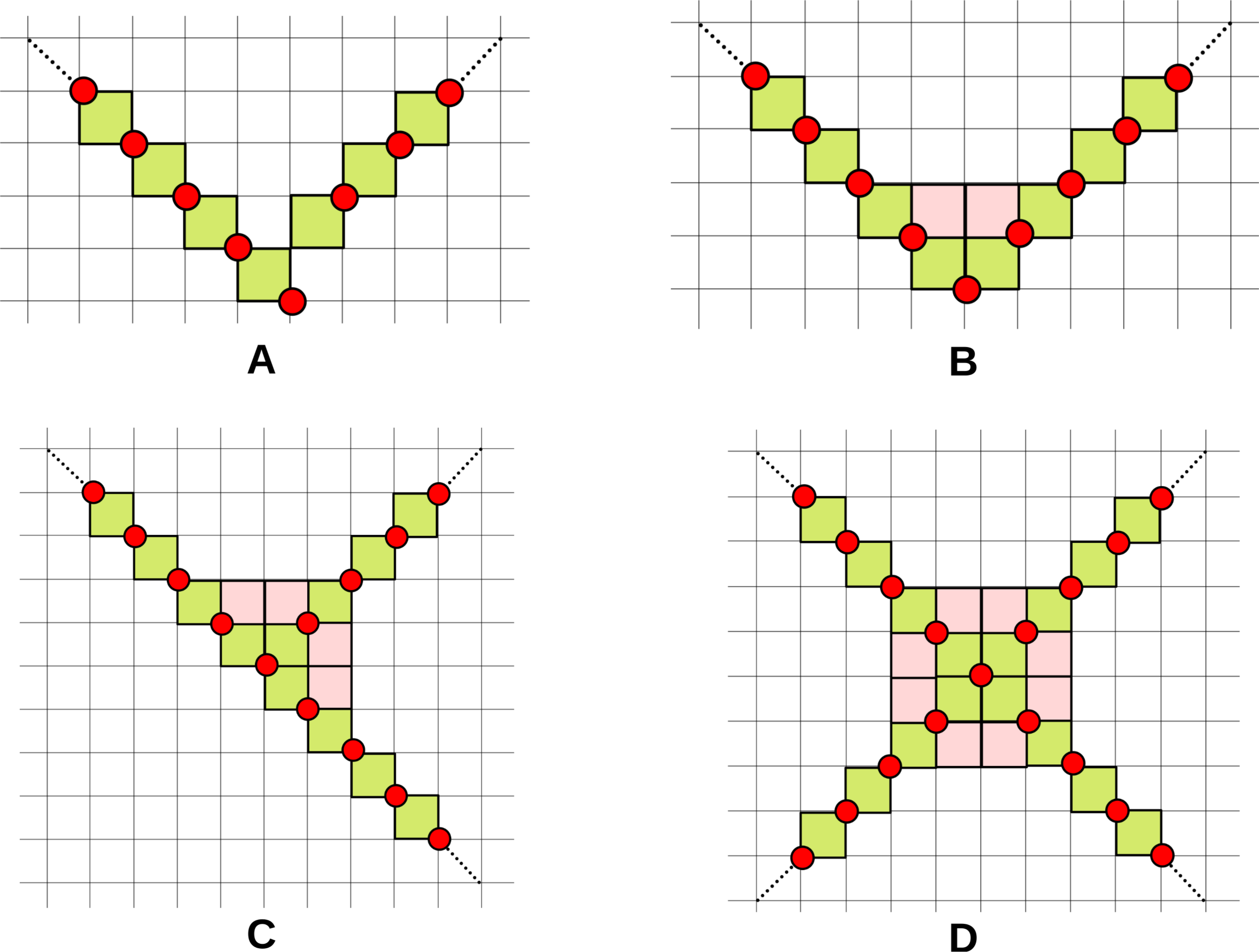}
    \caption{The quantum drums marked from A-D can be
      viewed as examples of different kinds of junctions of wires. A and B
      show examples of junctions of two wires while C (D) shows an
      example of a junction of three (four) wires.
      The
      green (pink) plaquettes in all the quantum drums
      follow the same convention as
    used in Fig.~\ref{drumsandshields}.
    The filled red dots indicate up-spins in all the panels
    and represent just one of the many
      possible Fock states of the corresponding
      drum.
    } \label{drumvarieties}
\end{center}
\end{figure}

Interestingly, one can create other quantum drums that resemble
different kinds of junctions of such wires. Examples of such
quantum drums are shown in Fig.~\ref{drumvarieties}.
In the top panel, the quantum drums marked by A and B can be
viewed as two different junctions of two wires, while in the bottom panel,
the quantum drum marked by C (D) can be viewed as a junction of three (four)
quantum wires.
Wires can be used to build still more intricate quasi-1D as well as
2D drums (see Sec.~\ref{wiresdecom} for details).
The fragment sizes for large quasi-1D (2D) drums scale as
$\alpha^l$ ($\beta^{l^2}$)
where $\alpha >1$ ($\beta >1$) as $l \gg 1$
where $l$ represents the linear dimension
of the drum and $\alpha$ ($\beta$) depend
on the nature of the quantum drum under consideration.
Each such quantum drum can be viewed as
an interesting example of an interacting quasi-1D/2D model with
a constrained Hilbert space that also satisfies an internal subsystem symmetry
of simultaneous conservation of magnetizations along each column and each row
of the drum,
where the columns/rows are defined with respect to the
$L_x \times L_y$ lattice in which the drum is embedded,
when only the sites that
belong to the drum are considered.

\subsection{Exact diagonalization of small lattices and deciphering spectrum using drums}
\label{modED}

The constrained nature of the Hilbert space reduces the number of allowed Fock
states from $2^{L^2}$ to $\kappa^{L^2}$ where $\kappa \approx 1.503\cdots$
is the hard square entropy constant~\cite{Baxter1999} for a square lattice
with $L \gg 1$. This
growth of the Hilbert space dimension with $L$ is, nonetheless,
still too large to perform ED for the full spectrum
for even moderately large values of $L$. However, analysing the
numerical results for small $L \times L$ lattices is already instructive.

\begin{figure}[!htb]
  \begin{center}
    \includegraphics[width=0.5\linewidth]{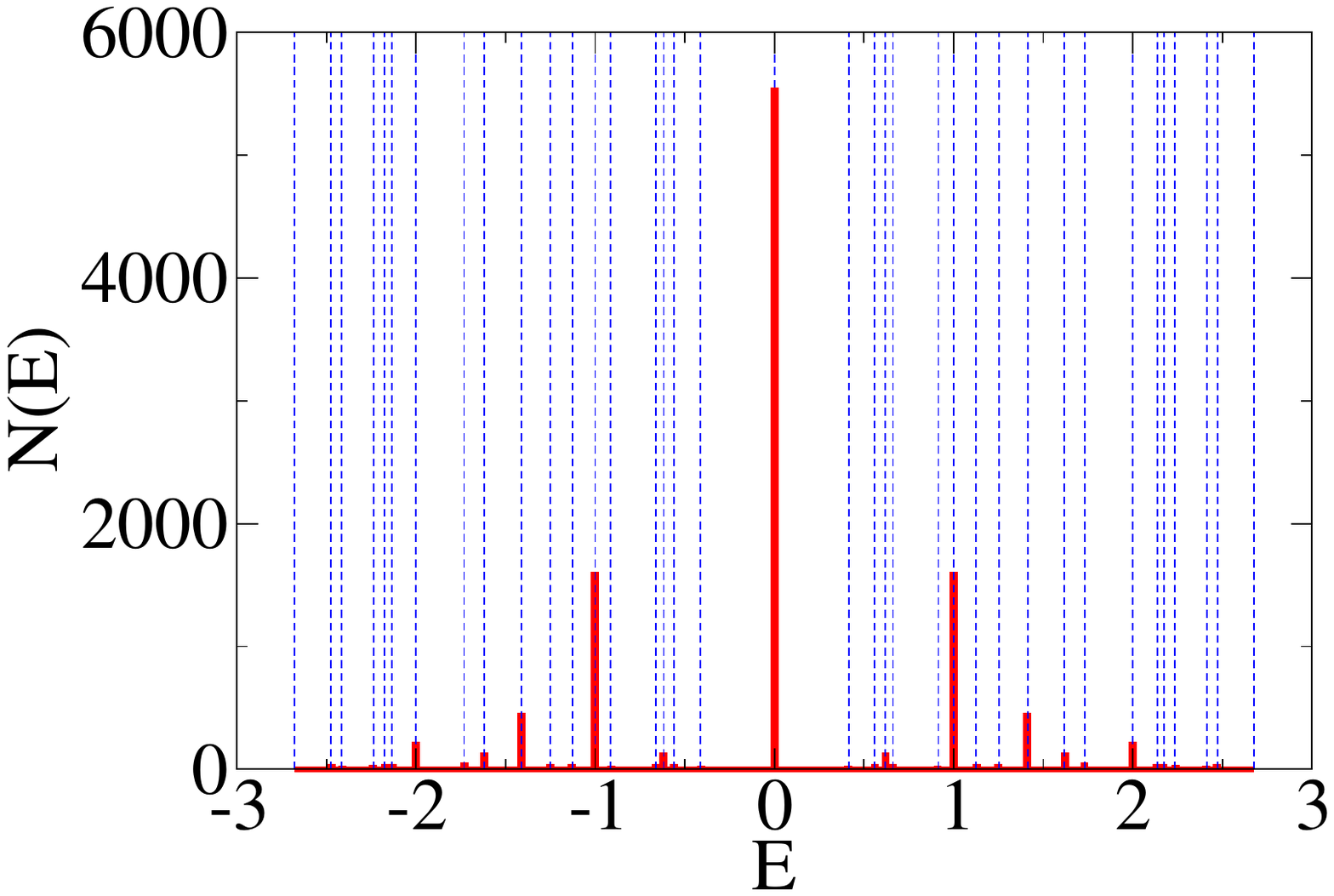}%
    \includegraphics[width=0.42\linewidth]{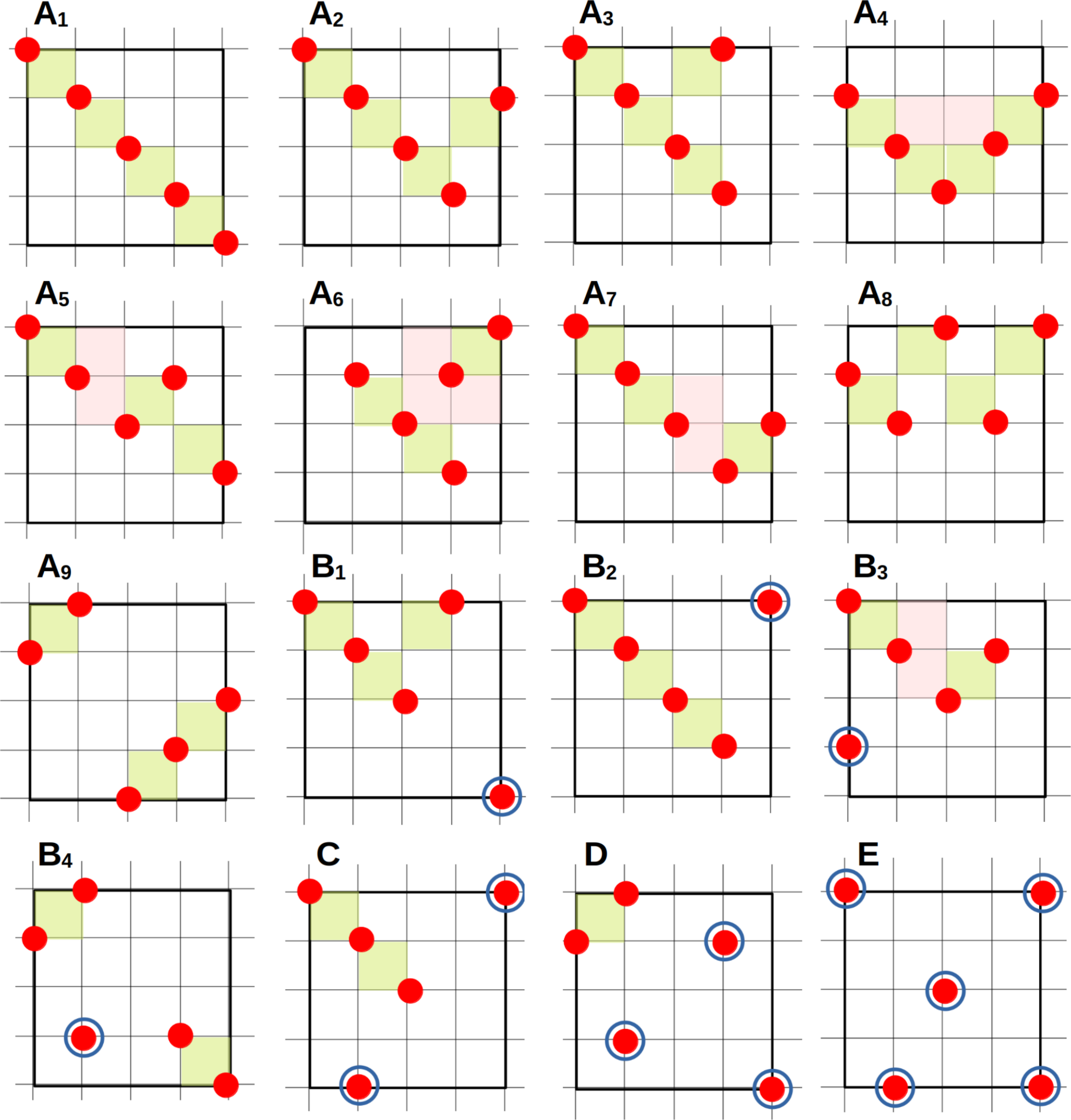}
    \caption{(Left panel) Histogram of the energy eigenvalues for a $5 \times 5$
      lattice with $5$ up-spins and OBC. The vertical dotted blue lines indicate
      the allowed eigenvalues from Eq.~\ref{smalldrums}. (Right panel) The
      allowed quantum drums in this system where the up-spins (bosons)
      are indicated
      by filled red dots and the plaquettes that belong to quantum drums are
      shaded. The
      green (pink) plaquettes in all the quantum drums
      follow the same convention as
      used in Fig.~\ref{drumsandshields}.
      The inert up-spins (bosons) are indicated by a blue circle
      around the
      filled red dot. Each quantum drum is consistent with more than one
      Fock state with only one representative Fock state shown
      here. The different eigenstates can be viewed as modes of
    these quantum drums.}
    \label{FIGfromED}
\end{center}
\end{figure}

Let us first consider a $5 \times 5$ lattice and focus on the total
magnetization sector with $5$ up-spins (bosons). This
gives a Hilbert space dimension of $10741$ from direct enumeration
taking the hard-core constraints in account.
Plotting the histogram of the energy eigenvalues obtained from full ED
reveals that the eigenvalues are clustered around only a {\it few}
special values (up to machine precision) (Fig.~\ref{FIGfromED}, left panel)
unlike what is expected of a
generic interacting system with a similar Hilbert space dimension. Furthermore,
while an explicit construction shows that there are $4559$ inert Fock states
that are trivially annihilated by $H$ (Eq.~\ref{hamdef})
in this magnetization sector, ED reveals
that there are a total of $5525$ zero modes (with zero eigenvalue within
machine precision) implying the presence of $966$
non-trivial zero modes. ED also shows the presence of $1580$ eigenmodes
with eigenvalue $+1$ ($-1$)
and $196$ eigenmodes with eigenvalue $+2$ ($-2$). Such non-zero integer
eigenvalues are unexpected in generic interacting models which have
highly irrational eigenvalues
that cannot be expressed in any simple closed form.
These and other features of the full ED data can be completely
understood in terms
of quantum drums (Fig.~\ref{FIGfromED}, right panel).
The $10741$-dimensional Hilbert space in the
  computational basis
gets fragmented into
$4559$ ($1$-dimensional), $1552$ ($2$-dimensional), $434$ ($3$-dimensional),
$324$ ($4$-dimensional), $32$ ($5$-dimensional), $32$ ($6$-dimensional), $16$
($7$-dimensional) and $2$ ($8$-dimensional) Hilbert space fragments. The
$1$-dimensional fragments simply correspond to the inert Fock states that
are annihilated by all local terms of $H$ (Eq.~\ref{hamdef})
(and are denoted collectively
by panel marked E in Fig.~\ref{FIGfromED}, right panel).
All the other fragments can be viewed as being generated from
a collection of appropriate
quantum drums and any remaining inert up-spins (bosons)
that are not part of
any drum.

Given the lattice dimensions and the number of up-spins (bosons),
only certain drums are allowed with specific degeneracies set by the lattice.
The different drum configurations
  along with one representative Fock state for each
  is shown in Fig.~\ref{FIGfromED} (right panel)
  and are marked as A$_1$, $\cdots$, A$_9$, B$_1$, $\cdots$, B$_4$, C and D.
  The configurations labeled from A$_1$ to A$_8$ consist of different
  types of single drums
  that contains all the $5$ up-spins (bosons) while A$_9$ consist of two
  independent drums, one with $2$ up-spins (bosons) and another with
  $3$ up-spins (bosons), respectively. The configurations B$_1$, B$_2$ and
  B$_3$ have a single quantum drum each with $4$ up-spins (bosons) while
  $1$ up-spin (boson) is inert as it does not belong to the drum. The
  configuration B$_4$ has two independent one-plaquette drums and $1$
  inert up-spin (boson) that does not belong to any of the $2$ drums.
  Configuration C (D) has a drum with $3$ ($2$)
  up-spins (bosons) and $2$ ($3$) inert up-spins (bosons).

The degeneracies associated with the different configurations
A$_1$ to E are indicated
inside $[]$ for each case in
Eq.~\ref{smalldrums} and arise from the number of distinct ways in
which the given drums and any inert up-spins (bosons)
can be placed on the $5 \times 5$ lattice with
OBC. For example, A$_1$ has a degeneracy of two because there are two diagonals
along which the associated drum may be placed.
Similarly, A$_2$ has a degeneracy of
sixteen since there are sixteen distinct ways to place a ``L'' composed
of four connected plaquettes that form the associated drum
on this lattice. The other degeneracies given in
Eq.~\ref{smalldrums} can be computed similarly. 
  The spectrum shown for
A$_1$ to E in Eq.~\ref{smalldrums} can be straightforwardly calculated
by solving for the spectra of the constituent drums.
The
eigenspectra of all the fragments that arise from these
quantum drums, barring the drum shown in A$_2$, can be expressed
in closed form and show a variety of eigenvalues including
zero modes, non-zero integer modes and
irrational modes (Eq.~\ref{smalldrums}). 
The extra non-trivial zero modes and their degeneracies
can also be understood as
zero modes of quantum drums shown in A$_1$, A$_2$, A$_3$, A$_4$, A$_6$,
A$_8$, B$_2$, B$_3$, B$_4$ and
C (Eq.~\ref{smalldrums}). It is useful to stress here that while
certain drums, e.g., the one contained in B$_3$ and the one contained
in C, are evidently different from each other, they
have identical spectra (Eq.~\ref{smalldrums}).

\begin{eqnarray}
  \mathrm{A}_1[2] &\rightarrow& (\pm \sqrt{4+\sqrt{10}}, \pm \sqrt{2}, \pm \sqrt{4-\sqrt{10}}, 0, 0 )  \nonumber \\
  \mathrm{A}_2[16] &\rightarrow& (\pm 2.47367\cdots, \pm 1.25235\cdots,\pm 0.559107\cdots, 0) \nonumber \\
  \mathrm{A}_3[16] &\rightarrow& (\pm \sqrt{3+\sqrt{3}}, \pm \sqrt{3-\sqrt{3}}, 0, 0) \nonumber \\
  \mathrm{A}_4[12] &\rightarrow& (\pm \sqrt{5}, \pm 1, 0, 0), \mbox{~~~~~~~~} \mathrm{A}_5[8], \mathrm{A}_7[8], \mathrm{B}_1[96]  \rightarrow \left(\pm \frac{1}{2}(1 \pm \sqrt{5})\right) \nonumber \\
  \mathrm{A}_6[16] & \rightarrow& (\pm \sqrt{3}, 0, 0), \mbox{~~~~~~~~} \mathrm{A}_8[12] \rightarrow  (\pm \sqrt{3}, \pm 1, 0)\nonumber \\
  \mathrm{A}_9[4] &\rightarrow& (\pm (1\pm \sqrt{2}), \pm 1), \mbox{~~~~~~~~} \mathrm{B}_2[20] \rightarrow \left( \pm \sqrt{\frac{1}{2}(5 \pm \sqrt{17})}, 0\right) \nonumber \\
  \mathrm{B}_3[128], \mathrm{C}[306] &\rightarrow& (\pm \sqrt{2},0), \mbox{~~~~~~~~} \mathrm{B}_4[196] \rightarrow (\pm 2, 0,0) \nonumber \\ 
  \mathrm{D}[1552] &\rightarrow& \pm 1, \mbox{~~~~~}\mathrm{E}[4559] \rightarrow 0
  \label{smalldrums}
  \end{eqnarray}

\subsection{Eigenstates with integer energies from packing of one-plaquette drums}
\label{intstates}

\begin{figure}[!htb]
  \begin{center}
    \includegraphics[width=0.4\linewidth]{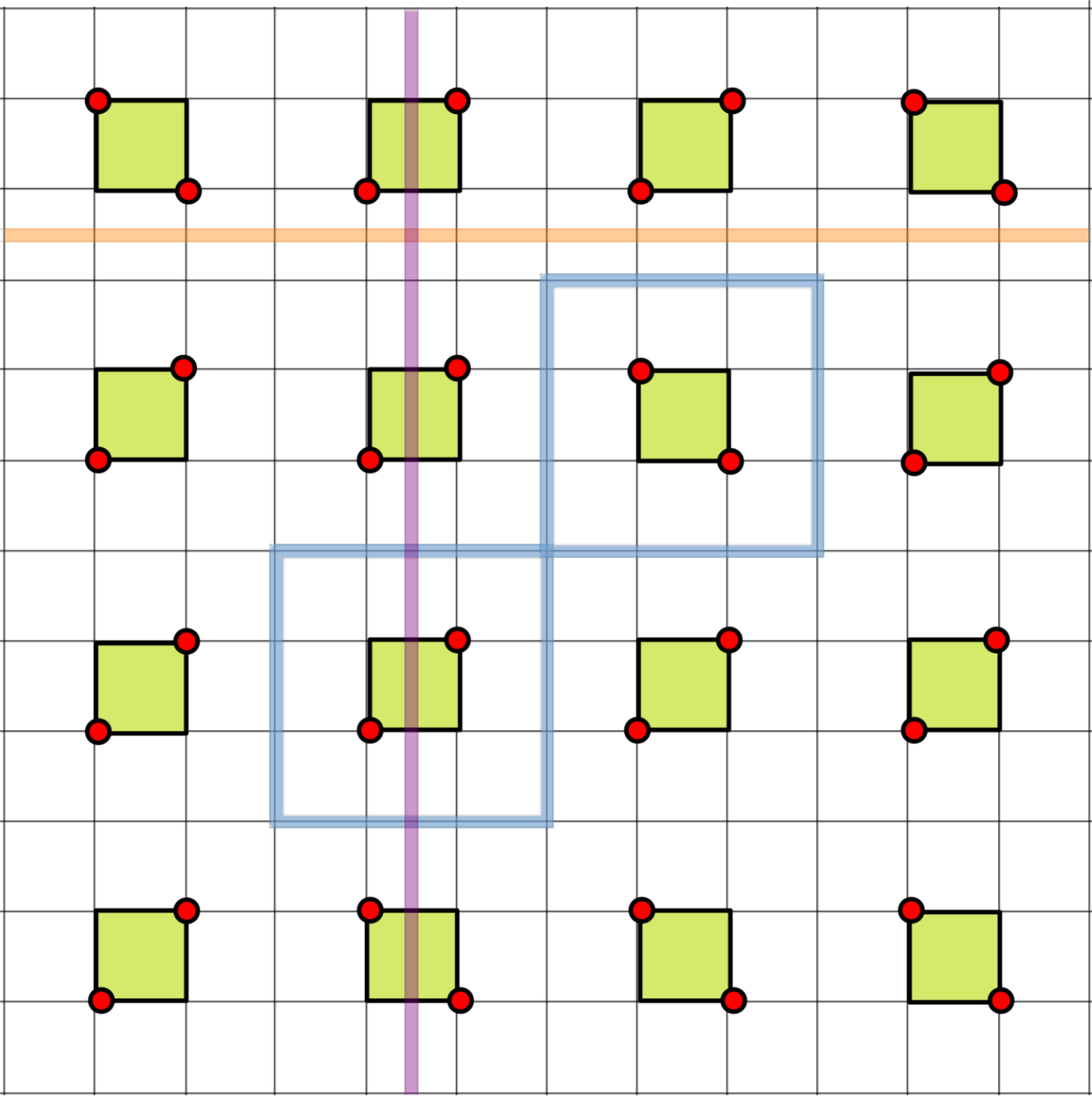}   
     \caption{Close packing of elementary one-plaquette quantum
       drums shown, where the drum plaquettes are shaded,
       alongwith the boundaries of the shielding regions for two such drums
       (shown as thick blue lines).
       An initial Fock state which is consistent with this
       arrangement of quantum drums and where up-spins (bosons) are indicated by
       filled red dots is also shown. Two bipartition cuts are also shown as thick orange and purple lines.}
    \label{closepacking}
 \end{center}
\end{figure}

Eigenstates composed of only elementary one-plaquette quantum drums and
inert spins already generate non-trivial zero modes and non-zero integer
eigenvalues. These can be viewed as the 2D generalization of
bubble eigenstates discussed in a 1D model of Hilbert space
fragmentation~\cite{Mukherjee2021fragment}.
Hilbert space fragments with $n_0$ such independent one-plaquette
drums have a dimensionality of $2^{n_0}$ since each such elementary
quantum drum is consistent with two configurations on the plaquette.
An extensive number of such elementary quantum drums are
needed to form finite energy-density eigenstates of $H$ with a macroscopic
number of up-spins (bosons) (Fig.~\ref{closepacking}). The closest packing
of these elementary quantum drums such that the boundaries of their
shielding regions do not overlap is shown in
Fig.~\ref{closepacking} which yields the
maximum possible value of $n_0$ $=$ $L^2/9$ for a $L \times L$ square lattice
when $L \gg 1$ thus fixing the
corresponding fragment's dimension to be equal to
\begin{eqnarray}
(2^{1/9})^{L^2} \approx
  (1.08006\cdots)^{L^2}.
  \label{HSDelementaryplaq}
\end{eqnarray}
and the density of up-spins (bosons) to be $n=2/9$. 
The corresponding matrix can be immediately
diagonalized by noting that the form of $H$ projected to any $n_0 \neq 0$
fragment produced solely by elementary one-plaquette quantum drums equals
\begin{eqnarray}
  H_{\mathrm{eff}} = \sum_{i=1}^{n_0} \tau_i^x
  \label{Heff}
  \end{eqnarray}
where $i$ denotes the center of an elementary drum plaquette, and $\tau^x_i$
locally flips an arrangement of $(+1,-1,+1,-1)$ to $(-1,+1,-1,+1)$ and
vice-versa on that drum in the computational basis. This
``non-interacting'' $H_{\mathrm{eff}}$ only leads to integer eigenvalues for
any $n_0$. If $n$ ($n_0-n$) of the elementary quantum drums are associated with
an eigenvalue $\tau_i^x=+1 (-1)$, the resulting eigenstate has energy
$E=2n-n_0$. Clearly, there are $\binom{n_0}{n}$ distinct eigenstates that have
the same energy $E=2n-n_0$. Assuming that both $n_0,n \gg 1$, the degeneracy
$\Omega(n)$ of such eigenstates is bounded below by
\begin{eqnarray}
  \Omega(n) > 2^{n_0} \sqrt{\frac{2}{\pi n_0}} \exp \left(2 n_0 \left( x-\frac{1}{2}\right)^2 \right)
\end{eqnarray}
where $n_0=L^2/9$ for the largest such fragment (Fig.~\ref{closepacking})
and $x=n/n_0$. This bound immediately shows that the number of such integer
eigenstates is exponentially large in the system size for integer energies
that range from $E=0$ to $|E| \sim O(L)$ (while the maximum value of the
integer energy $|E|=L^2/9$ when $L \gg 1$ for a $L \times L$ square lattice
with OBC). These high-energy eigenstates
satisfy a strict area law scaling of entanglement entropy with the
entanglement entropy of an arbitrary bipartition, $S_{\mathrm{bp}} = b L$,
where $b$ can range from $0$ to $\ln(2)/3$ (examples of two such bipartition
cuts which give the extreme values of $b$ are shown as thick lines in
Fig.~\ref{closepacking}), depending on the nature of the
bipartition.

Any Fock state consistent with $n_0$ independent one-plaquette drums (e.g.,
one such Fock state is shown in Fig.~\ref{closepacking} where the red
filled dots
represent up-spins (bosons)) shows persistent oscillations with a time-period
$T=\pi$ under unitary evolution under $H$ for a class of local operators.
This can be directly related to the non-interacting nature of
$H_{\mathrm{eff}}$ in Eq.~\ref{Heff} which leads to the following {\it emergent}
dynamical symmetry~\cite{Medenjak2020}:
\begin{eqnarray}
  \left[P_{\mathrm{eff}}H P_{\mathrm{eff}}, \frac{\tau_i^y+i\tau_i^z}{2}\right]=\omega \left(\frac{\tau_i^y+i\tau_i^z}{2} \right)
\end{eqnarray}
where $P_{\mathrm{eff}}$ is a projection operator to the Fock space with $n_0$
one-plaquette drums and $\omega=2$ given the form of $H_{\mathrm{eff}}$ in
Eq.~\ref{Heff}. Thus, for any such initial Fock state, any local operator with a
finite overlap with any of the $\left(\frac{\tau_i^y+i\tau_i^z}{2} \right)$
operators will show persistent oscillations with a time period $T=2\pi/\omega=
\pi$.

\subsection{Wire decomposition of quantum drums} 
\label{wiresdecom}

As introduced earlier, wires represent the basic
quantum drums that can be generated for any given number of
plaquettes, $N_p$, by arranging them in a vertex-sharing pattern
along any one of the two diagonal directions
of the parent $L_x \times L_y$ lattice.
A {\it reference Fock state} of the wire can be
taken to be all the $N_p+1$ up-spins (bosons) to be arranged along the
length of the drum.
The shielding region around a wire consists of all
plaquettes that share either an edge or a vertex with any of the $N_p$
plaquettes that belong to the drum.
We refer the reader to
Sec.~\ref{astwire} (Sec.~\ref{nstud}) where the spectrum for
wires with small (large) $N_p$ shall be discussed. For now, it is sufficient
to note that the number of Fock states
generated by a wire with $N_p$ plaquettes equals $F_{N_p+2}$ (a Fibonacci
number) where $F_0=0$,
$F_1=1$, and $F_n =F_{n-1}+F_{n-2}$ for $n>1$ (see Sec.~\ref{nslev} for the
derivation).

A reference Fock state for more complicated (non-wire) drums
can be constructed from two or more parallel wires with
differing lengths in general, with each wire being in its reference
state, and any remaining unpaired up-spins (bosons) [that do not belong
to any of the wires]
that lie at a minimum distance of $(3/2)\sqrt{2}$ (in lattice units)
from the nearest wire. Some of the Fock states displayed in
Fig.~\ref{drumsfromIS}
and Fig.~\ref{Figrecursion}
already serve to illustrate this concept of {\it
    wire-decomposed reference states}. The state in the top-middle
panel (Fig.~\ref{drumsfromIS}) can be viewed as a reference Fock state
composed of two parallel
wires with $N_p=2$ each for the drum in the top-right panel
(Fig.~\ref{drumsfromIS}). Similarly, the state in the center-middle
panel (Fig.~\ref{drumsfromIS}) can be viewed as a reference Fock state
composed of two parallel
wires, one with $N_p=2$ and another with $N_p=3$,
for the drum in the bottom-right panel (Fig.~\ref{drumsfromIS}). The
Fock states marked by \textcircled{$3$} and \textcircled{$4$} in
Fig.~\ref{Figrecursion} consist of unpaired up-spin(s) (boson(s)) at a distance
of $(3/2)\sqrt{2}$ from a wire with $N_p=3$. The parallel
wires that make up a wire-decomposed reference state cannot exist as
independent drums since either the shielding region of a wire overlaps with
that of other parallel wires or unpaired up-spins (bosons) exist at the
boundary of a shielding region of some wire(s).

In this section, we will see that
\begin{enumerate}
  \item Drums composed of only vertex-sharing plaquettes can
    be built from a reference Fock state where the
    parallel wires can fluctuate {\it simultaneously}
    to access {\it all their internal states} without violating any of the
    hard-core constraints of the model. Such wires are separated
    by a distance of $2\sqrt{2}$ or greater from each other. 
All the Fock states of such drums can be generated from the fluctuations of
these parallel wires and possibly, other sets of parallel wires in the same
direction or perpendicular to the direction of the original set of wires (with
all simultaneous fluctuations again allowed).

  \item Drums composed of only edge-sharing plaquettes can be built from a
    reference Fock state where the parallel wires {\it cannot} fluctuate
    simultaneously to access all their internal states being at a
  distance of $(3/2)\sqrt{2}$ from each other,
  but only do so if
alternate wires are kept in their reference state. The Fock states of
such drums can be generated from the fluctuations of the alternate
parallel wires and possibly, other sets of alternate parallel wires in the same
direction or perpendicular to the original set of wires.
However, not all simultaneous fluctuations of such consecutive wires
are disallowed by the hard-core constraints of the model and
these can be represented as {\it additional} excitations of
elementary plaquettes that
are separated by $3$ lattice units along either $x$ or $y$, or both,
such that these
plaquettes can fluctuate {\it independently}.

\item Drums with both edge-sharing as well as vertex-sharing plaquettes can
  be built from a reference Fock state that consists of
  parallel wires such that while all the wires cannot fluctuate
  simultaneously to access all their individual states,
  some consecutive wires can do so if the other wires
are kept in their reference state.
\end{enumerate}

\subsubsection{Calculating fragment dimension from wire decomposition}
\label{fragdim}
\begin{figure}[!tbh]
  \begin{center}
    \includegraphics[width=0.9\linewidth]{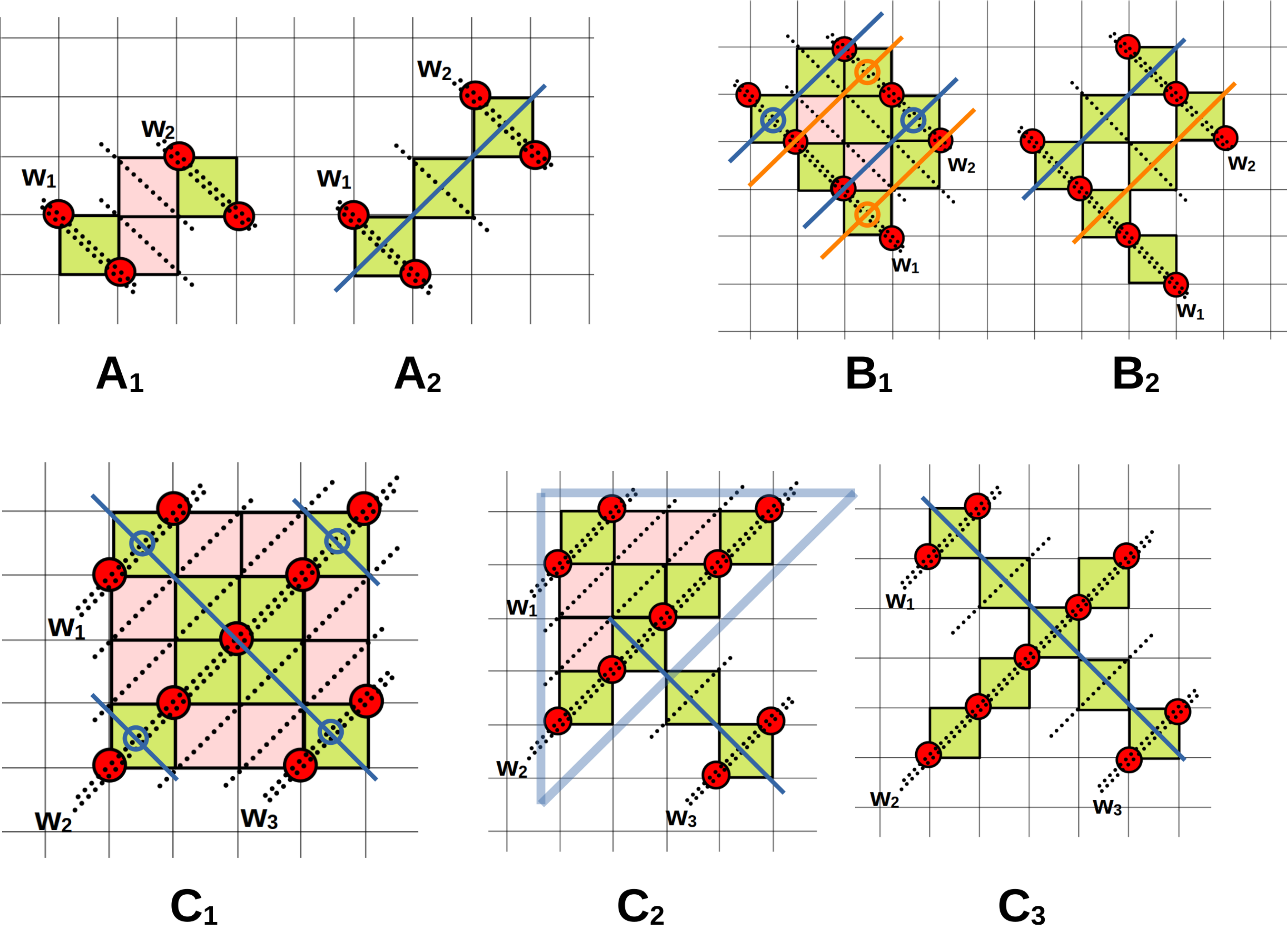}
    \caption{Examples of wire decomposition of quantum drums shown here.
      The drums labeled A$_1$ and A$_2$ can be constructed from two parallel
      wires both with $N_p=1$, the drums labeled B$_1$ and B$_2$ can be
      constructed from two parallel wires with $N_p=3$ and $N_p=2$ while the
      drums labeled C$_1$, C$_2$, and C$_3$ can be constructed from three
      parallel wires, where two of them have $N_p=1$ and one has $N_p=4$.
      The wires in all the panels are indicated by double dashed lines and
      also labeled by w$_1$, w$_2$, w$_3$. The
      red filled dots in all panels represent up-spins (bosons).
      The perpendicular wires that can be generated from the Fock state shown
      for each drum are indicated by bold lines in blue and orange.
      Additionally, the plaquettes marked by open blue (orange)
      circles at their centers in B$_1$ and C$_1$ represent
      the locations of one-plaquette excitations that generate
      additional Fock states that cannot be represented by
      excitations of alternate parallel wires in these two cases.
      In the drum C$_2$, a smaller drum made of $10$ edge-sharing plaquettes
      (indicated by thick blue region enclosing it) fluctuates simultaneously,
      with the wire w$_3$.
      The green (pink) plaquettes in the
      drums shown in all panels follow the same convention as
    used in Fig.~\ref{drumsandshields}.
    }
    \label{FIGw}
\end{center}
\end{figure}


We first demonstrate the aforementioned concepts for small
  quantum drums before considering macroscopic quantum drums and their
  corresponding fragment sizes in later Sections (Sec.~\ref{twowires} and
  Sec.~\ref{nonthermal}). We start with the simplest case of two $N_p=1$
wires in their reference state
that are placed parallel to each other. If such wires fluctuated independently,
these would have
produced a total of $F_3 \times F_3=4$ Fock states.
There are two distinct ways of placing these wires with respect to
each other such that they do not fluctuate independently and
no inert up-spins (bosons) are created.
These are shown as the quantum drums marked by A$_1$ and
A$_2$ in Fig.~\ref{FIGw}. While the drum A$_1$ generates a Hilbert space
fragment with $3$ Fock states, the drum A$_2$ generates one with $5$
Fock states.
In the drum indicated by A$_1$ (Fig.~\ref{FIGw}),
the wire w$_1$ (w$_2$) (indicated by double
dotted lines in Fig.~\ref{FIGw}) can
fluctuate to generate both its Fock states only if w$_2$ (w$_1$) is held
fixed in its reference state. Thus, the two wires w$_1$ and w$_2$ {\it cannot}
fluctuate simultaneously in A$_1$ and produce $2F_3-1=3$ states.
On the other hand, in the drum
A$_2$ (Fig.~\ref{FIGw}), both the wires w$_1$ and w$_2$ (indicated by double
dotted lines in Fig.~\ref{FIGw}) {\it can} fluctuate simultaneously
without producing a Fock state that violates the hard-core constraints.
Additionally, performing a ring-exchange from the reference state
on both the plaquettes
that represent w$_1$ and w$_2$ generates the reference state for another
wire with $N_p=3$ that is {\it perpendicular} to w$_{1}$ and w$_2$
(shown as a blue line in the drum marked A$_2$ in Fig.~\ref{FIGw}).
The Fock state obtained from a ring-exchange on
the middle plaquette from the reference state
of this $N_p=3$ wire cannot be represented
by combining any of the Fock states generated from the w$_1$ and
w$_2$ wires and accounts for the total $F_3^2+1=5$ Fock states for
the drum A$_2$. In the case of A$_1$ [A$_2$], the minimum distance between the
parallel wires w$_1$ and w$_2$ equals $(3/2)\sqrt{2}$ [$2\sqrt{2}$]
(Fig.~\ref{FIGw}).

The drums labeled B$_1$ and B$_2$ in Fig.~\ref{FIGw}
represent more complicated cases that arise when two parallel
wires (of unequal lengths)
in their reference states, one with $N_p=3$ and another with $N_p=2$, are
brought close to each other such that the minimum distance between the
wires equal $(3/2)\sqrt{2}$ and $2\sqrt{2}$ respectively.
If these two wires fluctuated independently, these
would have generated $F_5 \times F_4=15$ Fock states. However, the drum B$_1$
generates a fragment with $11$ Fock states while the drum B$_2$ generates
a fragment with $18$ Fock states. In the drum B$_1$ (Fig.~\ref{FIGw}),
the wire w$_1$ (w$_2$),
shown by double dotted lines in Fig.~\ref{FIGw},
can fluctuate to generate all its Fock states only if the other wire
w$_2$ (w$_1$) is held fixed in its reference state. Such wire fluctuations
lead to $F_5+F_4-1=7$ states. Two cases where fluctuations of a
perpendicular wire (indicated by the top blue
line and the bottom orange line respectively
in the drum marked B$_1$ in Fig.~\ref{FIGw}) when the
other wire parallel to it at a distance $(3/2)\sqrt{2}$
(indicated by the bottom blue line and the top orange line
respectively in Fig.~\ref{FIGw})
is kept fixed in its reference state generates an
additional $2$ Fock states. The remaining $2$ Fock states in the
fragment are
generated by two separate cases of
ring-exchanges on two plaquettes together [indicated by the plaquettes
with an open circle of the same color (blue and orange) at their centers]
that are separated by $3$ lattice units along $x$/$y$ as shown in
Fig.~\ref{FIGw} (panel marked B$_1$).
On the other hand, in the
drum B$_2$ (Fig.~\ref{FIGw}),
both the wires w$_1$ and w$_2$ (indicated by double dotted
lines in Fig.~\ref{FIGw}) can fluctuate simultaneously to
generate all their Fock states without violating the hard-core constraints.
Fluctuations of w$_1$ and w$_2$ in the drum B$_2$
cannot, however, generate any Fock state with two up-spins (bosons)
along the diagonal parallel to w$_1$, w$_2$ on any of
the two plaquettes that are not part of w$_1$ and w$_2$. The
extra $18-(F_5 \times F_4)=3$ Fock states are generated from
ring-exchange moves in
any one of these two
plaquettes starting with Fock states obtained from the fluctuations of
w$_1$ and w$_2$ that can be represented
as the reference state of a $N_p=3$ wire perpendicular to
both w$_1$ and w$_2$ (shown by a
blue and an orange line perpendicular to w$_1$, w$_2$
in Fig.~\ref{FIGw}) and containing one of these two plaquettes.

Finally, we consider a case where three parallel wires in their reference
states, w$_1$ with $N_p=1$, w$_2$ with $N_p=4$, and w$_3$ with $N_p=1$, are
brought close to each other to generate three different
drums labeled C$_1$, C$_2$, and
C$_3$ in Fig.~\ref{FIGw}. While independent fluctuations of
these three wires generate $F_3\times F_6 \times F_3=32$ Fock states, the
fragment generated by the drum C$_1$ contains $24$ Fock states,
by the drum C$_2$ contains $28$ Fock states, and by the
drum C$_3$ contains $42$ Fock states,
respectively. In the drum C$_1$ (Fig.~\ref{FIGw}), the wire w$_1$ (w$_3$)
can only access all its Fock states if w$_2$ is held fixed in its reference
state (with these wires indicated by double dotted lines in the drum C$_1$ in
Fig.~\ref{FIGw}).
Similarly, the wire w$_2$ can only access all its Fock states if
both w$_1$ and
w$_3$ are fixed to their reference states in C$_1$. This generates a total
$F_3^2+F_6-1=11$ Fock states. An additional $5$ Fock states of
drum C$_1$ are generated by similar wire fluctuations of
parallel wires separated by $(3/2)\sqrt{2}$ but perpendicular to
w$_1$, w$_2$, w$_3$ (indicated by blue lines in drum C$_1$ in
Fig.~\ref{FIGw}). Finally, the remaining $8$ Fock states in C$_1$
are generated by simultaneous
ring-exchanges on two/three of the four corner plaquettes (marked by
blue circles at the centres of the corresponding
plaquettes in C$_1$ in Fig.~\ref{FIGw}) that are separated from each
other/from a corner plaquette by $3$ lattice
units in the $x$ or $y$ direction.
On the other hand, in the drum C$_3$ (Fig.~\ref{FIGw}), all the wires,
w$_1$, w$_2$ and w$_3$ (indicated by double dotted lines in Fig.~\ref{FIGw}),
can fluctuate simultaneously without violating the
hard-core constraints. Furthermore, fluctuations in w$_1$, w$_2$ and
w$_3$ generates
a new open channel of fluctuations in the form of a wire with
$N_p=5$ plaquettes in the direction perpendicular to these wires (indicated
by a blue line in C$_3$ in Fig.~\ref{FIGw}) which
generates an additional $10$ Fock states besides the $F_3 \times F_6 \times F_3=
32$ Fock states generated from w$_1$, w$_2$, w$_3$.
The drum marked as
C$_2$ in Fig.~\ref{FIGw} represents an interesting
intermediate case between C$_1$ and
C$_3$ where the parallel wires w$_1$ and w$_2$ (w$_2$ and w$_3$)
are at a distance $(3/2)\sqrt{2}$ ($2\sqrt{2}$) from each other. The
smaller drum containing wires w$_1$, w$_2$ and composed of $10$
edge-sharing plaquettes (marked by the blue region in the drum C$_2$ in
Fig.~\ref{FIGw}) can fluctuate simultaneously with the wire w$_3$.
This leads to a total of $13 \times 2=26$ Fock states (see
Fig.~\ref{junctionthreewires} for the $13$ Fock states of the smaller drum
made by the wires w$_1$ and w$_2$).
The additional states are generated from an extra open channel for
fluctuations along a $N_p=3$ wire perpendicular to w$_1$, w$_2$, w$_3$
containing the w$_3$ plaquette as its right-most plaquette as indicated
by the blue line perpendicular to w$_1$, w$_2$, w$_3$ in the drum C$_2$ in
Fig.~\ref{FIGw}. This
results in $2$ Fock states where a ring-move is performed on the
reference state of this $N_p=3$ wire using the plaquette excluded from both the
smaller drum composed of $10$ edge-sharing
plaquettes and w$_3$.

These examples demonstrate the general principle that
given $n$ parallel wires (with unequal lengths in general)
in their reference state, it is optimal to place these such that all
the wires can fluctuate simultaneously and that these fluctuations
additionally generate the
maximum number of longest-possible wires perpendicular to the original wires
as extra open channels of fluctuations to maximize the fragment size generated
by the resulting drum. Both these conditions are satisfied by appropriate drums
composed of only vertex-sharing plaquettes as shown in Fig.~\ref{FIGw} (panels
marked A$_2$, B$_2$ and C$_3$).

\subsubsection{Constructing entire drums from wire-decomposed reference states}
\label{wirestodrum}
\begin{figure}[!tbh]
  \begin{center}
    \includegraphics[width=0.9\linewidth]{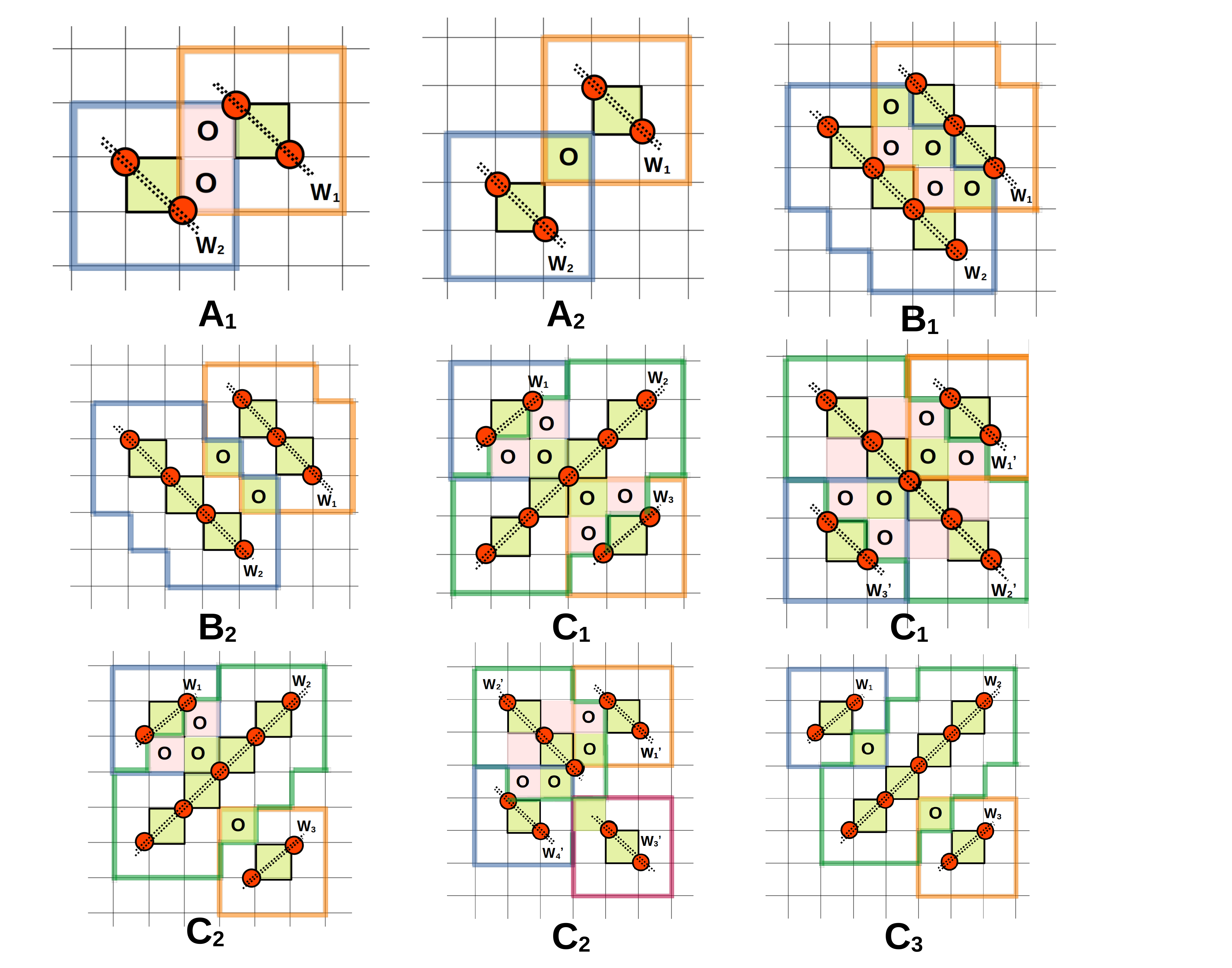}
    \caption{ The drums A$_1$ to C$_3$
      shown in Fig.~\ref{FIGw} can be constructed using
      the overlap of the shielding regions of parallel wires as shown. The
      wires are indicated by doble dashed lines in each panel. The
      red filled dots in all panels represent up-spins (bosons). The
      boundaries of the shielding region of each wire is shown using
      thick lines of different colors and the plaquettes formed by the
      overlap of the shielding regions are indicated by ``o''.
      The green (pink) plaquettes in the
      drums shown in all panels follow the same convention as
    used in Fig.~\ref{drumsandshields}.
    }
    \label{FIGdfs}
\end{center}
\end{figure}

The wire-decomposed reference states introduced here
  provide an efficient route to construct the entire quantum drums
  associated with them. For this, we note that the shielding region of a
  single wire consists of all external plaquettes that share an edge or a
  vertex with the plaquettes that belong to the wire. The different wires in a
  wire-decomposed reference state are coupled to each other
  because (i) shielding regions of different parallel wires
  with spacing $(3/2)\sqrt{2}$ or $2\sqrt{2}$ 
  have overlapping plaquettes
  and/or (ii) shielding regions of different parallel wires seperated by
  $3\sqrt{2}$ are coupled by unpaired up-spins (bosons).
  The overlapping plaquettes between different shielding
  regions (for (i)) and the plaquettes of the shielding
  region(s) that also contain any unpaired up-spin (boson) (for (ii)) provide
  the {\it remaining} plaquettes of the quantum drums starting from
  wire-decomposed Fock states.

Fig.~\ref{FIGdfs} illustrates the
  procedure for all the quantum drums shown
in Fig.~\ref{FIGw}. For the drums A$_1$, A$_2$,
B$_1$, B$_2$ and C$_3$, it is sufficient to consider the overlap of the
shielding regions of the parallel wires shown as double dashed lines
in Fig.~\ref{FIGw} as can be
seen from Fig.~\ref{FIGdfs}, where the overlapping plaquettes of
the shielding regions have been denoted by ``o'' in all panels of
Fig.~\ref{FIGdfs}. The drums
C$_1$ and C$_2$ present more interesting cases where this construction only
identifies a subset of plaquettes that belong to the corresponding
drum (Fig.~\ref{FIGdfs}). However, starting from the reference
wire-decomposed
state of
the original parallel
wires, it is easy to perform ring-exchange moves on a subset of the plaquettes
that belong to these wires to create another wire-decomposed
  Fock state that can be viewed as
parallel wires in their reference state, but in the perpendicular direction
to the original wires (Fig.~\ref{FIGdfs}).
The overlap of the shielding regions of these new
wires gives the remaining plaquettes that are part of the quantum drum for
both C$_1$ and C$_2$ (Fig.~\ref{FIGdfs}).
\begin{figure}[!tbh]
  \begin{center}
    \includegraphics[width=0.6\linewidth]{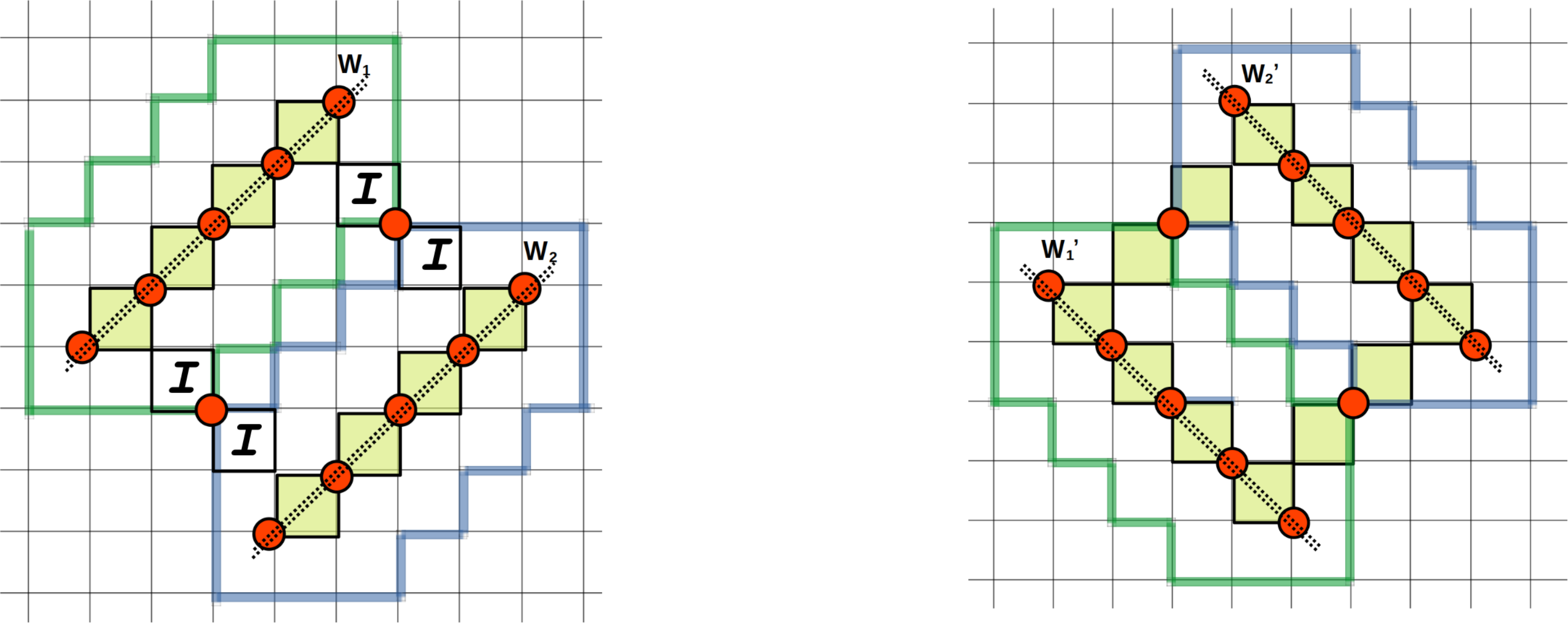}
    \caption{Two parallel wires w$_1$ and w$_2$ (see left panel)
      and two other parallel wires w$_1^{'}$
      and w$_2^{'}$ (see right panel) indicated by double dotted lines,
      both in their reference state and both with $N_p=4$,
      seperated by a distance of
      $3\sqrt{2}$. The wires w$_1$, w$_2$ are perpendicular
      to the wires w$_1^{'}$, w$_2^{'}$. In both panels, two up-spins
      (bosons) are located on two common sites between the
      boundaries of the shielding regions,
      shown using thick lines,
      of the wires. The red filled circles in both panels indicate
    up-spins.  The green plaquettes in the
      drums shown in both panels follow the same convention as
      used in Fig.~\ref{drumsandshields}, with the right panel showing
      all the plaquettes that belong to this drum. In the
        left panel, plaquettes of the shielding regions of w$_1$ and w$_2$ that
      contain unpaired up-spins (bosons) are indicated by ``I''.}
    \label{sparse}
 \end{center}
\end{figure}

In Fig.~\ref{sparse}, we show an example of a wire-decomposed
Fock state where two parallel wires are separated by a distance
of $3\sqrt{2}$. While the shielding regions of such
wires do not have any overlapping plaquettes due to the increased
distance between the parallel wires, these wires can still
be coupled to each other to make a larger drum
by placing unpaired up-spins (bosons) at a distance of
$(3/2)\sqrt{2}$ from both wires in a subset of
the common sites between the boundaries of the shielding regions of both the
wires. In Fig.~\ref{sparse}, two parallel wires w$_1$ and w$_2$,
  both with $N_p=4$ and in their reference state, are placed at a distance
  of $3\sqrt{2}$ from each other (Fig.~\ref{sparse}, left panel).
  Two unpaired up-spins (bosons) are placed on two of the
  common sites of the boundaries of the
shielding regions of both the wires (the boundaries of the shielding
regions are shown using thick green (blue) lines for w$_1$ (w$_2$) in
Fig.~\ref{sparse}, left panel). The plaquettes of the
  shielding regions that contain the unpaired up-spins (denoted by
  ``I'') then provide
  the remaining plaquettes of the entire quantum drum associated with this
  wire-decomposed reference state. Note that the wires w$_1$ and w$_2$ can
  fluctuate simultaneously to generate all their Fock states in spite of the
up-spins (bosons) on the boundaries of the shielding regions.
Performing appropriate ring-exchanges on
this reference state (Fig.~\ref{sparse}, left panel), it is easy to get a
Fock state that can be viewed as two parallel wires, w$_1^{'}$ and w$_2^{'}$,
that are both perpendicular to w$_1$, w$_2$ and again separated by
$3\sqrt{2}$ with two up-spins (bosons) on two of the common sites shared
by the boundaries of the shielding regions of  w$_1^{'}$ and w$_2^{'}$. This
shows that four other plaquettes (apart from the ones shaded in
Fig.~\ref{sparse}, left panel)
are part of the bigger drum (see Fig.~\ref{sparse}, right panel)
and that all the Fock states can be
generated from simultaneous fluctuations of either w$_1$, w$_2$ or
w$_1^{'}$, w$_2^{'}$, again generating a quantum drum with only
vertex-sharing plaquettes.


\section{Strong Hilbert space fragmentation}
\label{strongfrag}

In this section, we show strong Hilbert space fragmentation for
this kinematically constrained
2D model (Eq.~\ref{constr}) defined on a
$L_x \times L_y$ rectangular lattice with OBC as $L_x,L_y \gg 1$. We first
discuss numerical evidence from ED in Sec.~\ref{numfrag}.
The wire decomposition of quantum drums introduced earlier
is used in Sec.~\ref{twowires} to
calculate the scaling of the Hilbert space fragment for drums composed of two
long parallel wires.
The wire decomposition is further used to identify
  the nature of the quantum drums that define the largest Krylov subspaces
  as a function of the density of up-spins (bosons) in Sec.~\ref{nonthermal}.
  We then invoke the standard typicality argument~\cite{Gogolin_review}
  and show that typical initial states that belong to these large
  Krylov subspaces violate thermalization with respect to the full Hilbert
  space.

\subsection{Numerical evidence of strong fragmentation from exact diagonalization}
\label{numfrag}

One procedure to distinguish between weak and strong
fragmentation~\cite{Sala2020, Khemani2020} involves monitering the ratio of
the largest Hilbert space fragment
(denoted by $\mathrm{max}[\mathcal{D}_{f,n}]$) to the total Hilbert space
dimension (denoted by $\mathcal{D}_n$) in a sector with a
{\it fixed} density (denoted by $n$) of up-spins (bosons) for
different system sizes. We stress here that only the global symmetry of
total magnetization conservation and its associated density
is relevant for this analysis since internal symmetries like reflections etc
can always be removed by adding suitable diagonal terms to $H$ in the
computational basis that do not connect the different Hilbert space fragments.
If the ratio $\mathrm{max}[\mathcal{D}_{f,n}]$/$\mathcal{D}_n$
behaves as $\exp(-\gamma N)$ with $\gamma >0$
as the number of sites in the system, $N$, diverges, it implies strong
fragmentation; in contrast, if it approaches $1$ as $N \gg 1$, it implies
weak fragmentation.
\begin{figure}
  \begin{center}
    \includegraphics[width=0.50\linewidth]{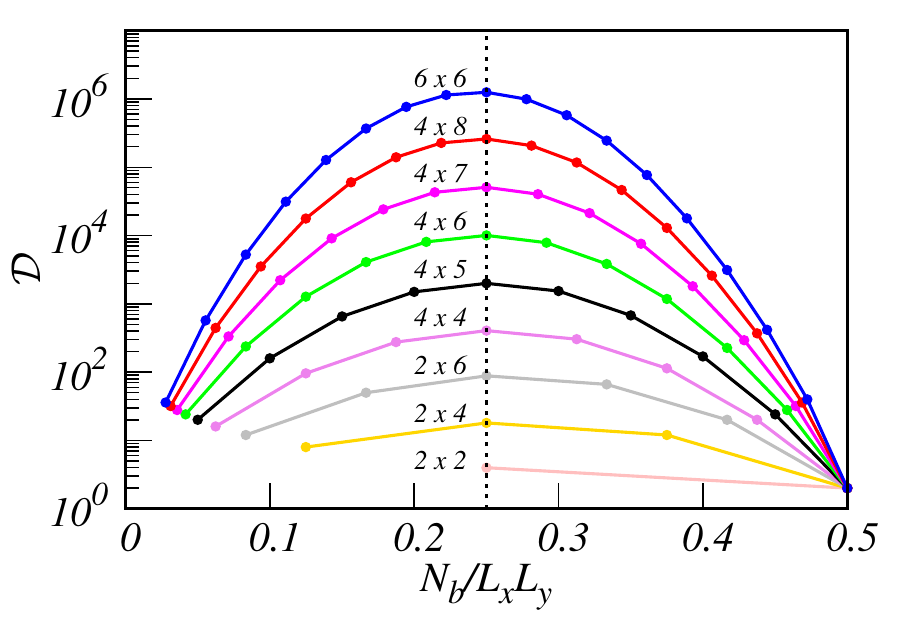}%
     \includegraphics[width=0.48\linewidth]{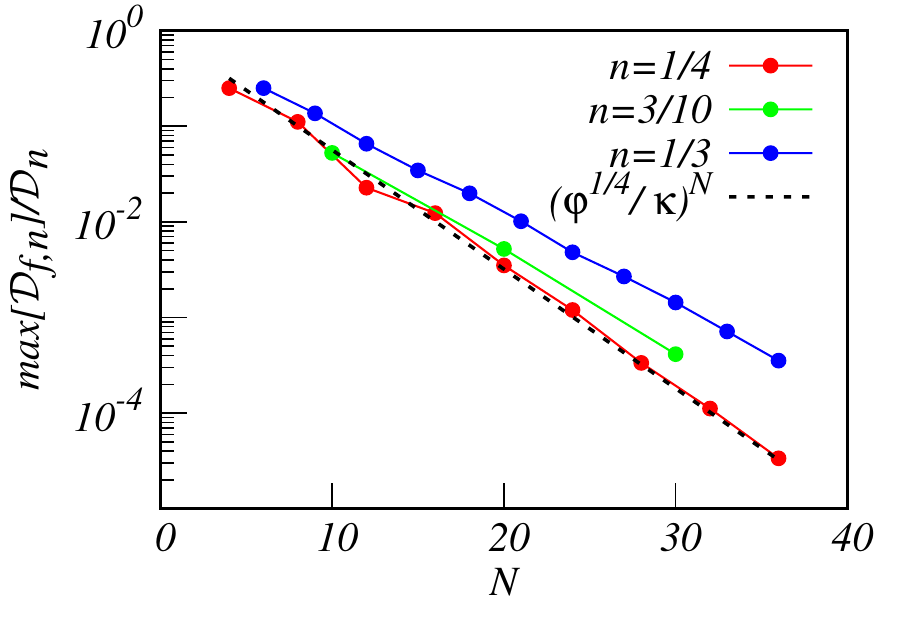}
     \caption{(Left panel) The total Hilbert space dimension $\mathcal{D}$
       as a function of the number of up-spins, $N_b$, for rectangular
       lattices $L_x \times L_y$ of various dimensions. The thin dotted vertical
       line is at $n=N_b/(L_x L_y)=1/4$. (Right panel) The behavior of the
       ratio of the dimension of the
       largest Hilbert space fragment and the total Hilbert space
       dimension for the magnetization sector for up-spin densities $n=1/4$ (blue),
       $n=3/10$ (green) and $n=1/3$ (red) shown for rectangular
       lattices with dimension $L_x \times L_y$ and OBC as a function of the
       system size $N=L_x L_y$. The dotted curve shows
         the function $(\varphi^{1/4}/\kappa)^N$ where $\varphi=(1+\sqrt{5})/2$
         (golden ratio) and $\kappa \approx 1.503\cdots$
         (hard square entropy constant). }
    \label{FIGnum}
 \end{center}
\end{figure}

Using exact enumeration techniques, we calculate the Hilbert space dimension,
$\mathcal{D}$, for a fixed number of up-spins (bosons), $N_b$, for a variety of
rectangular lattices of dimension $L_x \times L_y$ with OBC
(see Fig.~\ref{FIGnum}, left panel) which shows that $\mathcal{D}$ is
maximized when $n=N_b/(L_x L_y)=1/4$. We then focus on this particular
density of up-spins (bosons) $n=1/4$ as well as two other densities
$n=3/10$ and $n=1/3$ to
show the scaling of $\mathrm{max}[\mathcal{D}_{f,n}]/\mathcal{D}_n$ for
fixed $n$ as a function of $N=L_xL_y$ in Fig.~\ref{FIGnum}, right
panel using data from
exact enumeration. The data for these
limited system sizes already clearly indicate that
$\mathrm{max}[\mathcal{D}_{f,n}]$/$\mathcal{D}_n$ $\sim$ $\exp(-\gamma N)$
with $\gamma$ depending on the density of up-spins (bosons), $n$,
and thus points towards strong Hilbert space fragmentation in this
2D model. In Sec.~\ref{nonthermal}, we will show that the
  dimension of the largest Hilbert space
  fragment generated in this model scales as
  $(\varphi^{1/4})^{L_xL_y}$ for $L_x,L_y \gg 1$ when $n=1/4$, where
  $\varphi=(1+\sqrt{5})/2$ (golden ratio). Given that the
  total Hilbert space dimension
  scales as $\kappa^{L_xL_y}$ for $L_x,L_y \gg 1$~\cite{Baxter1999}, it is
  reassuring to see that $(\varphi^{1/4}/\kappa)^{L_xL_y}$ (dotted curve in
  Fig.~\ref{FIGnum}, right panel) closely follows
  the data for $\mathrm{max}[\mathcal{D}_{f,n}]$/$\mathcal{D}_n$ at $n=1/4$
  (Fig.~\ref{FIGnum}, right panel) since the density $n=1/4$ gives the
  dominant contribution to the total Hilbert space (Fig.~\ref{FIGnum}, left
  panel) at these system sizes.

\subsection{Quantum drums generated from two long parallel wires}
\label{twowires}

\begin{figure}[!tbh]
  \begin{center}
    \includegraphics[width=0.7\linewidth]{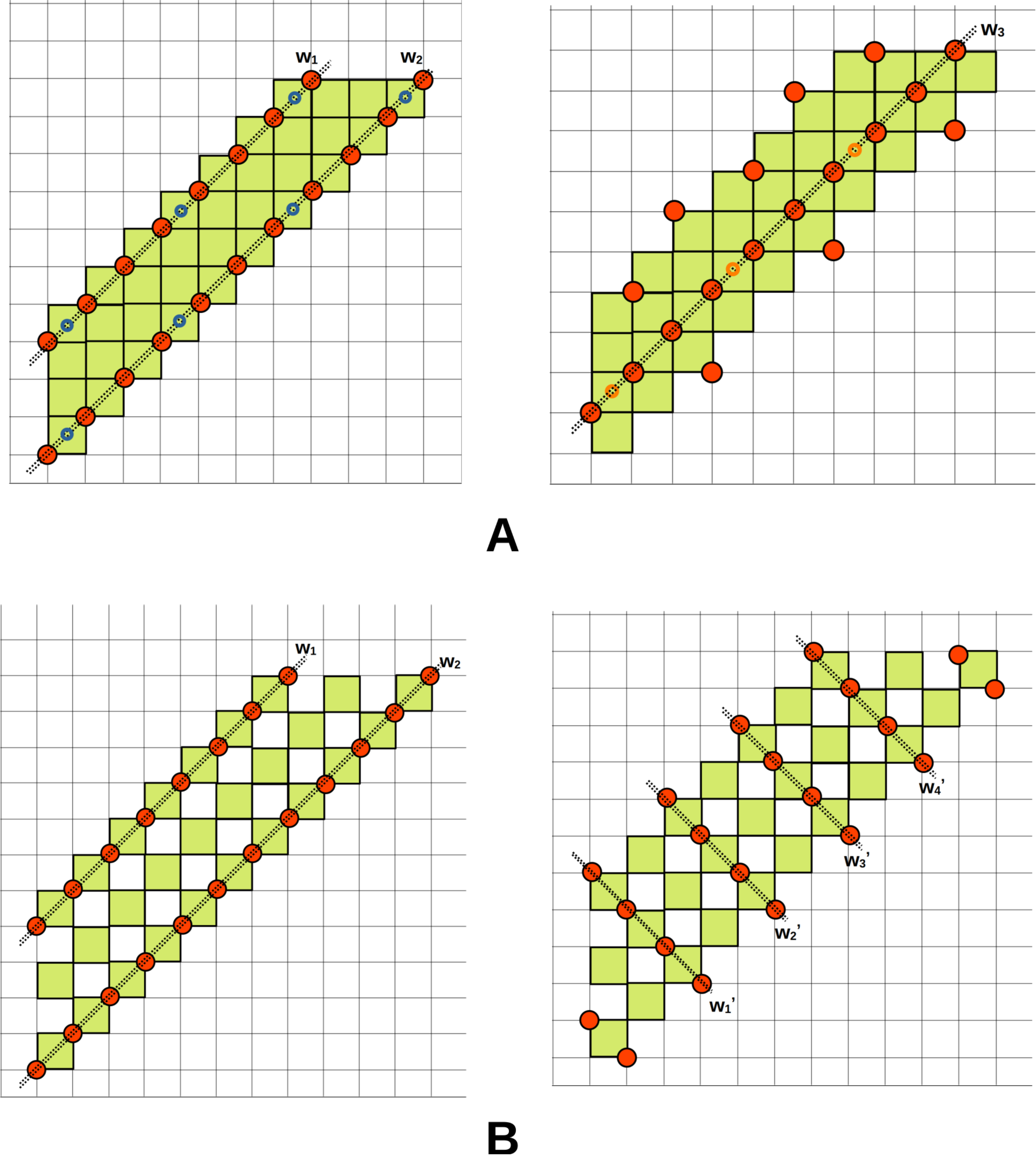}
    \caption{A section of two quantum drums that are
      generated by placing two long parallel wires, denoted by
      w$_1$ and w$_2$, in their reference states at a mutual separation of
      $(3/2)\sqrt{2}$ (panel A) and $2\sqrt{2}$ (panel B) respectively,
      shown here . The
      green plaquettes in the quantum drums
      follow the same convention as
    used in Fig.~\ref{drumsandshields}.
    The filled red dots indicate up-spins (bosons) in all the panels
    and represent just one of the many
      possible Fock states of the corresponding
      drum. The elementary plaquettes indicated by open blue circles (open
      orange circles) in the top-left figure (top-right figure) in panel A allow
      for independent ring-exchange moves starting from the Fock state shown
      in the same figure. Note that only these plaquettes allow for
      ring-exchange moves for the wire w$_3$ in the top-right figure of panel
      A. In panel B, the Fock states can be generated either from the
      simultaneous fluctuations of the wires, w$_1$ and w$_2$, as shown in
      bottom-left figure (panel B) or from the simultaneous fluctuations
      of parallel short wires (four of them,
      w$_1^{'}$, w$_2^{'}$, w$_3^{'}$, w$_4^{'}$ shown in the bottom-right figure
      (panel B)) that are perpendicular to w$_1$ and w$_2$. }
    \label{longwires}
 \end{center}
\end{figure}
While Sec.~\ref{wiresdecom} illustrated the wire decomposition
  using small drums, this concept becomes most efficient to calculate the
  scaling of the resulting fragment size when macroscopic drums
  are considered.
  Since the wire-decomposed reference Fock states of such
  drums typically contain long parallel wires, we will consider the warm-up
  exercise of wire-decomposed Fock states with two such wires in this 
  section. More precisely, we consider two parallel wires, each with
$N_p$ plaquettes such that
$N_p \gg 1$, where both the wires are in their reference Fock states. Bringing
these wires (denoted by w$_1$ and w$_2$ in top-left and bottom-left panels
of Fig.~\ref{longwires})
at a distance of $(3/2)\sqrt{2}$ [$2\sqrt{2}$] 
from each other generates
a drum with only edge-sharing [vertex-sharing] plaquettes as shown in
panel A [panel B] of Fig.~\ref{longwires}. Putting two such parallel
wires closer than $(3/2)\sqrt{2}$ (further than $2\sqrt{2}$) leads to
both the wires being inert (independent of each other). As we will show
below, the number of Fock states in the corresponding drum scales as
$\varphi^{N_p}$ [$\varphi^{2N_p}$], where $\varphi=(1+\sqrt{5})/2$ denotes the
golden ratio, for $N_p \gg 1$ when the two parallel wires,
w$_1$ and w$_2$, are at a distance of $(3/2)\sqrt{2}$ (Fig.~\ref{longwires},
panel A) [$2\sqrt{2}$] (Fig.~\ref{longwires}, panel B) from each other.

Focusing on the case where w$_1$ and w$_2$ are in their reference Fock states
and separated from each other by a distance $(3/2)\sqrt{2}$ (
Fig.~\ref{longwires}, top left panel), the wire w$_1$ [w$_2$] can access all
its allowed Fock states if and only if
w$_2$ [w$_1$] is kept fixed in its reference
state. This immediately shows that $2\varphi^{N_p} < \mathcal{N}_A$ for
$N_p \gg 1$, where
$\mathcal{N}_A$ denotes the total number of Fock states for the quantum
drum shown in panel A of Fig.~\ref{longwires}. Crucially, not all simultaneous
fluctuations of w$_1$ and w$_2$ are disallowed. E.g., starting from the
Fock state shown in Fig.~\ref{longwires} (top left), independent fluctuations
of elementary plaquettes that are arranged in a regular pattern generated
by the primitive vectors $3\hat{x}$ and $3\hat{y}$, as indicated by open blue
dots at the centers of such plaquettes in Fig.~\ref{longwires} (top left),
are allowed. The number of Fock states generated from such 
  independent one-plaquette excitations scale as
  $3 (2^{N_p/3})(2^{N_p/3})$ where the factor of $3$ is due to the inequivalent
  arrangements of this regular pattern. Moreover, the Fock state shown in
Fig.~\ref{longwires} (top right)
can be generated using a sequence of ring-exchange moves from the reference
Fock state shown in Fig.~\ref{longwires} (top left), which again allows for
independent one-plaquette excitations along the wire w$_3$, where such
plaquettes are again arranged in a regular pattern generated
by the primitive vectors $3\hat{x}$ and $3\hat{y}$, as indicated by open orange
dots at the centers of such plaquettes in Fig.~\ref{longwires} (top right).
The number of such Fock states scale as
  $6 (2^{N_p/3})$ where the factor of $3+3$ is due to the $3$ inequivalent
  arrangements of such a regular pattern along w$_3$ and $2$ inequivalent
  ways of placing w$_3$.

A combination of all these independent one-plaquette excitations and excitations
of w$_1$ (w$_2$) keeping w$_2$ (w$_1$) fixed in its reference Fock state
generates all the Fock states of the quantum drum shown in panel A of
Fig.~\ref{longwires} with certain Fock states being produced multiple times.
Thus, in the limit $N_p \gg1$, we get that
\begin{eqnarray}
  2(\varphi)^{N_p} < \mathcal{N}_A <  2(\varphi)^{N_p} + 3(2^{2/3})^{N_p}+6(2^{1/3})^{N_p} \Rightarrow \mathcal{N}_A \sim (\varphi)^{N_p}
  \label{twoclosewires}
  \end{eqnarray}
where the final result in Eq.~\ref{twoclosewires} follows
since $\varphi > 2^{2/3} > 2^{1/3}$.

We now consider the case of the quantum drum
shown in Fig.~\ref{longwires} (bottom left) which is
generated from two parallel wires, w$_1$ and w$_2$, both in their reference
Fock states such that these wires are seperated by a distance of $2\sqrt{2}$.
Since simultaneous fluctuations of both w$_1$ and w$_2$ are allowed in this
case without violating any hard-core constraints, the total number of
Fock states, $\mathcal{N}_B$, in the quantum
drum shown in panel B of Fig.~\ref{longwires} is bounded below by
$(\varphi^2)^{N_p} < \mathcal{N}_B$. The remaining Fock states are
generated from a subset of the Fock states generated from the simultaneous
fluctuations of parallel wires that are perpendicular to the original
w$_1$ and w$_2$ and separated by $2\sqrt{2}$ lattice units from each
other (e.g., a subset of such wires are
marked as w$_1^{'}$, w$_2^{'}$, w$_3^{'}$ and w$_4^{'}$ in the bottom right panel of
Fig.~\ref{longwires}). Since each of these short wires are composed of $3$
plaquettes (Fig.~\ref{longwires}, bottom right), it follows that
each such wire contributes $F_5$ states with there being
  $N_p/2$ of them which fluctuate independently. The resulting number of
  Fock states equal $2(F_5)^{N_p/2}$ where the factor of $2$ is due to the
  $2$ inequivalent arrangements of such short parallel wires. Thus, in the
limit $N_p \gg1$, we get that 
\begin{eqnarray}
  (\varphi^2)^{N_p} < \mathcal{N}_B <  (\varphi^2)^{N_p} + 2(\sqrt{F_5})^{N_p}  \Rightarrow
  \mathcal{N}_B \sim (\varphi^2)^{N_p}
  \label{twosimwires}
\end{eqnarray}
where the final result in Eq.~\ref{twosimwires} follows 
since $\varphi^2 > \sqrt{F_5}$.

Eq.~\ref{twoclosewires} and Eq.~\ref{twosimwires} show
that an important simplification emerges when macroscopically
  long parallel wires are involved in the construction of the reference
  Fock state of a drum. The
correct scaling of the Hilbert space fragment dimension
in both drums is obtained
simply from the fluctuations of those
long wires that can access {\it all} their states simultaneously
while the rest of the fluctuations can be considered as subdominant.
When the wires w$_1$ and w$_2$ are at a
distance of $(3/2)\sqrt{2}$, w$_1$ (w$_2$) can access all its states only
when w$_2$ (w$_1$) is fixed to its reference Fock state implying that
$\mathcal{N}_A \sim (\varphi)^{N_p}+(\varphi)^{N_p} \sim (\varphi)^{N_p}$.
When w$_1$ and w$_2$ are at a distance
of $2\sqrt{2}$ from each other, both wires can access all
their states simultaneously to give $\mathcal{N}_B \sim (\varphi)^{N_p} \cdot
(\varphi)^{N_p} \sim (\varphi^2)^{N_p}$.

\subsection{Large Krylov subspaces and absence of ETH-predicted
  thermalization}
\label{nonthermal}

 In constrast to systems that display
  weak Hilbert space fragmentation, typical initial states in strongly
  fragmented systems {\it do not} thermalize with respect to the full Hilbert
  space due to the absence of a single dominant Krylov subspace in the
  thermodynamic limit. We now consider the fate of typical unentangled
 initial states for a large system, say a $L \times L$
 lattice with OBC where $L \gg 1$, under unitary time
   evolution with $H$. Given the $E$ to $-E$ symmetry of the
 many-body spectrum, typical initial states
   at any fixed density of up-spins (bosons), $n$,
   will have an average energy per site equal to $\langle E \rangle/L^2 =0$
   for $L \gg 1$. Thermalization in the full Hilbert space (ETH)
   with fixed $n$
will imply that such an initial state with a
macroscopic number of up-spins (bosons) should thermalize to the
infinite temperature ensemble (ITE) with $n$
fixed by the initial condition
as far as local operators are concerned. Thus, ETH-predicted
  thermalization implies that local operators lose {\it all} memory of the
  initial state, except its conserved $n>0$, under unitary evolution
  with $H$.

Given a typical initial state with an extensive number of
up-spins (bosons), it can be categorized in one of the following five classes:
\begin{enumerate}
\item The initial state is an inert Fock state which forms a $1$-dimensional
  fragment on its own.
\item The initial state is consistent with a finite number of finite-sized
  drums
  when $L \gg 1$.
\item The initial state is consistent with an extensive number of
  finite-sized drums when $L \gg 1$.
\item The initial state is consistent with the presence of one or more
  (subextensive)
  quasi-1D drums with a typical linear dimension of $O(L)$ as $L \gg 1$.
\item The initial state is consistent with the presence of one or more
  2D drums with a typical linear dimension of $O(L)$ as $L \gg 1$.
\end{enumerate}
Initial states in class $1$ and $2$ clearly belong to
  Krylov subspaces that remain of size $O(l)$, where $l$ stays finite
  even in the thermodynamic
  limit and cannot thermalize with respect to the full Hilbert space.
  Initial states in class $3$, $4$, $5$ belong to large Krylov spaces
  whose size scale exponentially with $L^2$ when $L \gg 1$. Thus, it is
  not immediately obvious whether such states evade ETH-predicted
  thermalization with respect to the full Hilbert space or not.

Initial states in class $3$ contain an extensive number of inert
down-spins when all the boundary sites of the shielding regions of
the drums are considered together. For initial states in class $4$, one
simply needs to consider local operators that have support from sites on
opposite sides of a quasi-1D drum of linear dimension $O(L)$.
Such local operators evade ETH-predicted thermalization with
  respect to the full Hilbert space
since all the sites that compose such a local operator cannot be part of the
same quantum drum and are, therefore, dynamically disconnected and retain
memory of the initial state.

\begin{figure}[!tbh]
  \begin{center}
    \includegraphics[width=\linewidth]{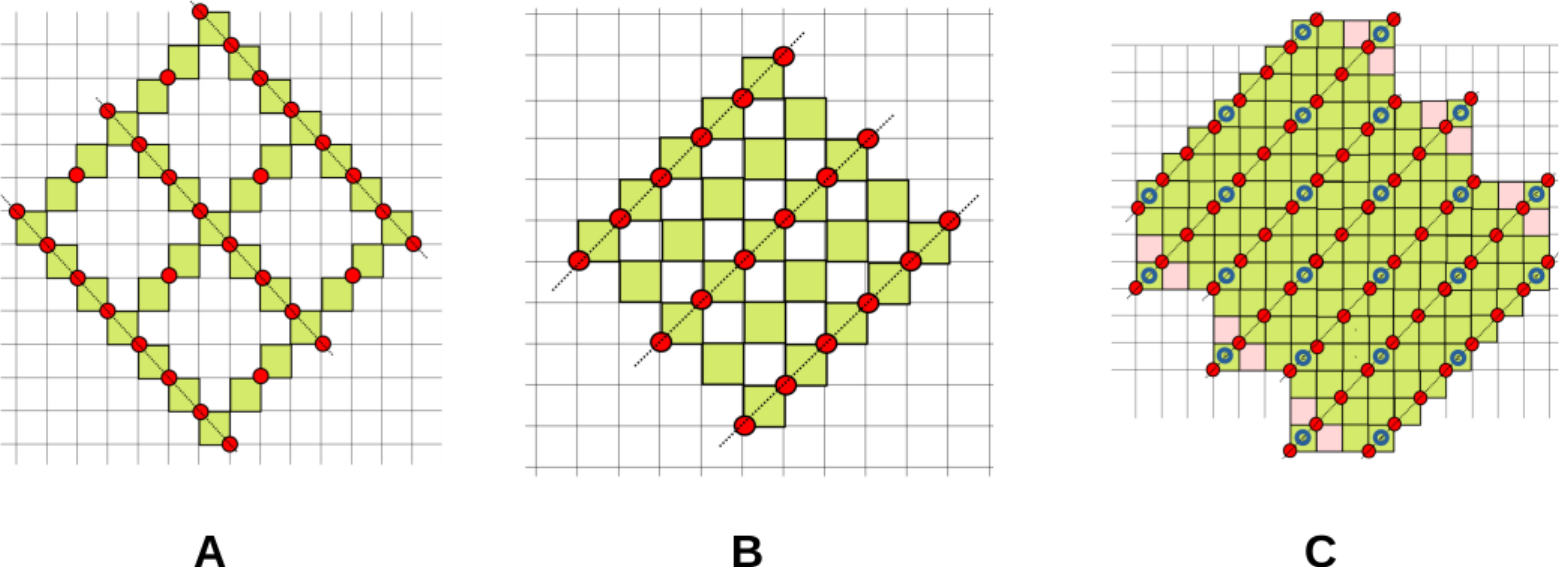}
     \caption{Three examples of 2D drums shown here. The
      green (pink) plaquettes in all the quantum drums
      follow the same convention as
    used in Fig.~\ref{drumsandshields}.
    The filled red dots indicate up-spins in all the panels
    and represent just one of the many
      possible Fock states of the corresponding
      drum. The wire decomposition of each drum is indicated by
      thin dotted lines in each of the three panels. Panels A and B are
      composed of parallel wires that can fluctuate simultaneously
      and hence contain only vertex-sharing plaquettes. The
      checkerboard drum in panel B represents the maximum packing of such wires
      that gives the density of up-spins (bosons) to be $1/4$.
      The close packed drum in panel C represents the maximum packing of
      wires such that none of them are inert that gives the density of
      up-spins (bosons) to be $1/3$. The plaquettes where ring-exchanges
      can be simultaneously carried out are indicated by open
     blue circles at their centres in panel C.  }
    \label{2dDrums}
 \end{center}
\end{figure}

\begin{figure}[!tbh]
  \begin{center}
    \includegraphics[width=0.3\linewidth]{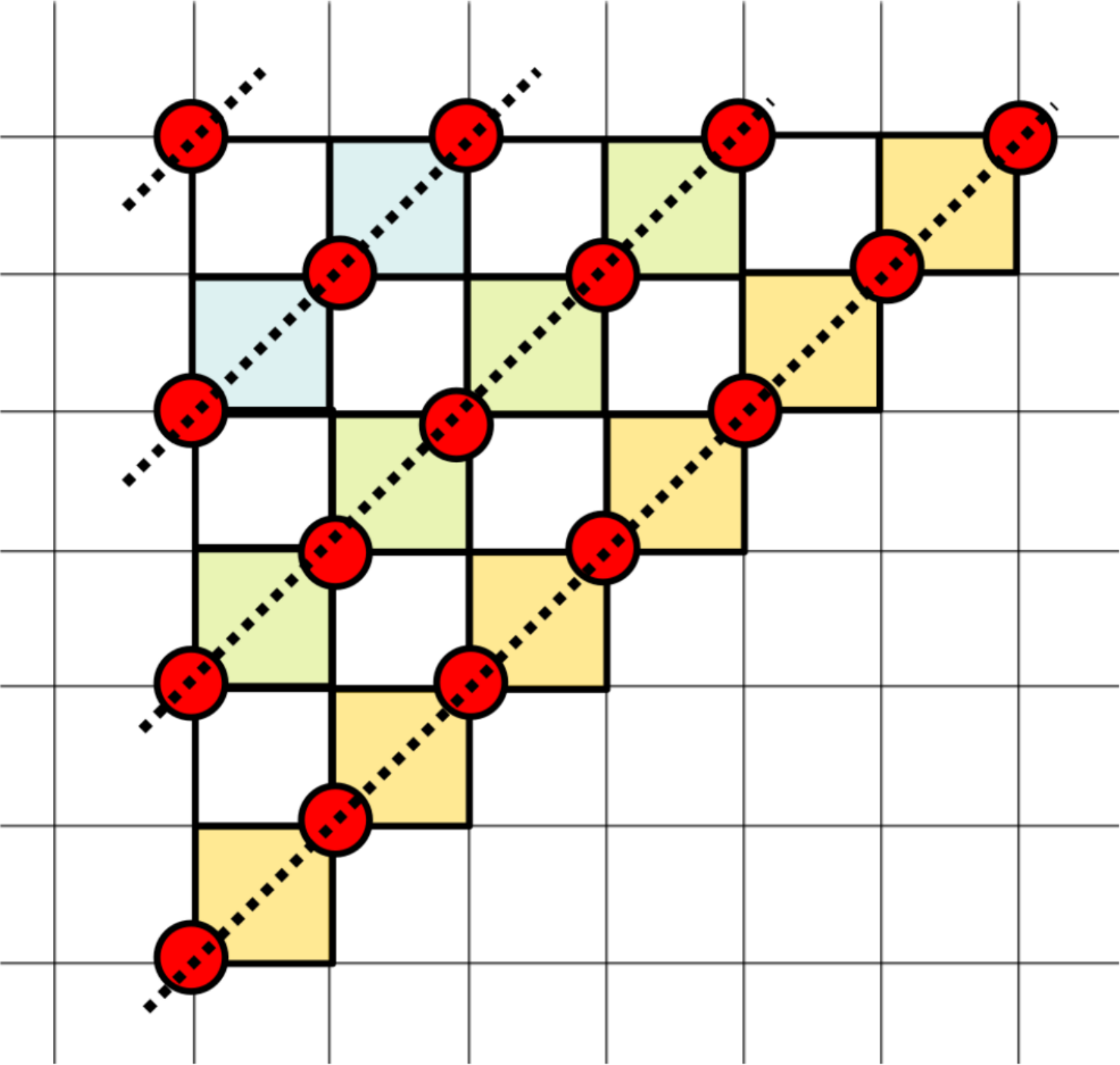}%
    \includegraphics[width=0.35\linewidth]{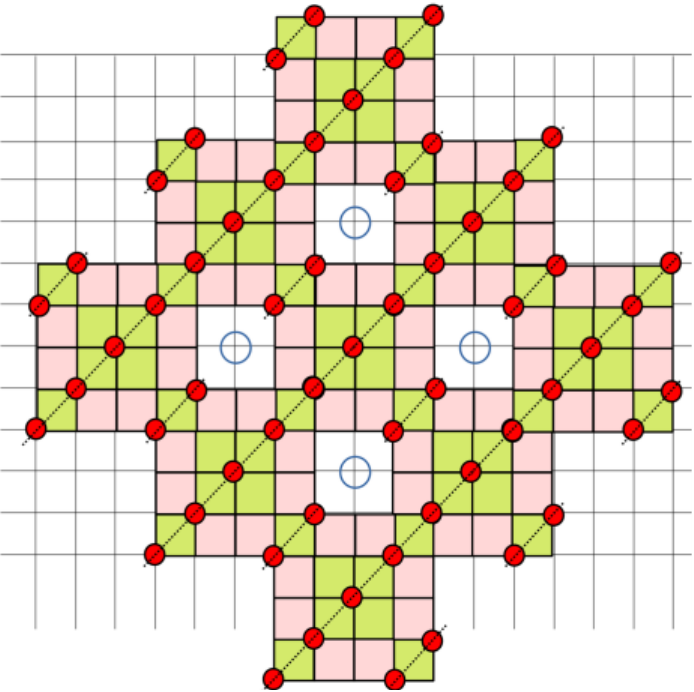}
    \caption{(Left panel) A section of the inert state
      with the maximum density
      of up-spins (bosons) equal to $1/2$ shown with the filled red dots
      indicating up-spins. This state can be
      viewed in terms of parallel wires (indicated by dotted lines and shaded
      plaquettes of different colors) that are placed so close that they
      cannot fluctuate out of their reference states.
      (Right panel) Example of a 2D quantum drum with a density
      of up-spins (bosons) between $1/4$ and $1/3$. In this particular
      example, the density equals $5/18$. The
      green (pink) plaquettes in this quantum drum
      follows the same convention as
    used in Fig.~\ref{drumsandshields}.
    The filled red dots indicate up-spins
    and represents just one of the many
     possible Fock states of the corresponding
      drum. The wire decomposition of the drum is indicated by
      thin dotted lines. The blue
      circles indicate inert down-spins that are not part of this quantum
      drum.
    }
    \label{inert}
 \end{center}
\end{figure}

Initial states in class $5$ are more subtle since almost all local operators
contain sites in the interior of a single 2D quantum drum and
  it is not immediately clear which local operators evade ETH-predicted
  thermalization unlike initial states in class $3$ and $4$.
We take such a
2D quantum drum to cover the entire $L \times L$ lattice without
any loss of generality.
Following Sec.~\ref{wiresdecom}, a reference wire-decomposed
Fock state
for such a 2D drum can be composed by bringing $O(L)$ parallel wires together,
typically of length $O(L)$, in their reference states.
Examples of some 2D drums are shown in
Fig.~\ref{2dDrums} (three panels) and in Fig.~\ref{inert} (right panel).
When macroscopic 2D drums {\it can} be formed, their
  fragment dimensions dominate over those generated by a collection of an
  extensive number of finite-sized drums or a subextensive number of quasi-1D
  drums at the same density of up-spins (bosons) since
  new channels of fluctuations open up in these 2D drums. However,
  at low densities of up-spins (bosons), $n \ll 1$, it
  is clear that macroscopic 2D drums cannot be formed purely from geometric
  considerations and there must exist some critical $n_c$, only
  above which these
  2D drums start dominating statistically.

  In a wire-decomposed Fock state,
  the macroscopic parallel wires cannot be farther than a distance of
  $3\sqrt{2}$ from
  each other so that such wires can be at least coupled to
  each other using unpaired up-spins (bosons). One such example of a 2D drum is
  shown in Fig.~\ref{2dDrums} (panel A) at $n=2/9$.
  While the parallel wires contain $3/4$ of the up-spins (bosons), the
  unpaired up-spins (bosons) account for the rest of the $1/4$ up-spins
  (bosons) contained in the 2D drum. Note that the density of the unpaired
  up-spins (bosons) can be reduced to an arbitrarily low number to still
  give a 2D, though highly anistropic, drum. This immediately gives
  $n_c=1/6$. Thus, typical initial states for $n \in (0,1/6)$ evade
  ETH-predicted thermalization simply from the presence of local operators
  that retain the memory of their initial condition by being a
  class $3$ or class $4$ state.

  For $n>1/6$, it becomes important to be able to
    calculate the scaling of the fragment size of a macroscopic 2D drum
    given its wire-decomposed reference
    state. Given that the number of Fock states accessible to a single wire of
  length
$l$ equals $F_{l+2}$ (Eq.~\ref{wireHSD}) where $F_n$ are the
Fibonacci numbers, the fragment size scales exponentially with increasing $l$
as $\varphi^{l}$ for $l \gg 1$, where
$\varphi$ $=$ $(1+\sqrt{5})/2$ $\approx$ $1.618\cdots$ is the golden ratio.
Thus, if wire-decomposed reference states of 2D drum
consist of parallel wires that can fluctuate simultaneously
to access all their internal states,
then the corresponding number of Fock states generated equals
\begin{eqnarray}
  \varphi^{(L_1+L_2+L_3+L_4+\cdots)} 
  \label{scalingHSD}
  \end{eqnarray}
where $L_1, L_2, L_3, L_4$ etc denote the lengths of these wires, which are
typically $O(L)$ for macroscopic drums. Eq.~\ref{scalingHSD} already shows
that the corresponding
Hilbert space fragment grows exponentially with the
system size.
Just like the case of quantum drums formed out of two
  long parallel wires as
discussed in Sec.~\ref{twowires}, the scaling of the total number of Fock states
in 2D drums composed of a macroscopic number of long parallel wires can be
simply estimated by considering the number of ways in which such wires can
fluctuate simultaneously as given in Eq.~\ref{scalingHSD}. The
{\it missing Fock states} that cannot be accounted for by the wire
fluctuations represented in Eq.~\ref{scalingHSD} only give a
subdominant correction for macroscopic drums. We will show this
explictly by considering two important cases of such 2D quantum drums
below.

We first consider a ``checkerboard'' drum
(see panel B in Fig.~\ref{2dDrums}) which represents the closest packing of
parallel wires (indicated by dotted lines in panel B of
Fig.~\ref{2dDrums}) in their reference state such that all
the wires can fluctuate
simultaneously to access all their states. Comparing the
representative Fock state of this drum shown in Fig.~\ref{2dDrums} (panel B)
to the inert state with the maximum density of up-spins (bosons) that
equals $n=1/2$
(Fig.~\ref{inert}), we see that the former may be obtained
from the latter
by removing the up-spins (bosons) on alternate parallel wires from
the inert state. This fixes the
density of up-spins (bosons) to be $n=(1/2) \times (1/2)=1/4$ for the
checkerboard drum. All the Fock states of this drum can be
generated (in fact, overcounted) by considering simultaneous fluctuations of
wires along either of the diagonal directions of the square lattice
with mutual separation of $2\sqrt{2}$ and also their shifted counterparts
with a shift of $\sqrt{2}$ perpendicular to the direction of the wires.
This immediately establishes that
\begin{eqnarray}
(\varphi^{1/4})^{L^2} <
  \mathcal{N}_\mathrm{HSD,ch} < 4(\varphi^{1/4})^{L^2}
\label{boundch}
\end{eqnarray}
where
$\mathcal{N}_\mathrm{HSD,ch}$ equals the number of Fock states for this drum when
$L \gg 1$. Thus, we get that
\begin{eqnarray}
  \mathcal{N}_\mathrm{HSD,ch} \sim (\varphi^{1/4})^{L^2} \approx (1.12784\cdots)^{L^2}
  \label{checkerboardHSD}
  \end{eqnarray}
for the 2D checkerboard drum that accomodates the maximum density of
simultaneously fluctuating parallel wires, resulting in a density of
up-spins (bosons) that we denote as $n_{\mathrm{ch}}=1/4$ henceforth.

As shown in Sec.~\ref{wiresdecom}, the closest distance of approach between two
parallel wires in their reference state equals $(3/2)\sqrt{2}$
such that these do not become inert. Extending this to 2D, one
    gets a ``close packed drum'' as shown in Fig.~\ref{2dDrums} (panel C)
    where the parallel wires are indicated by dotted lines.
    Unlike the checkerboard drum, this 2D drum is composed of only
    edge-sharing plaquettes and its interior has no unshaded plaquettes that
    do not belong to the drum.
    The density of up-spins (bosons)
    for this close packed drum equals $n=1/3$ which can be seen
    by comparing the reference Fock state shown in
  panel C of Fig.~\ref{2dDrums} to the
  inert state with the maximum density of up-spins (bosons), $n=1/2$,
  (Fig.~\ref{inert}). We see that the former may be obtained
from the latter
by deleting the up-spins (bosons) on every two parallel wires and keeping every
third wire intact from the inert state in a $1-0-0$
pattern. This gives one set of
simultaneously flippable wires of the close packed drum, implying that
$n=(1/6+1/6)=1/3$ since two sets of such wires are needed to make up the
close packed drum. As discussed in Sec.~\ref{wiresdecom}, all Fock states
of such structures where the consecutive parallel wires are at a distance of
$(3/2)\sqrt{2}$ can be generated from two types of
excitations: (a) simultaneous
fluctuations of every alternate parallel wire and (b) excitations of
elementary plaquettes that are separated by $3$ lattice units along $x$ or
$y$ and thus simultaneously flippable. One set of such parallel wires
(indicated by parallel dotted lines) and elementary plaquettes (indicated by
open blue circles in the centers of the corresponding plaquettes) are shown in
Fig.~\ref{2dDrums} (panel C). The scaling of the number of
Fock states associated with the wire
fluctuations can be simply calculated using Eq.~\ref{scalingHSD} and
gives $(\varphi^{1/6})^{L^2}$. The
number of states generated from the simultaneously flippable
elementary plaquettes can be
calculated by simply noting that the result is identical to the one
discussed in Sec.~\ref{intstates} since the elementary plaquettes
have the same spatial arrangement in Fig.~\ref{closepacking} as
the marked plaquettes
in Fig.~\ref{2dDrums} (panel C) thus giving the number of such
excitations as $(2^{1/9})^{L^2}$ (Eq.~\ref{HSDelementaryplaq}).
Furthermore, all the Fock states can
be generated (in fact, overcounted) by considering all combinations of
such parallel wires as well as their perpendicular counterparts
and the simultaneously flippable elementary plaquettes and their lattice
translations. This
gives that
\begin{eqnarray}
(\varphi^{1/6})^{L^2}<
  \mathcal{N}_\mathrm{HSD,cp} < 4[(\varphi^{1/6})^{L^2}+(2^{1/9})^{L^2}]
  \label{boundcp}
\end{eqnarray}
where
$\mathcal{N}_\mathrm{HSD,cp}$ equals the fragment dimension for this drum when
$L \gg 1$. Importantly, since $2^{1/9}/\varphi^{1/6}$ $\approx$ $0.996819
\cdots$, the above
equation can be simplified to give
\begin{eqnarray}
  \mathcal{N}_\mathrm{HSD,cp} \sim (\varphi^{1/6})^{L^2} \approx (1.08351\cdots)^{L^2}
  \label{closepackedHSD}
  \end{eqnarray}
for the close packed drum that accomodates the maximum density of non-inert
parallel wires, resulting in a density of
up-spins (bosons) that we denote as $n_{\mathrm{cp}}=1/3$ henceforth.

Now that we have established that Eq.~\ref{scalingHSD}
  gives the correct scaling of Hilbert space fragment dimension
  for 2D quantum drums using these two examples, a
  straightforward approach to maximize the number of states produced in a
  fragment, given a certain density of
up-spins (bosons), $n >1/6$, is to consider 2D drums
where {\it all} the parallel wires that compose the drums can fluctuate
simultaneously. This automatically lead to drums
made of only vertex-sharing
plaquettes. Since such parallel
wires can only have a minimum separation of $2\sqrt{2}$,
this sets an upper bound on the density of
up-spins (bosons) to be $n_{\mathrm{ch}}=1/4$ (the density for the
checkerboard drum).
Since the checkerboard drum
  maximizes Eq.~\ref{scalingHSD}, it generates the largest Hilbert space
  fragment for this model.
For $n \in (1/6,1/4]$, typical initial states belong to
fragments generated by 2D
  drums composed of only vertex-sharing
  plaquettes since these dominate statistically.

  An example of such a 2D drum at a lower density compared to $n=1/4$
  using unpaired up-spins (bosons) at a distance of
    $(3/2)\sqrt{2}$ to couple consecutive parallel wires that are $3\sqrt{2}$
    distance apart
  is shown in
  panel A of Fig.~\ref{2dDrums}. Comparing the reference Fock state shown in
  panel A of Fig.~\ref{2dDrums} to the
  inert state with the maximum density of up-spins (bosons), $n_b=1/2$,
  (Fig.~\ref{inert}), we see
  that $n=(1/6)\times(1+1/3)=2/9$ for this quantum drum.
  The parallel wires in panel A of Fig.~\ref{2dDrums} can be
  obtained by deleting every two wires in the inert state and keeping every
  third wire intact in a $1-0-0$ pattern while the up-spins that do not
  belong to any wire in panel A of Fig.~\ref{2dDrums} can again be obtained
  by taking the same $1-0-0$ pattern of wires and deleting every two up-spins
  (bosons) and keeping every third up-spin (boson) in the surviving wires.
  The wire decomposition of the drum
    (shown in panel A, Fig.~\ref{2dDrums}) shows that the fragment size
    scales as
    $(\varphi^{1/6})^{L^2}$ $\approx$ $(1.0835\cdots)^{L^2}$ which dominates
    over the fragment produced by closed-packed one-plaquette drums
    $(2^{1/9})^{L^2} \approx (1.08006 \cdots)^{L^2}$ (see Sec.~\ref{intstates})
    at the same density $n=2/9$.

  Crucially, any
  2D drum composed of vertex-sharing plaquettes alone contain an extensive
  number of unshaded plaquettes in its interior (Fig.~\ref{2dDrums}, panels
  A and B) that do not belong to the drum. By definition, the two-spin
  local correlators
$\langle (\sigma^z_{j_x,j_y}+1)(\sigma^z_{j_x+1,j_y+1}+1)\rangle$ and
$\langle (\sigma^z_{j_x,j_y+1}+1)(\sigma^z_{j_x+1,j_y}+1)\rangle$ stay pinned to
zero for any such unshaded plaquette
during the time evolution induced by $H$ as two up-spins
  (bosons) cannot occupy the diagonals for these unshaded plaquettes. Thus, the
  intial states that belong to such 2D drums {\it retain} an extensive amount
  of local memory during time evolution with $H$ and {\it do not} satisfy
  ETH-pedicted thermalization for this model.

Furthermore, while the checkerboard drum (Fig.~\ref{2dDrums}, panel B)
does not contain any inert spin
in its interior, drums made of vertex-sharing plaquettes
with a lower density of up-spins (bosons)
(e.g., Fig.~\ref{2dDrums}, panel A)
also contain an extensive number of sites in their interior
that do not belong to the drum and
thus harbor inert spins.
It is useful to note here that if
{\it all} unentangled initial states that satisfy the hard-core
constraints in Eq.~\ref{constr}
are considered uniformly, then the average density of up-spins (bosons)
equals $\langle n \rangle =
0.226570\cdots$ in the thermodynamic limit~\cite{Baxter1999}. Thus,
drums composed of only vertex-sharing plaquettes
dominate statistically if typical initial
states are considered {\it without putting any
further restriction on $n$}. 

2D drums with the largest fragment dimension for
  $n\in (1/4,1/2]$ cannot be determined by just considering drums
  composed of vertex-sharing plaquettes.
  We will first show that typical initial states that
    belong to large Krylov subspaces always contain an
  extensive number of inert up-spins (bosons) for $n \in (1/3,1/2]$.
    As already discussed in this section, the close packing of long
    parallel wires such that
neighboring wires can still fluctuate generates the close packed drum
(Fig.~\ref{2dDrums}, panel C) and a corresponding density of
$n_{\mathrm cp}=1/3$. Adding an excess amount of up-spins (bosons) in the
system such that $n=(1/3)+\delta$, where $\delta \in (0,1/6)$,
necessarily leads to an extensive number of inert wires
locked in their reference state, with the density of such wires
scaling as $\delta/6$. Thus, the number of inert
  up-spins (bosons) in typical initial states should scale as
  $\delta L^2/6$ for $n=(1/3)+\delta$ with $\delta \in (0,1/6)$ when
  $n\in (1/4,1/2]$. Thus,
  these states also evade ETH-predicted thermalization.

\begin{figure}[!tbh]
  \begin{center}
    \includegraphics[width=0.55\linewidth]{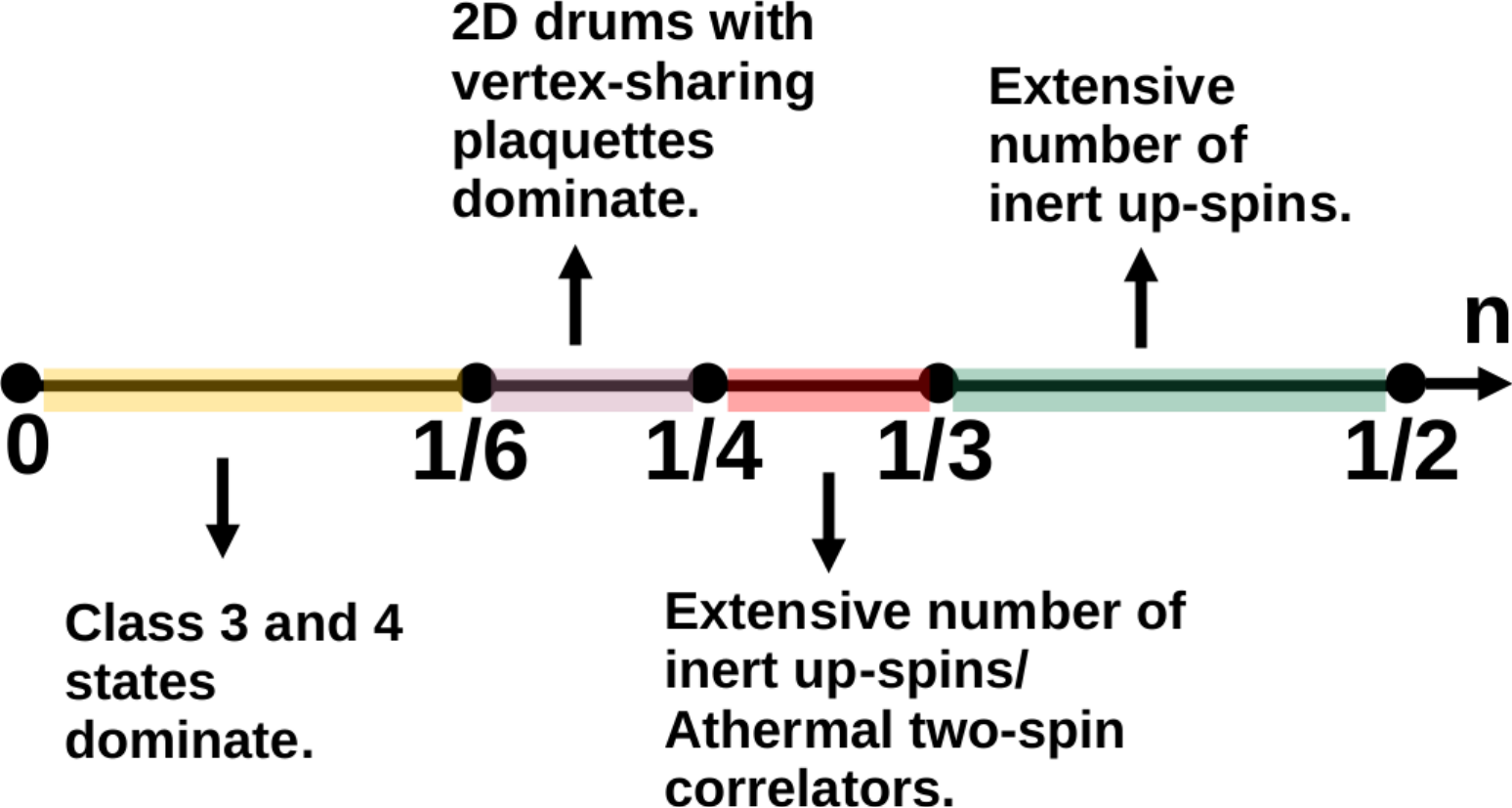}%
    \caption{The fate of a typical
      initial state with a macroscopic number of up-spins (bosons), $n$,
      under unitary time evolution with $H$ is
      shown. Below $n_c=1/6$, a typical initial state is either consistent
      with an extensive number of finite-sized drums (class $3$) or a
      subextensive number of quasi-1D drums (class $4$). For $n \in (1/6, 1/4)$,
      typical initial states belong to fragments produced by macroscopic
      2D drums with vertex-sharing plaquettes. For $n \in (1/4, 1/3)$,
      a typical initial state contains either an extensive number of inert
      up-spins (bosons) or an extensive number of athermal next-nearest
      neighbor two-spin correlators. For $n>1/3$, a typical initial state
      always contains an extensive number of inert up-spins. The
      largest Hilbert space fragment is produced by the 2D checkerboard
      drum (panel B of Fig.~\ref{2dDrums}) at $n=1/4$. 
    }
    \label{phasedia}
 \end{center}
\end{figure}

We now come to the nature of large Krylov subspaces where the
  density of up-spins (bosons) equals $n \in (1/4,1/3)$ such that we
  can write $n=(1/4)+\gamma$ with $\gamma \in (0,1/12)$. 
    Three different possibilities exist here. \\
  (i) One can start with the reference state of the checkerboard drum
  in Fig.~\ref{2dDrums} (panel B) and insert extra parallel
  wires in their reference state such that the distance between parallel wires
  equals $\sqrt{2}$ for a linear extent of $4\gamma L$ of the system while
  the rest of the system still has parallel wires that can fluctuate
  simultaneously. However, this immediately produces $O(4\gamma L^2)$ inert
  up-spins (bosons) in the system and thus such initial states evade
  thermalization. Since the size of the spatial region that harbors
  wires that can simultaneously fluctuate reduces from $L^2$ to
  $(1-4\gamma)L^2$, the fragment size scales as
  $\left(\varphi^{\frac{1}{4}-\gamma}\right)^{L^2}$ using Eq.~\ref{checkerboardHSD}
  in this case.\\
  (ii) One can have ``phase-separated'' drums, with the phase separation in
  one direction, such that different macroscopic regions are
  composed of close packed parallel wires in their reference
  state. One of these sets have parallel wires at a distance
  $(3/2)\sqrt{2}$ (with a local density $n=1/3$)
  for a total linear extent of $(12\gamma)L$. The other regions comprise of
  parallel wires in their reference state
  at a distance $2\sqrt{2}$ (with a lower local density of $n=1/4$)
  for the rest of the system
  to get the correct up-spin (boson) density. We refer the reader to the
  drum marked as C$_2$ in Fig.~\ref{FIGw} for an illustration of
  this principle for a small drum where the parallel
  wires w$_1$ and w$_2$ (w$_2$ and w$_3$) are at a distance $(3/2)\sqrt{2}$
  ($2\sqrt{2}$) with respect to each other. The leading scaling for
  the Hilbert space size of such drums can simply be obtained by
  considering independent fluctuations
  of these ``checkerboard drum'' and ``close packed drum'' regions. Using
  Eq.~\ref{checkerboardHSD} and Eq.~\ref{closepackedHSD}, this immediately
  gives
  \begin{eqnarray}
    \mathcal{N}_{\mathrm{HSD,phase-sep}} \sim [\varphi^{1/6}]^{12 \gamma L^2} [\varphi^{1/4}]^{(1-12\gamma)L^2} \sim \left(\varphi^{\frac{1}{4}-\gamma}\right)^{L^2}.
    \label{phasesepHSD}
    \end{eqnarray}
  (iii) One can
  start with the close packed drum reference state and delete up-spins (bosons)
  in such a manner that another 2D drum with a lower density of
  up-spins (bosons) is created. We refer the reader to the 2D drum shown
  in the right panel of Fig.~\ref{inert} where the reference Fock state
  of this particular drum can be produced from the reference Fock state
  of the close packed drum by deleting every third up-spin (boson) from
  alternate wires in their reference state. This gives a reduced
  density of up-spins (bosons) to be $(1/6)(1+2/3)$ $=$ $5/18$ $\approx$
  $0.2777\cdots$ as well as an extensive density of inert down-spins
  from sites that are not part of the drum. However, the scaling of the
  fragment size for such drums is upper-bounded by
  $\left(\varphi^{\frac{1}{6}}\right)^{L^2}$.\\

  Fragments in (i) and
    (iii) dominate statistically
  for $n \in (1/4,1/3)$ and typical initial states thus either contain
  an extensive number of inert spins or an extensive number
  [$O((1-12\gamma)L^2)$] of 
  unshaded plaquettes in the interior of ``phase-separated'' drums
  where two-spin
  local correlators
  $\langle (\sigma^z_{j_x,j_y}+1)(\sigma^z_{j_x+1,j_y+1}+1)\rangle$ and
  $\langle (\sigma^z_{j_x,j_y+1}+1)(\sigma^z_{j_x+1,j_y}+1)\rangle$ stay pinned to
  zero, thus evading ETH-predicted thermalization.
  
This completes our analysis for the lack of ETH-predicted thermalization
  in typical initial states for all $n \neq 1/3$.
  We see that typical initial states have either an
  extensive number of inert
  spins or an extensive number of two-spin next-nearest neighbor correlators
  that are pinned to athermal values or both. The situation
    is summarized in Fig.~\ref{phasedia} as a function of the density of
    up-spins (bosons), $n$. 
  
The case of the close packed drum (Fig.~\ref{2dDrums}, panel C)
  with density of up-spins (bosons)
  $n=1/3$ with a Hilbert space fragment whose dimension scales as
  $\left(\varphi^{\frac{1}{6}}\right)^{L^2}$ seems more subtle since it contains
  neither inert spins nor unshaded plaquettes (that harbor athermal next-nearest
  neighbor spin correlations) in its bulk.
  However, initial states that arise from a wire pattern
  composed of $(2/3)L^2$ of the system in a
  checkerboard drum pattern, with a local density $n=1/4$,
  and the rest of the system being fully inert
  with a local $n=1/2$ also yields the same leading scaling of its fragment
  size as $\left(\varphi^{\frac{1}{6}}\right)^{L^2}$.
  Thus, it is clear that there is no single dominant
    Krylov subspace even at $n=1/3$. We leave the issue of thermalization, or
    lack of it, or of an even more exotic feature like a behavior intermediate
    to both weak and strong Hilbert space fragmentation for the particular
    density of up-spins (bosons), $n=1/3$, as an interesting open problem.
    For completeness, we note that ED results on small systems points towards
    a strong Hilbert space fragmentation scenario even at $n=1/3$
    (Fig.~\ref{FIGnum}, right panel).

\section{Analytical study of small quantum drums}
\label{astud}

In this section, we shall study the spectrum of some of the simplest
quantum drums of the model analytically. We first discuss how the connection
diagrams between different Fock states of a drum, where the connections are
generated by $H$, can be represented by unidirectional trees in
Sec.~\ref{nstree}. Such tree structures turn out to be particularly useful
in finding the non-zero matrix elements of $H$ for large drums
(e.g., see Sec.~\ref{nstud}).
We will subsequently study the
case of a wire with $N_p$ plaquettes for small $N_p$ (Sec.~\ref{astwire}),
and then
consider some other examples of small quantum drums that can be
viewed as building blocks of the different kinds of wire junctions
shown in Fig.~\ref{drumvarieties} (Sec.~\ref{astjunc}).

\subsection{Tree structure}
\label{nstree}
It is convenient to represent
the connection diagram between different Fock states in the Hilbert
space of a drum as nodes of a tree with the non-zero matrix elements of
$H$ (which equals $1$ due to the form of $H$ in Eq.~\ref{hamdef})
between two such states denoted as a link between the corresponding nodes.
Such a tree can be build in the forward direction with the different levels
being denoted by integers starting from level-$0$ for the top level and
being incremented by $1$ for each of the levels below. The top level consists
of a single node that can be represented by any ``reference state''.
It is optimal to choose a reference state such that the Fock state maximizes
the number of flippable plaquettes for the corresponding drum, but this is not
a necessary
condition.
A single application of $H$ on this reference
state at level-$0$ generates all the nodes at level-$1$, where the corresponding
Fock states have exactly $1$ flipped plaquette with respect to the
reference state with the location of the flipped plaquette uniquely
identifying the corresponding
Fock state.
Links are then formed between nodes at level-$0$ and level-$1$.
Applying $H$ on each of the level-$1$ nodes generates level-$2$ nodes where the
corresponding Fock state has another flipped plaquette with respect to the
level-$1$ state with the locations of the two flipped plaquettes
characterizing the generated Fock state uniquely.
The possibility that different level-$i$ nodes may generate the same
level-$(i+1)$ node first arises at $i=1$. New links are then drawn between the
appropriate nodes at level-$1$ and level-$2$. This process is continued
recursively at each subsequent level-$i$ to go forward to level-$(i+1)$ during
which the links between appropriate level-$i$ and
level-$(i+1)$ nodes are also generated.
Carrying out this forward construction of the tree, one
also encounters ``dead nodes'' which are Fock states at level-$i$ from which no
other Fock states with an extra flipped plaquette can be generated to go to
the next level ($i+1$). The forward construction of the tree terminates when the
last level is reached which is characterized by all its nodes being dead nodes.
 A plaquette, once flipped, cannot be unflipped in the tree
construction which makes the construction unidirectional.

\subsection{Wire}
\label{astwire}
\begin{figure}[!tbh]
  \begin{center}
    \includegraphics[width=0.6\linewidth]{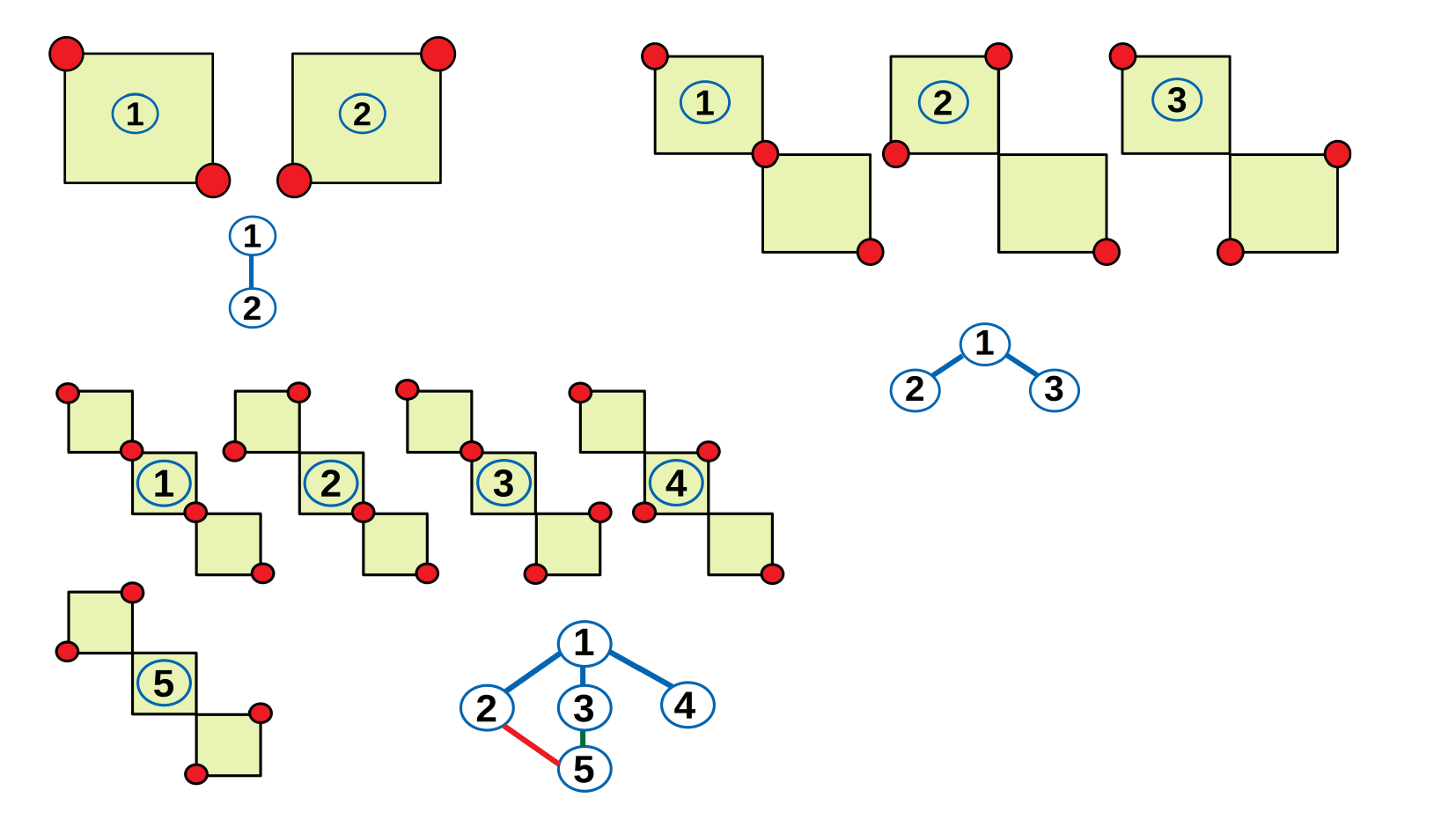}
    \caption{Left
Panel:(a) Schematic representation of the basis states of the
simplest fragment with $N_p=1$ and (b) the corresponding tree
between states in the Hilbert space. Center Panel: Same as
the left panel but corresponding to $N_p=2$. Right Panel: Same as
the left panel but corresponding to $N_p=3$. In all plots, the red
filled dots indicates sites with up-spins (bosons).
The green plaquettes in the drums follow the same convention as
    used in Fig.~\ref{drumsandshields}.}
    \label{smallwire}
\end{center}
\end{figure}
The simplest quantum drum
of the Hamiltonian given by Eq.\ \ref{hamdef}
constitutes a single plaquette ($N_p=1$) with two up-spins (bosons) as shown
in the left panel of Fig.\ \ref{smallwire}. The Hilbert space of this drum
constitutes
two states; the Hamiltonian in the space of these two states,
$|\psi_a\rangle \equiv |a\rangle$ for $a=1,2$, can be written as
(Fig.\ \ref{smallwire})
\begin{eqnarray}
H_{1\ell} &=& \tau^x, \quad  H_{1\ell} |\psi_{1(2)}\rangle =
|\psi_{2(1)}\rangle \label{oneline}
\end{eqnarray}
where $\tau^x$ denotes Pauli matrix in the space of the states in
the Hilbert space. This yields integer eigenvalues $E=\pm 1$.

The next set of quantum drums that we discuss constitutes two elementary
square plaquettes ($N_p=2$) with three up-spins (bosons) in total
as shown in the central panel of Fig.\ \ref{smallwire}.
The Hilbert space consists of three
states, $|\phi_a\rangle \equiv |a\rangle$ for $a=1,2,3$, as shown in
the panel. The action of the Hamiltonian is summarized by the
tree given in the central panel of Fig.\ \ref{smallwire}.
In the space of these states, the Hamiltonian can be represented as
\begin{eqnarray}
H_{2\ell} &=& \left(\begin{array}{ccc} 0 & 1 & 1 \\ 1 & 0 & 0 \\ 1 &
0 & 0 \end{array} \right) \label{twoline}
\end{eqnarray}
This yields eigenvalues $E=0, \pm \sqrt{2}$. Thus these fragments
leads to eigenenergies which can be represented by simple irrational
numbers as well as a zero mode.

Finally, we consider a wire with $N_p=3$ where the states have 4
up-spins (bosons). The basis states spanning the $5$-dimensional Hilbert
space of such a fragment is charted in the right panel of Fig.\
\ref{smallwire} and the corresponding tree is shown in the
bottom panel of the figure. As can be read off from the tree,
in the space of these states, the $5 \times 5$ Hamiltonian
matrix can be written as
\begin{eqnarray}
H_{3 \ell} &=&  \left(\begin{array}{ccccc} 0 & 1 & 1 & 1 & 0\\ 1 & 0 & 0 & 0 & 1 \\
1 & 0 & 0 & 0 & 1\\ 1 & 0 & 0 & 0 & 0 \\ 0 & 1 & 1 & 0 & 0
\end{array} \right) \label{threeline}
\end{eqnarray}
The corresponding eigenvalues are given by $E=0, \pm \sqrt{5\pm
\sqrt{17}}$ leading to eigenvalues represented by non-trivial
irrational numbers and a zero mode. The spectrum of these wires for
larger $N_p$
gets complicated and
these shall be studied in details numerically in Sec.\ \ref{nstud}.

\subsection{Junction units}
\label{astjunc}
\begin{figure}[!tbh]
  \begin{center}
\includegraphics[width=0.7\linewidth]{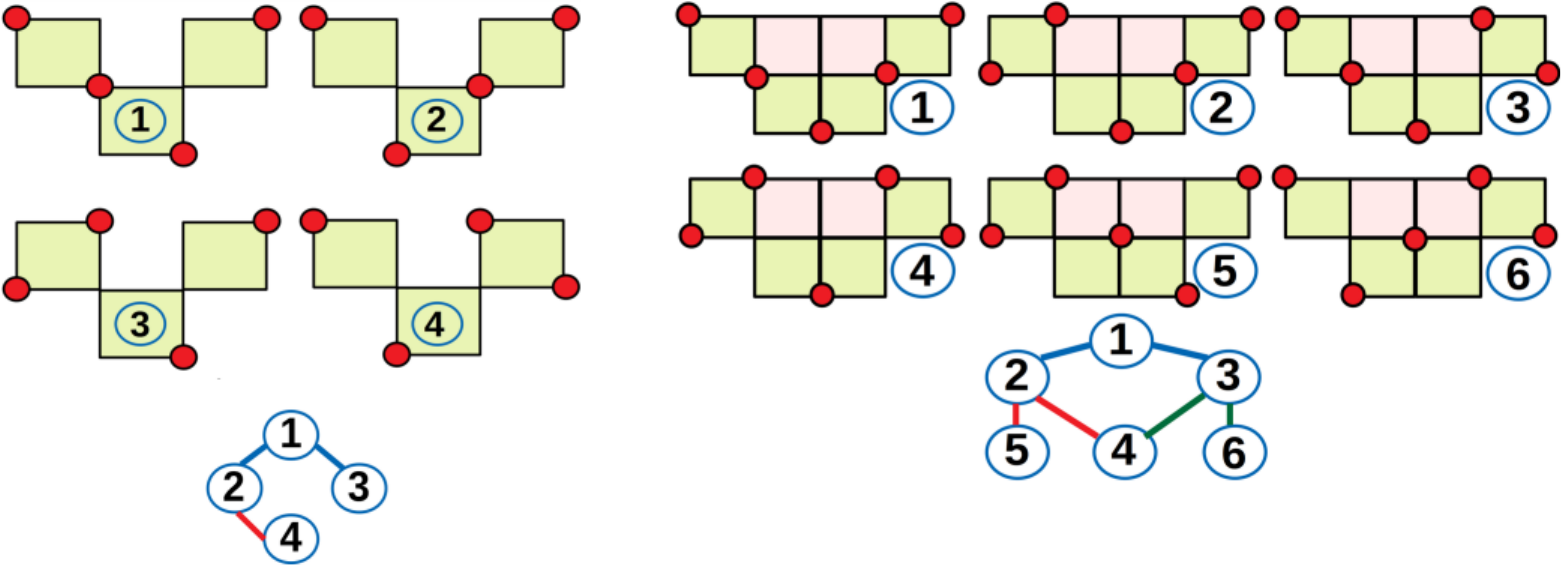}
 \caption{Left
Panel:(a) Schematic representation of the basis states of the
quantum drum corresponding to a junction of two wires
as shown in Fig.~\ref{drumvarieties} (A) with $N_p=3$
and (b) the corresponding
tree between the four states in the Hilbert space.
Right Panel: Same as the left panel but corresponding
to the simplest quantum drum (with $N_p=4$) that can be treated as the
junction unit to generate another junction of two wires as shown in
Fig.~\ref{drumvarieties} (B). In all plots, the red filled dots
indicates sites with up-spins (bosons). The green (pink) plaquettes in the
      drums follow the same convention as
    used in Fig.~\ref{drumsandshields}.} \label{junctiontwowires}
 \end{center}
\end{figure}
In this section, we shall study small quantum drums corresponding to the
simplest junction units that are building blocks of the different
junctions of wires shown in Fig.~\ref{drumvarieties} (A, B, C, D)
and calculate their spectra analytically. Larger quantum drums that
resemble a junction of two wires as shown in Fig.~\ref{drumvarieties} (A)
shall be studied numerically in Sec.\ \ref{nstud}.

We begin with the quantum drum corresponding to a junction of two wires
as shown in Fig.~\ref{drumvarieties} (A) with $N_p=3$ elementary plaquettes. The
basis states corresponding to such a junction is shown in the left
panel of Fig.\ \ref{junctiontwowires}. The Hilbert space, as can be seen from
this figure, is four dimensional. The tree for the
states in the Hilbert space is shown in the bottom of the left panel
Fig.\ \ref{junctiontwowires}. A straightforward analysis shows that $H$ admits a
four-dimensional matrix representation in the space of these states
which can be written in terms of outer product of two sets of Pauli
and identity  matrices ($\vec \tau_a$ and $I_a$ for $a=1,2$) as
\begin{eqnarray}
H_{1j} &=& \tau_1^x \otimes (I_2 +\tau_2^z)/2 + I_1 \otimes \tau_1^x
\label{ham1jn}
\end{eqnarray}
The corresponding eigenvalues satisfy the characteristic equation
$E^4-3E^2+1=0$ and yields a solution $E= \pm (1\pm \sqrt{5})/2$.
These eigenvalues therefore yield the golden ratio for this particular
drum.

Next, we consider the simplest quantum drum that can be treated as the
junction unit to generate another junction of two wires as shown in
Fig.~\ref{drumvarieties} (B). This unit corresponds
to a drum with $N_p=4$ as shown in the right panel of
Fig.\ \ref{junctiontwowires}. The
basis states spanning the six-dimensional Hilbert space is shown in
the top of the right panel of Fig.\ \ref{junctiontwowires} while the tree
for these states is shown in the bottom of this figure. We
find that the Hamiltonian has a $6 \times 6$ matrix representation
given by
\begin{eqnarray}
H_{3 \ell} &=&  \left(\begin{array}{cccccc} 0 & 1 & 1 & 0 & 0 & 0\\ 1 & 0 & 0 & 1 & 1 & 0 \\
1 & 0 & 0 &  1 & 0 & 1 \\ 0 & 1 & 1 & 0 & 0 & 0 \\  0 & 1 & 0 & 0 &
0 & 0 \\ 0 & 0 & 1 & 0 & 0 & 0
\end{array} \right) \label{ham2jn}
\end{eqnarray}
The characteristics equation for the eigenvalues simplifies to
$E^2(E^2-1)(E^2-5)=0$ and yields eigenvalues $E= 0$ (doubly
degenerate) and $E=\pm 1, \pm \sqrt{5}$.
\begin{figure}[!tbh]
  \begin{center}
    \includegraphics[width=0.8\linewidth]{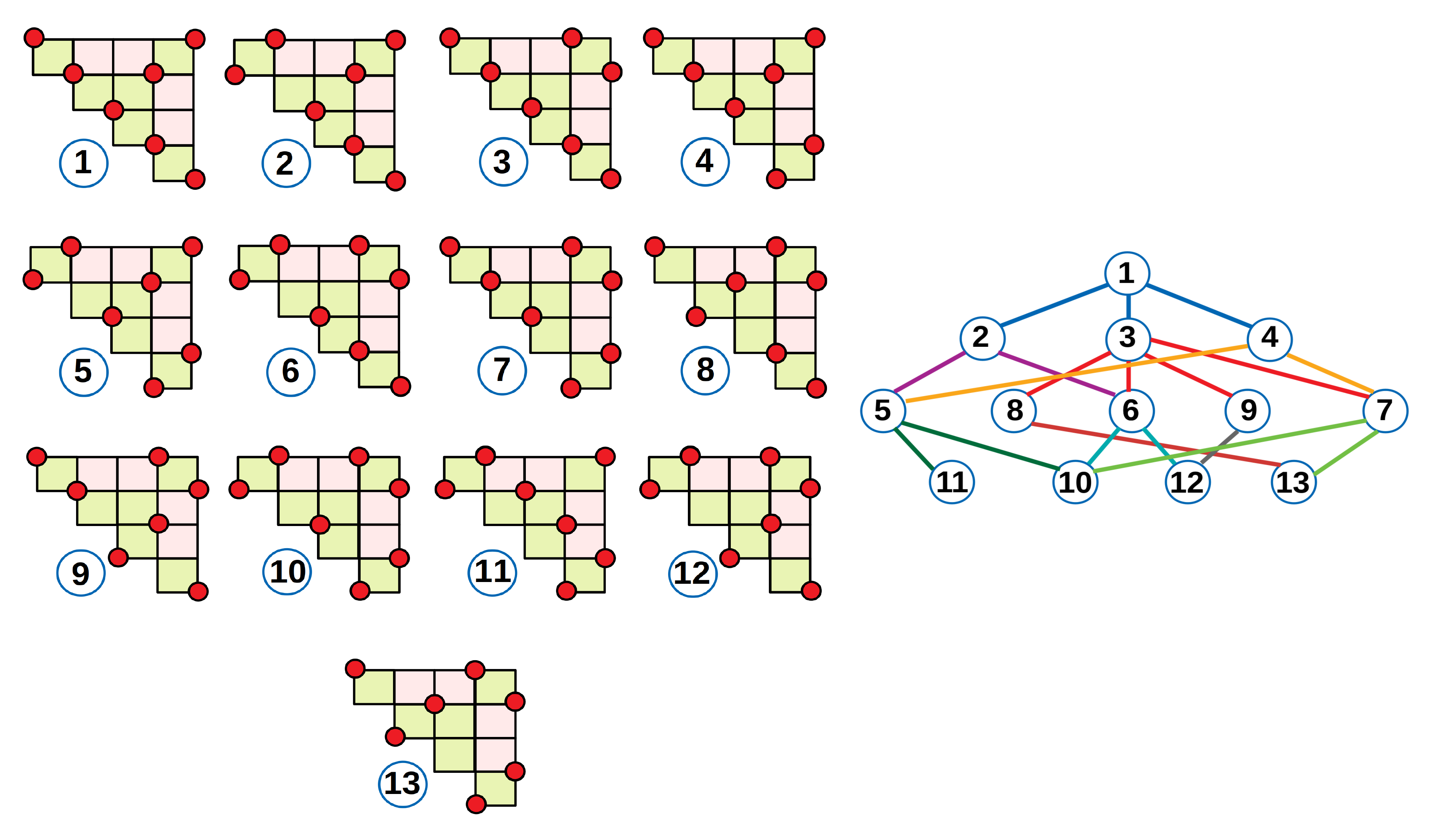}
 \caption{Left
   Panel: Schematic representation of the basis states of the
simplest quantum drum that can be viewed
as the junction unit of the junction of three wires shown
in Fig.~\ref{drumvarieties} (C) with $N_p=10$ plaquettes.
The red filled circles indicates sites with
up-spins (bosons). Right Panel: The corresponding tree between the
$13$ states in the Hilbert space. The green (pink) plaquettes in the
      drum follows the same convention as
    used in Fig.~\ref{drumsandshields}.} \label{junctionthreewires}
 \end{center}
\end{figure}

Next, we consider the simplest quantum drum that can be viewed
as the junction unit of the junction of three wires shown
in Fig.~\ref{drumvarieties} (C). This drum has $N_p=10$ plaquettes in it.
The basis states spanning the Hilbert space is shown
in the left panel of Fig.\ \ref{junctionthreewires}. The Hilbert space is $13$
dimensional; the tree for these states is shown in the
right panel of Fig.\ \ref{junctionthreewires}. This allows a $13 \times 13$
dimensional matrix representation of $H$. We do not write this
matrix explicitly here since it can be easily constructed from the
tree shown in Fig.\ \ref{junctionthreewires}. This matrix needs to
be numerically diagonalized and yields eigenvalues $E=$
$\pm 3.18259\cdots$, $\pm 1.91182\cdots$, $\pm (1 \pm \sqrt{5})/2$,
$\pm 1$, $0$,  and $\pm 0.464856\cdots$.

\begin{figure}[!tbh]
  \begin{center}
    \includegraphics[width=0.8\linewidth]{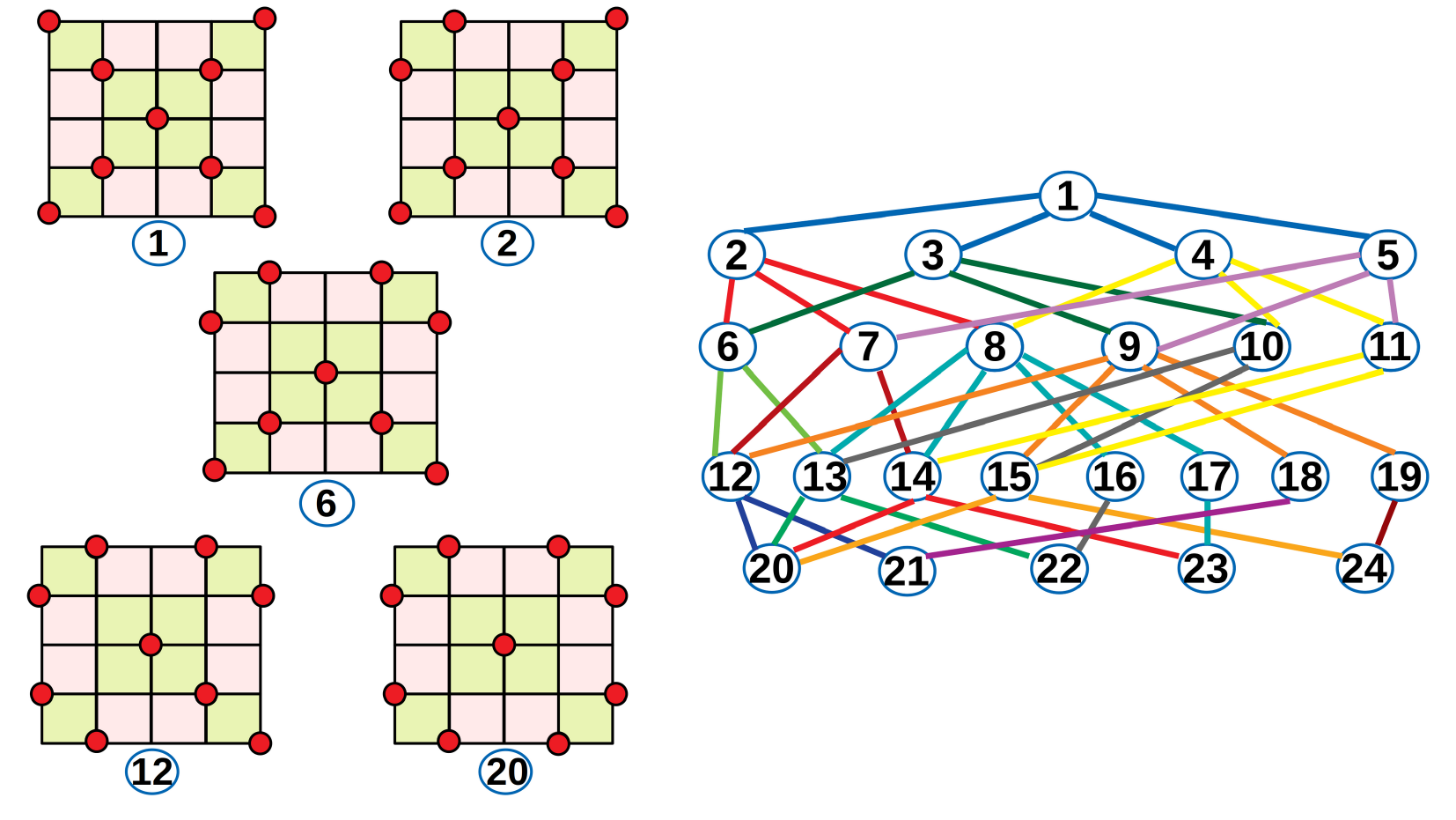}
 \caption{Left
Panel: Schematic representation of a few basis states of the
drum that can be viewed as the junction unit
of a junction of four wires as shown in Fig.~\ref{drumvarieties} (D).
The drum contains $N_p=16$ plaquettes. The red filled circles indicates sites
with up-spins (bosons). Right Panel: The corresponding tree
between the $24$ states in the Hilbert space. Each of the states can
be obtained from a state connected to it by application of $H$ on
it. The green (pink) plaquettes in the
      drum follows the same convention as
    used in Fig.~\ref{drumsandshields}. See text for details} \label{junctionfourwires}
 \end{center}
\end{figure}
Finally, we study a quantum drum that is the junction unit
of a junction of four wires as shown in Fig.~\ref{drumvarieties} (D).
This junction unit is shown in Fig.\ \ref{junctionfourwires} and corresponds to
$N_p=16$ with nine
up-spins (bosons) and $24$ basis states. A few representative
such states are shown in the left panel of Fig.\ \ref{junctionfourwires}.
Each of these states belong to a different level in the tree
starting from the state $|1\rangle$ as shown in the right panel of
Fig.\ \ref{junctionfourwires}; they can be obtained from the state in the
preceding level shown by application of $H$. The other states can be
analogously obtained following the tree; we do not
show them explicitly to avoid clutter. The tree shows
that $H$ admits a $24 \times 24$ matrix representation. Remarkably,
this matrix can be analytically diagonalized; its eigenvalues
satisfies the characteristics equation
\begin{eqnarray}
E^6 (E^2-2 E - 2) (E^2-2) (E^4 - 22 E^2 + 80) (E^4 - 6 E^2 +6)^2
(E^2+2 E-2) &=& 0  \label{n4char}
\end{eqnarray}
These leads to the $24$ eigenvalues given by $0$ (six fold
degenerate), $\pm \sqrt{ 3 \pm \sqrt{3}}$ (each two fold
degenerate), $\pm \sqrt{11\pm \sqrt{41}}$, $\pm \sqrt{2}$, and $\pm
1\pm \sqrt{3}$.



\section{Numerical study of two quasi-1D quantum drums}
\label{nstud}
\begin{figure}[!tbh]
  \begin{center}
    \includegraphics[width=0.6\linewidth]{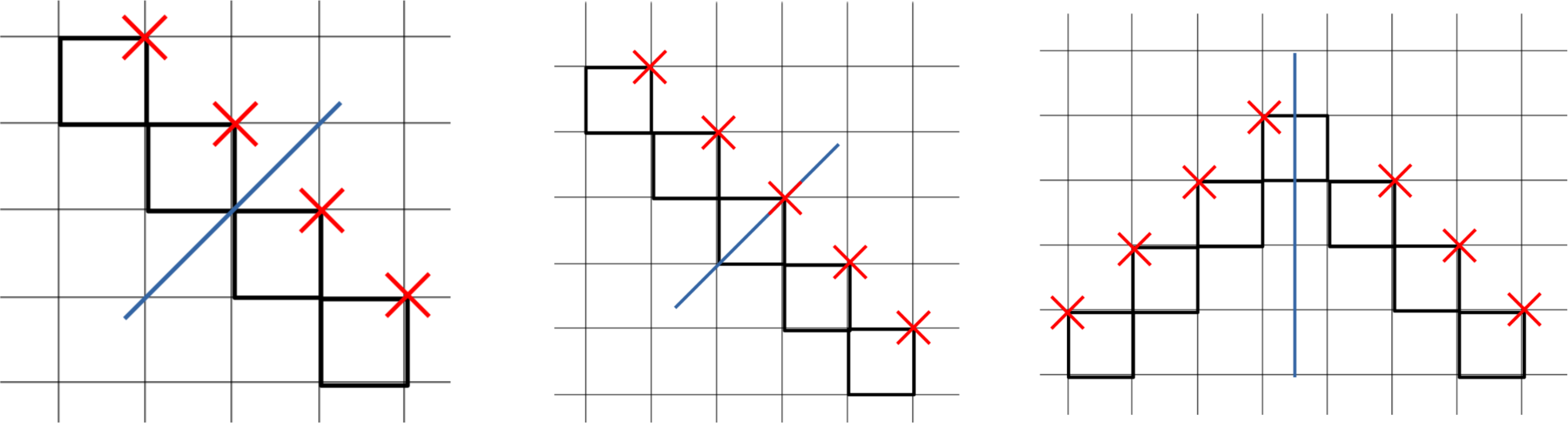}
    \caption{Two wires with $N_p=4$ (left panel) and $N_p=5$ (middle panel)
      respectively
      and a junction of two wires with $N_p=7$ (right panel)
      are illustrated here. These quantum drums have a discrete reflection
      symmetry with the thick blue line in all the three panels indicating the
      corresponding axis of reflection. A chiral operator can be constructed
      from the product of $\sigma^z$ on all sites indicated by $\times$ (in red)
    for these drums.} \label{FIG10}
 \end{center}
\end{figure}

In this section, we will numerically calculate the spectrum of large
quantum drums with $N_p$ elementary plaquettes
using the examples of a wire (Fig.~\ref{FIG10}, left and middle panels
have $N_p=4$ and $5$, respectively) and a particular junction of
two equal length wires (Fig.~\ref{FIG10}, right panel with $N_p=7$).
We refer to this latter case as ``junction of two wires''
henceforth. Using exact diagonalization (ED), we could calculate the
spectrum up to
$N_p=22$ for the wire and $N_p=23$ for the junction of two wires.
We will show that the spectrum of a wire with $N_p$ plaquettes is
identical to the paradigmatic 1D PXP chain~\cite{Turner2018a, Turner2018b}
with $N_p$ sites on a chain with
OBC. This equivalence allows us to extract several features of the high-energy
spectrum of the wire from known results in the
literature~\cite{Turner2018a, Turner2018b}.
However, our numerical studies also reveal enhanced fidelity revivals
from a period-$3$ initial state for $N_p=3n+1$, where $n$ is an integer,
without the need of adding any optimal perturbations to the Hamiltonian which
was not pointed out earlier in the literature.
While the junction of two wires differs from the wire by only a
``surface term'' when $N_p$ is large, the structure of the Hilbert space
is completely different and gives a different constrained model compared
to the 1D PXP chain. Thus, the presence or absence
of a single junction leads to interesting differences in the high-energy
spectrum that persist for large drums.

\subsection{Tree generating algorithm and equivalence of wire to
1D PXP chain}
\label{treelargedrums}
\begin{figure}[!tbh]
  \begin{center}
    \includegraphics[width=0.4\linewidth]{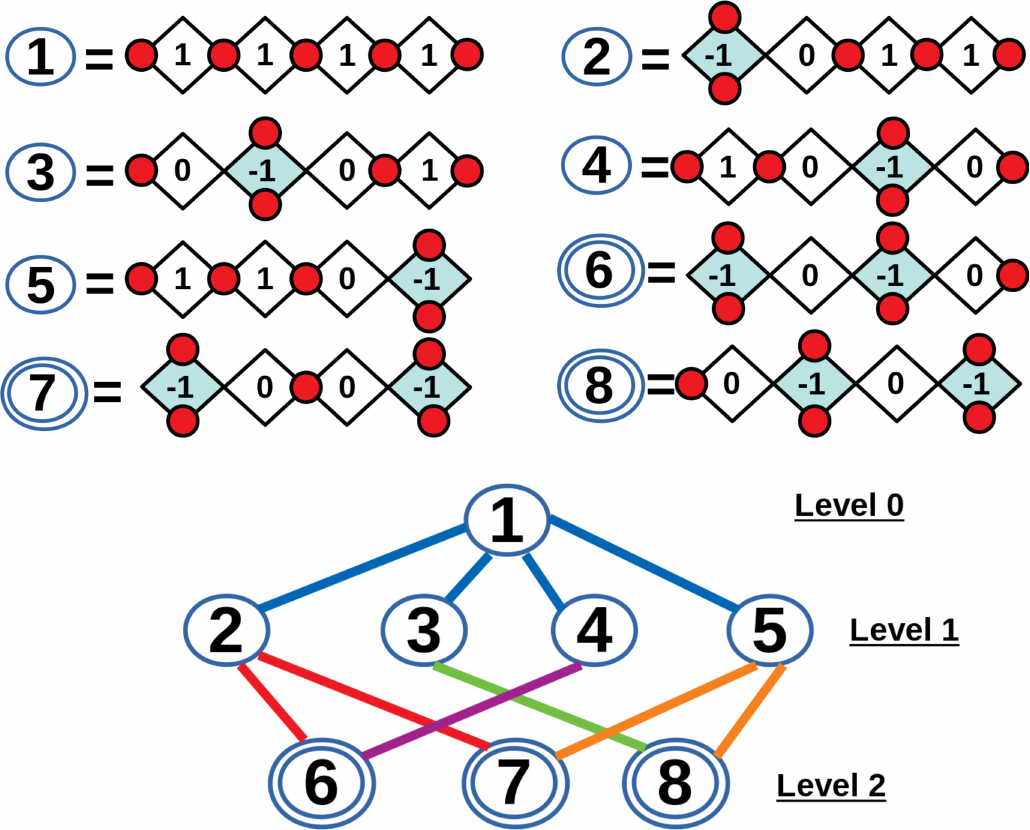}
    \caption{Tree structure (bottom panel)
      for a wire with $N_p=4$ shown here. The $i$th
      level contains Fock states with $i$ flipped plaquettes
      (indicated by shaded plaquettes in the top panel)
      with respect to the reference state defined
      in level-$0$. The nodes between two Fock states
      imply that these are connected with a single application of $H$.
      The Fock states enclosed by double circles represent dead
      ends of the tree. The corresponding value
      of the pseudospin variable $(\pm 1,0)$ is also shown at the center of
      each plaquette for each of the Fock states in the top panel. }
    \label{FIG11}
 \end{center}
\end{figure}
\begin{figure}[!tbh]
  \begin{center}
     \includegraphics[width=0.9\linewidth]{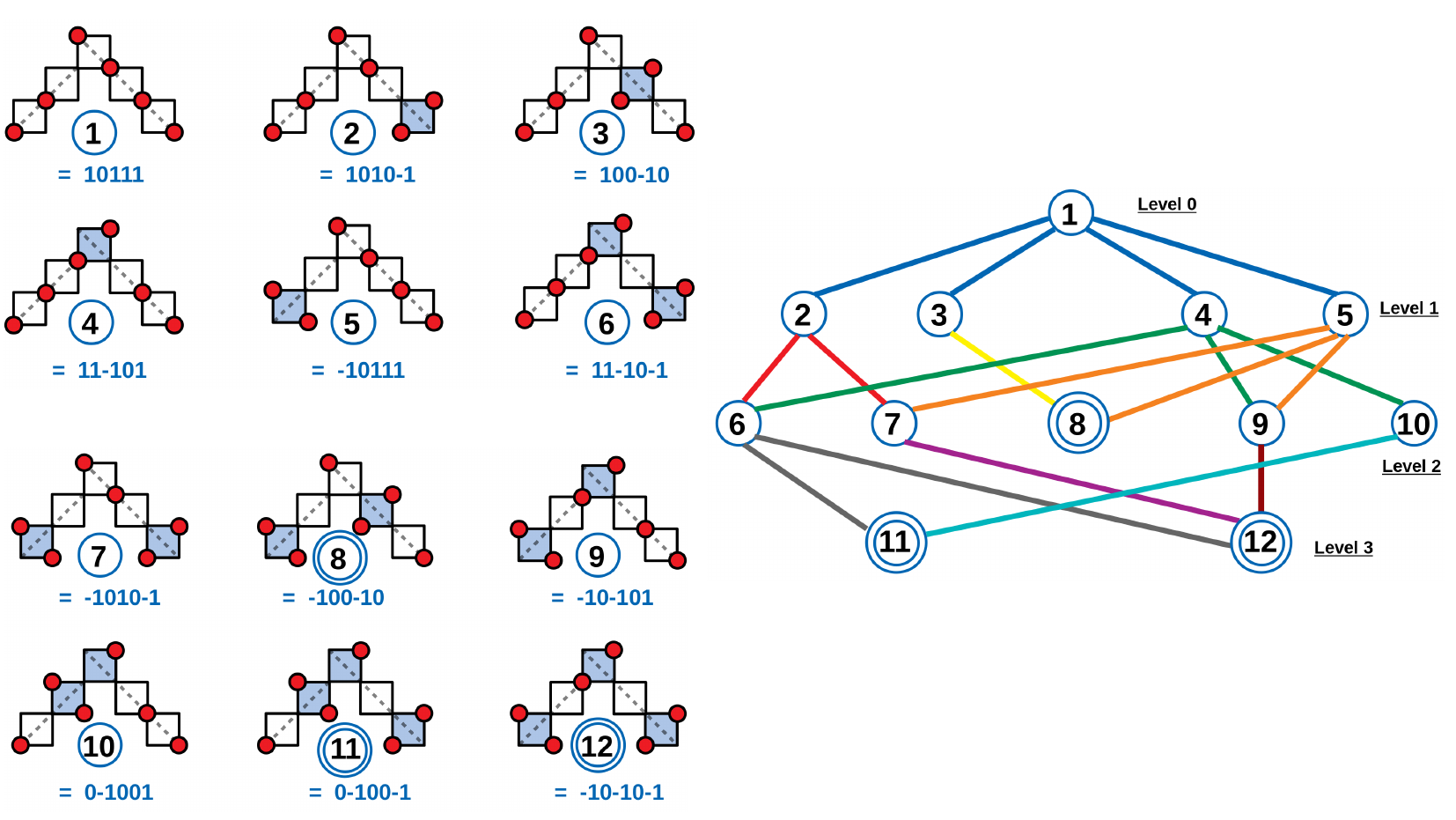}
    \caption{Tree structure for a junction of two wires with $N_p=5$ shown here. The $i$th level contains Fock states with $i$ flipped plaquettes
      (indicated by shaded plaquettes)
      with respect to the reference state defined
      in level-$0$.  The nodes between two Fock states (where the links imply
      the forward construction of the tree) imply that these are connected
      with a single application of $H$.
      The Fock states enclosed by double blue circles represent dead
      ends of the tree. The pseudospins on an open chain with
      $N_p=5$ sites is also shown for each of the Fock states in the
      left panel where red filled circles denote up-spins (bosons). } \label{FIG12}
 \end{center}
\end{figure}

The concept of a unidirectional tree starting from a reference state has
already been introduced in Sec.~\ref{nstree}. The reference state for a wire
with $N_p$ plaquettes can be taken to be the Fock state with all
$N_p+1$ up-spins to be along the wire length as previously
done in Sec.~\ref{wiresdecom}. For the junction of two wires with
$N_p=2x+1$ plaquettes, we take the
reference state to be the one where $x+2$ [$x$] up-spins (bosons)
are arranged along
the length of the wire of length $x+1$ [$x$] to the
right [left] of the central junction plaquette,
including [excluding] the junction plaquette (e.g., see an example of
the reference state marked as \textcircled{$1$} in the left panel
of Fig.~\ref{FIG12}.).
For both the wire and
the junction of two wires, computationally it is convenient to adopt a
one-to-one map from a spin
configuration on the wire or a junction of two wires to another defined on an
open chain with $N_p$ sites in 1D where each site of the chain
can have a pseudospin
variable $\tau^z_i=\pm 1$ or $0$, where $i=1$ to $N_p$.
The pseudospins on the chain represent the
plaquettes of the drum sequentially from left to right in both the cases.
For the wire, these variables take the value $+1$ ($-1$) for elementary
plaquettes that have two up-spins along (perpendicular to) the wire
direction and $0$ otherwise (Fig.~\ref{FIG11}, top panel). For the junction of
two wires, we follow the same
convention and remove the ambiguity at the central plaquette by associating
it to the wire to the right of the junction plaquette (Fig.~\ref{FIG12},
left panel).
While a pseudospin with
$0$ has multiple possibilities associated with an elementary plaquette
involving states with zero or one up-spin, specifying the locations of the
$\pm 1$ pseudospins also fixes the spin state of the other plaquettes on the
drum.

The tree generating algorithm then proceeds as follows. One starts with the
reference state which has $\tau_i^z$ equal to $111\cdots1$ for the wire
and $11\cdots1011\cdots1$
for the junction of two wires where the $0$ in the latter case represents the
plaquette to the immediate left of the $1$-junction plaquette. For the wire,
the states at subsequent levels are generated by flipping a $1$ to $-1$ and
replacing the pseudospins at neighboring site(s) of the flipped
pseudospin by $0$ (Fig.~\ref{FIG11}).
For the junction of two wires, the rules are practically
the same except at the $1$-junction plaquette denoted by the site $i_0$ on
the open chain. When $1$ is flipped to
$-1$ at $i_0$, while the pseudospin at $i_0+1$ is replaced by $0$ as usual, the
pseudospin at $i_0-1$ is replaced by $0$ if the pseudospin at $i_0-2$ equals
$-1$, else it is replaced by $+1$ (Fig.~\ref{FIG12}).
Following this algorithm, we generate the
tree and the corresponding $H$ matrix for both the quantum drums being
discussed here.

We now show that the wire with $N_p$ plaquettes has the same spectrum as
that of the 1D PXP chain with $N_p$ sites and OBC, whose Hamiltonian
is defined as follows:
\begin{eqnarray}
  H_{\mathrm{PXP}} = \sum_{i=2}^{N_p-1}P_{i-1}\mu_i^x P_{i+1}+\mu_1^x P_2+ P_{N_p-1}\mu^x_{N_p}
  \label{pxp}
\end{eqnarray}
where $\mu^\alpha_i$ for $\alpha=x,y,z$ represents spin-$1/2$ Pauli matrices
at site $i$ of the open chain with $N_p$ sites, and $P_i=(1-\mu_i^z)/2$ is a
local projection operator. The constrained Hilbert space of the PXP chain
is defined by the
condition that no two nearest neighbor sites $i,i+1$ can have $\mu^z_i=+1$ and
$\mu^z_{i+1}=+1$ together. We now make the following correspondence between the
pseudospins $\tau^z_i$ for the wire and the spins $\mu_i^z$ for the PXP chain:
\begin{eqnarray}
  \tau_i^z=-1 &\Rightarrow& \mu_i^z=+1, \nonumber \\
  \tau_i^z=+1 &\Rightarrow& \mu_i^z=-1_\mathrm{f}, \nonumber \\
  \tau_i^z=0 &\Rightarrow& \mu_i^z=-1_\mathrm{uf}
  \label{mapping}
\end{eqnarray}
where $\mu_i^z=-1_\mathrm{f}$ ($-1_\mathrm{uf}$) implies that flipping $\mu_i^z$
from $-1$ to $+1$ is allowed (disallowed) due to the hard-core constraints
of the 1D PXP chain. In this language, the reference state of the wire
with $\tau^z_i=+1$ for all $i$
corresponds to the ``Rydberg vacuum'' state of the PXP chain with no Rydberg
excitations, i.e., $\mu_i^z=-1_\mathrm{f}$ for all $i$. The tree generating
algorithm then constructs a unidirectional tree starting from the reference
state at level-$0$ by flipping a $\tau^z_i=1$ to $\tau^z_i=-1$ and
replacing the pseudospins at neighboring site(s) of the flipped
pseudospin by $\tau_{i+1}^z=\tau_{i-1}^z=0$ for $i \neq 1, N_p$ and
$\tau_{i+1}^z=0$ ($\tau_{i-1}^z=0$) for $i = 1$ ($i=N_p$) at each subsequent
level of the tree. The action of $H_{\mathrm{PXP}}$ in Fock space can also
be represented by the same tree structure as the wire
using Eq.~\ref{mapping} since
flipping any $\mu_i$ from $-1$ to $+1$ starting from the Rydberg vacuum state
automatically makes
the previously flippable nearest neighbor site(s) with $\mu=-1$ unflippable
due to the hard-core constraints of the PXP chain.

We note that this equivalence immediately breaks down for the junction of
two wires since flipping a pseudospin $\tau^z_{i_0}=+1$ to $\tau^z_{i_0}=-1$
on the central junction plaquette, denoted by $i_0$,
starting from the reference state produces
a flippable $\tau^z_{i_0-1}=+1$ to its immediate left (see Fock states
marked by $1$ and $4$ in the left panel of
Fig.~\ref{FIG12} for an example) which implies that the junction of two wires
cannot be represented by the same constrained Hilbert space as the PXP
chain by this mapping.

\subsection{Hilbert space dimension and level statistics}
\label{nslev}
Let us calculate the Hilbert space dimension for both these drums for an
arbitrary $N_p$ which will justify their interpretation as effective quasi-1D
models since the dimensionality scales exponentially with $N_p$ as
$N_p \gg 1$ in both cases. Let us denote the number of possible Fock states
for a wire with $N_p$ plaquettes to be $\mathcal{N}_w(N_p)$.
All the Fock states for such a wire can
be built in either one of the following two ways. Consider starting from the
reference state (Fig.~\ref{FIG11}, level-$0$ state) and
building all possible Fock
states using the first $N_p-1$ plaquettes starting from the top. The number of
generated states then equals $\mathcal{N}_w(N_p-1)$ and it is easy to
see that the last plaquette will then either have the pseudospin to be $+1$
or $0$. The remaining states of the wire with $N_p$ plaquettes can be
generated by
starting from the reference state and fixing the pseudospin of the
last plaquette to be $-1$ (i.e., flipping this last plaquette). The first
$N_p-2$ plaquettes from the top can then be used to generate the missing
Fock states whose number equals $\mathcal{N}_w(N_p-2)$. Thus, we get that
\begin{eqnarray}
  \mathcal{N}_w(N_p) &=& \mathcal{N}_w(N_p-1) + \mathcal{N}_w(N_p-2) \nonumber \\
  &=& F_{N_p+2}
  \label{wireHSD}
  \end{eqnarray}
By construction, $\mathcal{N}_w(1)=2$ and $\mathcal{N}_w(2)=3$ which implies
that $\mathcal{N}_w(N_p) = F_{N_p+2}$ as written above,
where $F_n$ are the Fibonacci numbers defined
by the recurrence relation $F_0=0$, $F_1=1$ and $F_n=F_{n-1}+F_{n-2}$ for
$n>1$.

Similarly, for the junction of two wires with $N_p=2x+1$ plaquettes,
all the Fock states can again be built in
one of the following two ways. Consider starting from the
reference state (Fig.~\ref{FIG12}, level-$0$ state) and
building all possible Fock states of the left wire with $x-1$ plaquettes
starting from the left-bottom plaquette and the right wire with $x+1$
plaquettes starting from the right-bottom plaquette. The number of such
states equal $\mathcal{N}_w(x-1)\mathcal{N}_w(x+1)$ and the plaquette to the
immediate left of the central junction plaquette can have a
pseudospin of either be $+1$ or $0$.
To generate the remaining configurations, we start from the reference state
again and make the pseudospin of this particular plaquette to be $-1$ by
first flipping
the central junction plaquette and then flipping the plaquette to the
immediate left of the junction. The number of Fock states generated
from the rest of the plaquettes then equals
$\mathcal{N}_w(x-1)\mathcal{N}_w(x-2)$, thus giving the relation
\begin{eqnarray}
  \mathcal{N}_j(N_p=2x+1)=\mathcal{N}_w(x-1)\left[\mathcal{N}_w(x-2)+\mathcal{N}_w(x+1)\right] =F_{x+1}\left(F_{x}+F_{x+3}\right)
  \label{junctionHSD}
  \end{eqnarray}
where $\mathcal{N}_j(N_p=2x+1)$ refers to the number of Fock states in a
junction of two wires composed of $N_p=2x+1$ elementary plaquettes.
Eq.~\ref{wireHSD} and Eq.~\ref{junctionHSD} show that the number of
allowed Fock states scale exponentially for large $N_p$ for both the drums.
Note that while Eq.~\ref{wireHSD} is identical to the Hilbert space
dimension of a 1D PXP chain with $N_p$ sites and OBC, as should be the
case from the equivalence of both models shown in Sec.~\ref{treelargedrums},
the Hilbert space dimension of the junction of two wires (Eq.~\ref{junctionHSD})
cannot be expressed as $F_m$ with an integer $m$ in general showing that the
structure of the constrained Hilbert space of this drum is different
from that of the 1D PXP chain.

We can then ask whether these large quantum drums satisfy a Krylov-restricted
version of the ETH, i.e., whether these quasi-1D models are non-integrable.
We check this using the method of level statistics that can be obtained
directly using the eigenspectrum from ED (e.g., see Ref.~\cite{Oganesyan2007}).
To calculate the level statistics for large quantum drums, it is important to
first project to a sector where all the commuting global symmetries have been
resolved. Since both the wire and the junction of two wires (Fig.~\ref{FIG10})
are quasi-1D
structures with open boundaries, momentum is not a good quantum number.
The total magnetization in the computational basis, $S^z_{\mathrm{tot}}$,
represents a conserved quantity for these drums. However, all nodes of a
tree (Fig.~\ref{FIG11} and Fig.~\ref{FIG12})
already have the same $S^z_{\mathrm{tot}}$ by construction.
The only remaining non-trivial
global symmetry turns out to be a reflection symmetry, denoted by
$\mathcal{R}_w$ ($\mathcal{R}_j$) for the wire (junction of two wires),
which takes a Fock state $|\alpha\rangle$ to another Fock
state $|\beta \rangle = \mathcal{R}_{w/j}|\alpha\rangle$ with the axis of
reflection
shown in Fig.~\ref{FIG10} for both the drums.
For a wire with even (odd) number of
plaquettes, the
axis passes through a site (the diagonal of a square) (Fig.~\ref{FIG10}, left
and middle panels) whereas for a junction
of two wires, it passes through the central junction plaquette as shown
in Fig.~\ref{FIG10}, right panel.
Since $\mathcal{R}_{w/j}^2|\alpha\rangle=|\alpha\rangle$ for any Fock state, the
basis states $(|\alpha\rangle \pm \mathcal{R}_{w/j}|\alpha\rangle)/\sqrt{2}$
define
states with $\mathcal{R}_{w/j}=\pm 1$ respectively.
If $\mathcal{R}_{w/j}|\alpha\rangle=
|\alpha\rangle$ for some Fock state(s), then such Fock state(s) only contribute
to the $\mathcal{R}_{w/j}=+1$ sector.
This happens in the case of the wire, where,
the reference state provides one example of such a Fock state. Thus, the
number of basis states in $\mathcal{R}_{w}=+1$ always exceeds the corresponding
number for $\mathcal{R}_w=-1$ for a wire whereas these two numbers are
equal to each other
for a junction of two wires.
\begin{figure}[!tbh]
  \begin{center}
     \includegraphics[width=0.45\linewidth]{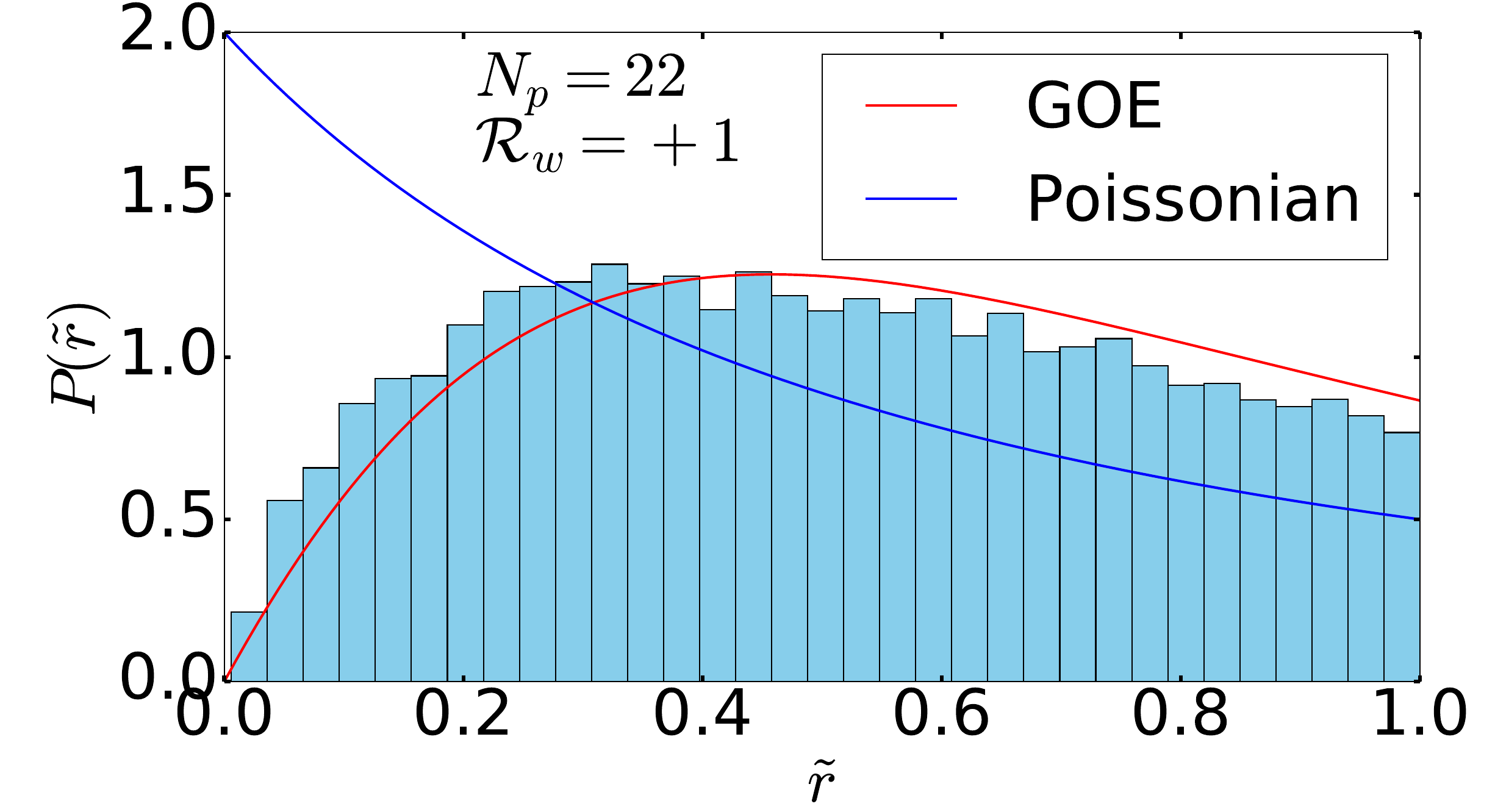}%
    \includegraphics[width=0.45\linewidth]{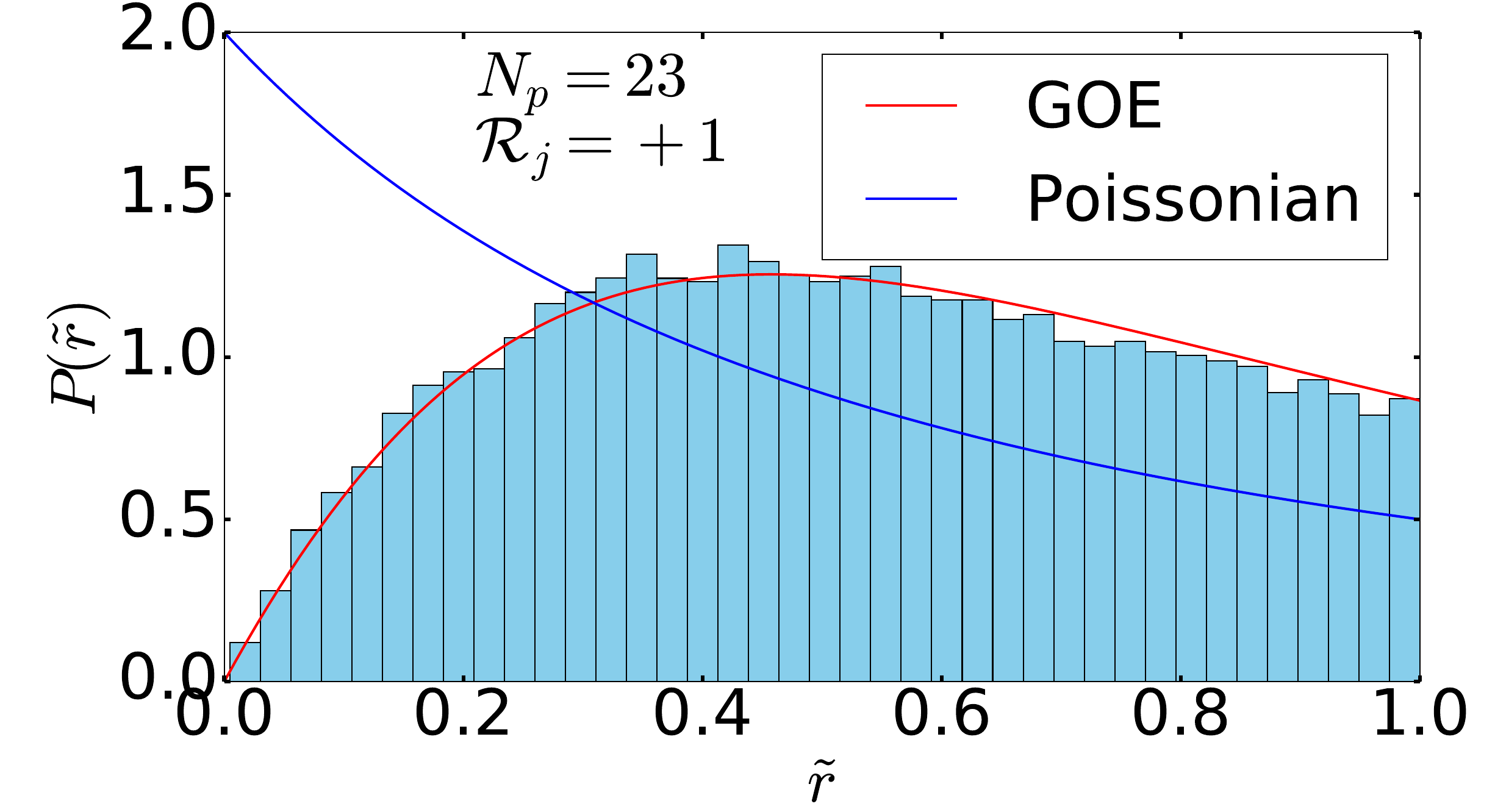}
    \caption{Level spacing ratio distribution $P(\tilde{r})$ versus $\tilde{r}$ for a wire with $N_p=22$ (left panel) and a junction of two wires with $N_p=23$
      (right panel), with both the data taken in the symmetry sector with
      $\mathcal{R}_{w/j}=+1$.
      The histograms indicate the non-integrability of both
    the quasi-1D models.} \label{FIG13}
 \end{center}
\end{figure}

Restricting to the larger sector with $\mathcal{R}_{w/j}=+1$,
we construct the distribution of
consecutive level spacing ratios $\tilde{r}$ (with support in $[0,1]$) where
$\tilde{r}$ is defined as follows:
\begin{eqnarray}
  \tilde{r}=\mathrm{min} \left\{r_n,\frac{1}{r_n} \right \} \le 1, \mbox{~~~} r_n =\frac{s_n}{s_{n-1}}, \mbox{~~~}s_n=E_{n+1}-E_n
\end{eqnarray}
where $E_n$ represent the energies of the eigenvectors obtained from ED. For a
non-integrable model, one expects a Gaussian orthogonal ensemble (GOE)
distribution, while an integrable system leads to a Poisson distribution
for $P(\tilde{r})$~\cite{Atas2013},
where the two distributions have the following forms:
\begin{eqnarray}
  P_{\mathrm{GOE}}(\tilde{r}) =\frac{27}{4} \frac{\tilde{r}+\tilde{r}^2}{(1+\tilde{r}+\tilde{r}^2)^{5/2}}; \mbox{~~~} P_{\mathrm{P}}(\tilde{r}) =\frac{2}{(1+\tilde{r})^2}.
\end{eqnarray}
The numerically generated data for $P(\tilde{r})$ versus $\tilde{r}$ is shown
for a wire with $N_p=22$ plaquettes and a junction of two wires with $N_p=23$
plaquettes in Fig.~\ref{FIG13}. The data clearly indicates that $P(\tilde{r})$
follows $P_{\mathrm{GOE}}(\tilde{r})$ much more closely than
$P_{\mathrm{P}}(\tilde{r})$ for these system sizes giving strong evidence for the
non-integrable nature of both these quasi-1D models.

\subsection{Zero modes and index theorem}
\label{nsind}
While both the wire and the junction of two wires have a symmetric
eigenspectrum of $H$ around $E=0$ as is expected for any quantum drum,
the ED data further reveals the presence of
an ever-increasing number of exact zero modes (up to machine precision)
with increasing $N_p$ for the former case
and the absence of any zero mode for the latter case.

\begin{figure}[!tbh]
  \begin{center}
     \includegraphics[width=0.6\linewidth]{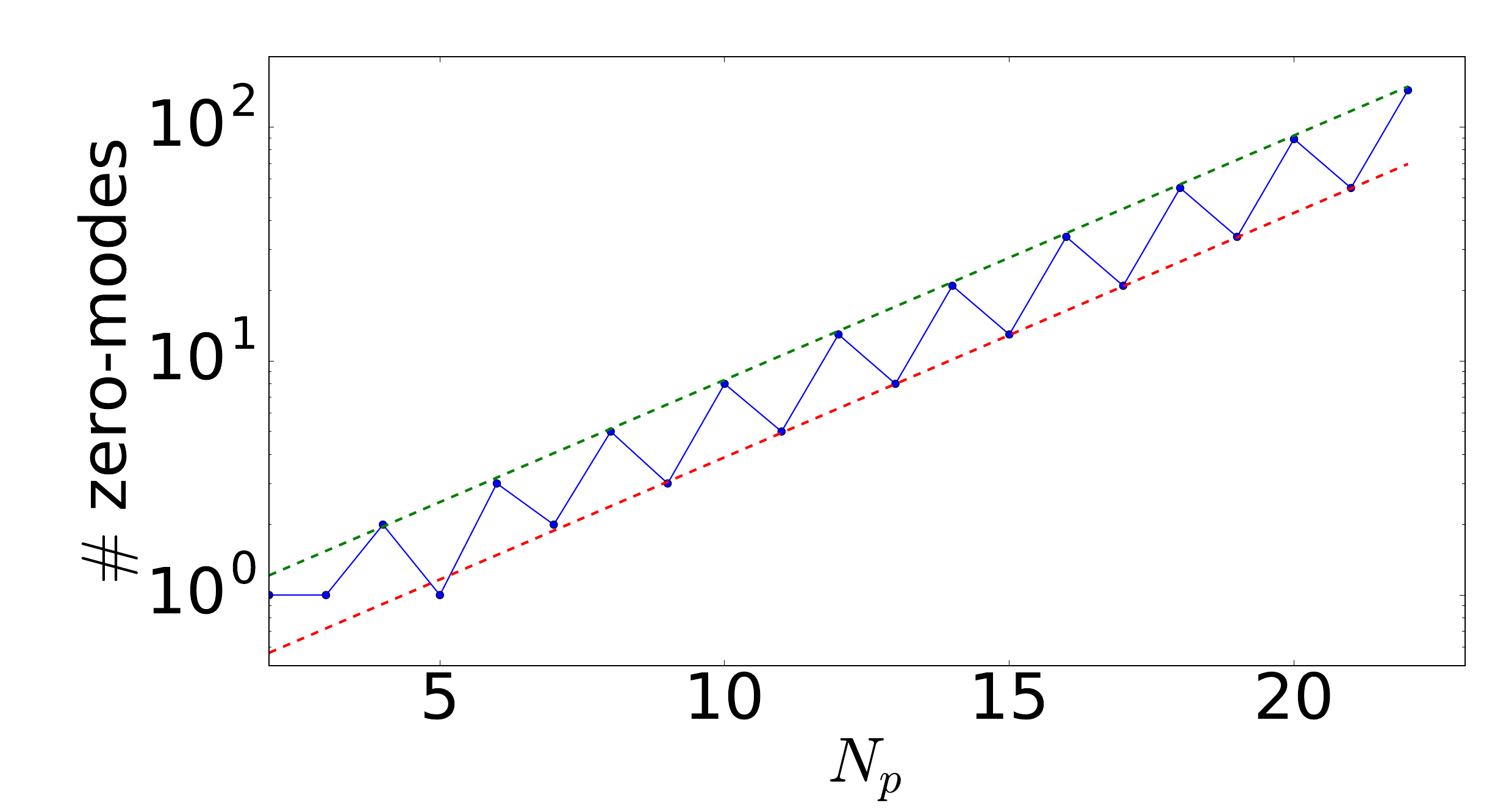}
     \caption{Scaling of the total number of zero modes versus $N_p$ for a
       wire. For even (odd) values of $N_p$, the number of zero modes
       grow as $\mu_e (\sqrt{\varphi})^{N_p}$ ($\mu_o (\sqrt{\varphi})^{N_p}$)
     where $\mu_e \approx 0.75$ ($\mu_o \approx 0.35$) [indicated by dotted lines].} \label{FIG14}
 \end{center}
\end{figure}

This striking difference between the two drums can be understood in terms of
the index theorem of Ref.~\cite{Schecter2018index}. Firstly, a chiral
operator $\mathcal{C}_{w/j} =\prod_{\Box_{j}}\sigma^z_{j_x,j_y}$ (where the
subscript $w (j)$ refers to the wire (junction of two wires)) can be defined
in both cases, which involves one site $(j_x,j_y)$ per elementary plaquette
contained in the drum (these sites are indicated by crosses in red in all
panels of Fig.~\ref{FIG10}). This operator satisfies $\{H,\mathcal{C}_{w/j}\}=0$
for the Hilbert space fragments generated by these drums, thus ensuring the
$E \rightarrow -E$ symmetry of the spectrum. Furthermore, as already discussed,
these two drums have a global reflection symmetry, $\mathcal{R}_{w/j}$, that
commutes with $H$ (Fig.~\ref{FIG10}). Importantly, while $[\mathcal{R}_w,
  \mathcal{C}_w]=0$, it turns out that $[\mathcal{R}_j,
  \mathcal{C}_j] \neq 0$ which means that the index theorem of
Ref.~\cite{Schecter2018index} applies to the wire but not to
junction of two wires. This leads to a macroscopically large number
of protected zero modes in the former case and also explains our
numerical data (Fig.~\ref{FIG14}). The number of zero modes in the
wire show an interesting even-odd effect as a function of $N_p$
(Fig.~\ref{FIG14}) with the even values of $N_p$ showing a higher
number of zero modes. This even-odd effect stems from the fact that
the axis that defines the reflection symmetry, $\mathcal{R}_w$,
passes through a single site shared by two elementary plaquettes for
even values of $N_p$; in contrast, it passes through two sites along
a diagonal of an elementary plaquette for odd values of $N_p$
(Fig.~\ref{FIG10}, left and middle panels). The number of zero modes
scale as $\mu_{e/o}(\sqrt{\varphi})^{N_p}$ (with $\varphi$ $=$
$(1+\sqrt{5})/2$ being the golden ratio as defined before) where
$\mu_e \approx 0.75$ ($\mu_o \approx 0.35$) for even (odd) values of
$N_p$ (see Fig.~\ref{FIG14}). Identical scaling behavior was also
observed for the number of zero modes in the 1D PXP
model~\cite{Turner2018a, Turner2018b, Buijsman2022}.


It is useful to
point out here that a different type of junction of two
equal-length wires (Fig.~\ref{drumvarieties}, panel B) instead of
this junction being studied here will again have an exponentially
large number of exact zero modes. Similarly, a junction of  three
equal-length wires (Fig.~\ref{drumvarieties}, panel C) as well as a
junction of four equal-length wires (Fig.~\ref{drumvarieties}, panel D)
will also have a
macroscopic number of zero modes due to the index theorem of
Ref.~\cite{Schecter2018index}.

\subsection{QMBS and related diagonastics}
\label{nsscars}
While the level statistics distribution of both the wire and the junction of
two wires (Fig.~\ref{FIG13}) is consistent with these quasi-1D models
being non-integrable for large $N_p$ and thus satisfying Krylov-restricted
ETH, both quantum drums also harbor QMBS that give rise to observable
dynamical signatures like periodic revivals from certain simple initial
states.

\begin{figure}[!tbh]
  \begin{center}
    \includegraphics[width=0.8\linewidth]{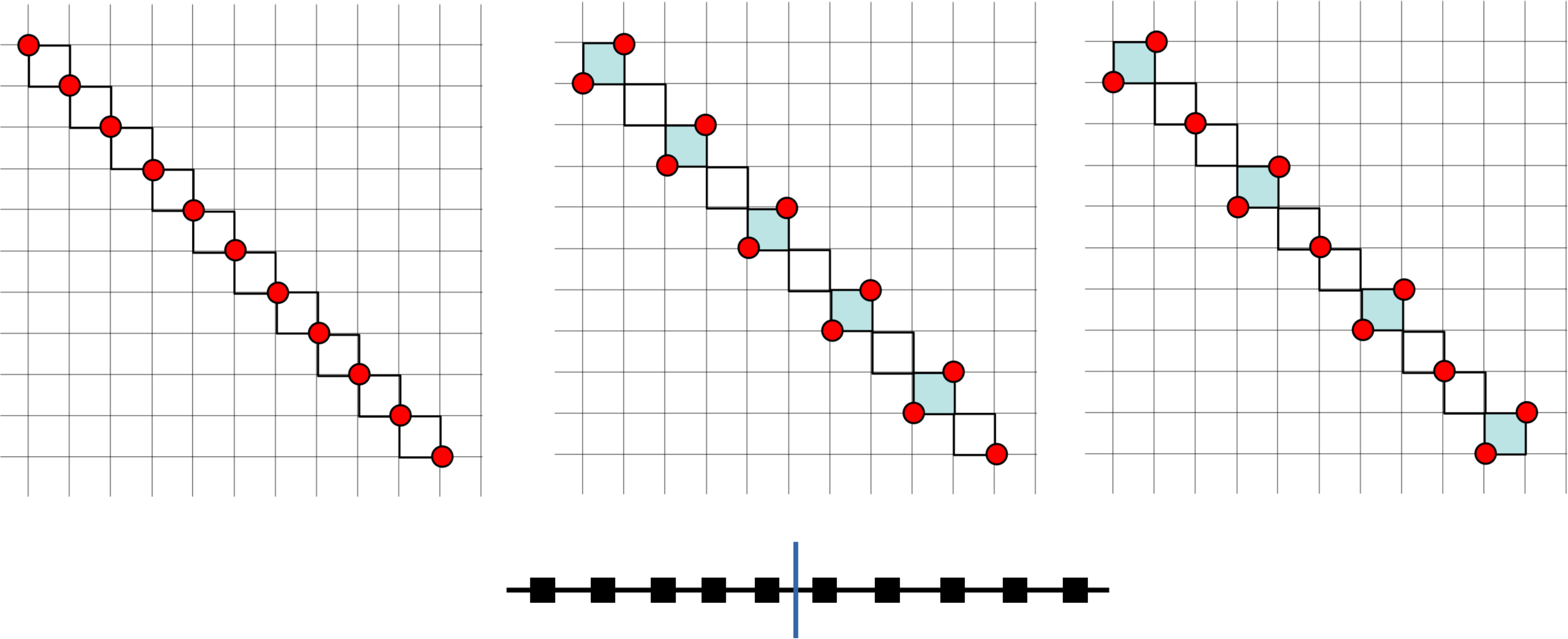}
    \caption{Three Fock states shown for the wire with $N_p=10$.
      (Left panel)
      Reference
      state denoted by $|r \rangle_w$ (Middle panel)
      A $|fu \rangle_w$ state created
    by flipping every alternate elementary plaquette in the reference state
    (Right panel) A $|fuu \rangle_w$ state created by flipping every
    third elementary plaquette of the reference state. Flipped plaquettes
    with respect to the reference state are shown shaded in blue.
    The
    entanglement cut used to calculate the bipartite entanglement entropy
    of the system after mapping it to an open chain of pseudospins is
    shown below; such a cut divides
    the system into two equal halves with $N_p/2$ pseudospins each.
    }
    \label{FIG15}
  \end{center}
  \end{figure}

\begin{figure}[!tbh]
  \begin{center}
    \includegraphics[width=0.7\linewidth]{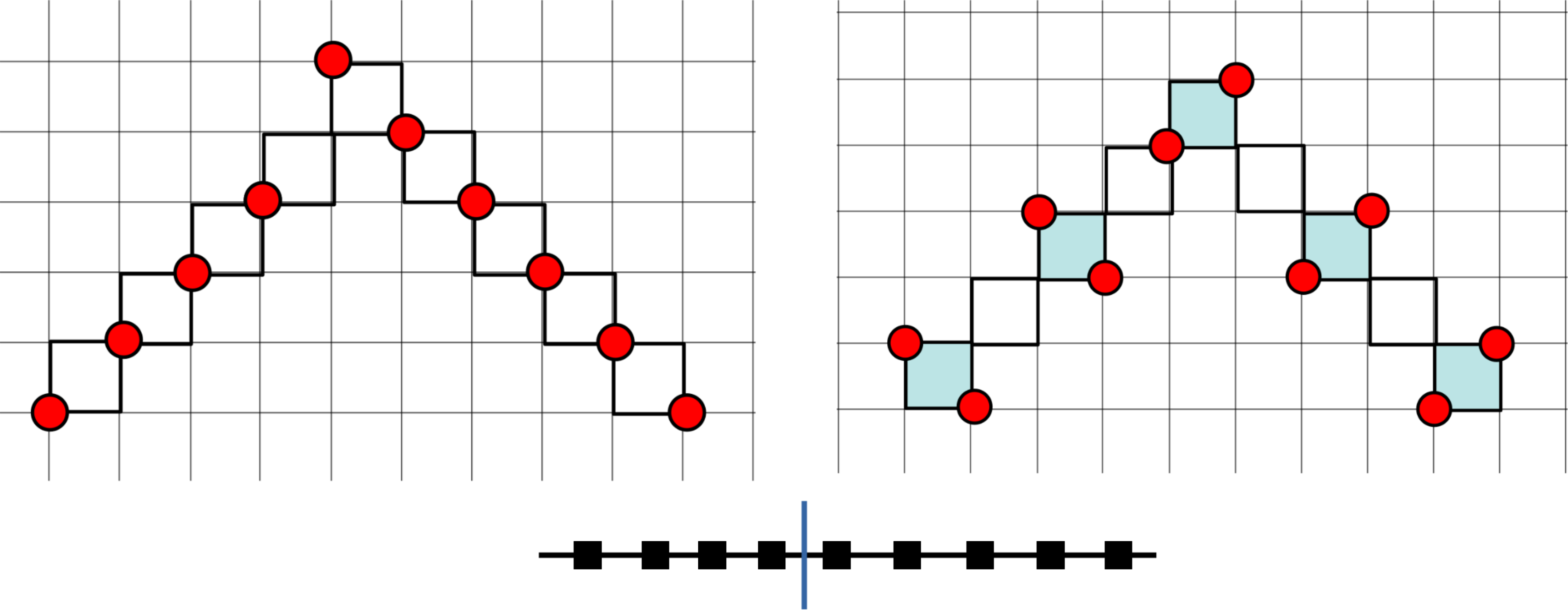}
    \caption{Two Fock states shown for the junction of two wires with
      $N_p=9$.
      (Left panel)
      Reference
      state denoted by $|r \rangle_j$ (Right panel)
      A $|fu \rangle_j$ state created
      by flipping every alternate elementary plaquette in the reference state.
      Flipped plaquettes
    with respect to the reference state are shown shaded in blue.
     The
    entanglement cut used to calculate the bipartite entanglement entropy
    of the system after mapping it to an open chain of pseudospins divides
    the system into two halves with $(N_p/2) \pm 1$ pseudospins as
    shown below. } \label{FIG16}
  \end{center}
  \end{figure}

Let us consider three such Fock states for the wire as shown in
Fig.~\ref{FIG15}. Fig.~\ref{FIG15} (left panel) shows the reference state
which we denote as $|r\rangle_w$, Fig.~\ref{FIG15} (middle panel) shows a
Fock state obtained by flipping every alternate elementary plaquette in the
reference state which we denote as $|fu\rangle_w$, and Fig.~\ref{FIG15}
(right
panel) shows a Fock state obtained by flipping every third elementary plaquette
in the reference state which we denote as $|fuu\rangle_w$. Similarly, two
representative Fock states are shown in Fig.~\ref{FIG16} for the junction of
two wires. Fig.~\ref{FIG16} (left panel) shows the reference state
which we denote as $|r\rangle_j$ (note that there are two such reference states
possible for the junction of two wires) and Fig.~\ref{FIG16} (right panel)
shows a Fock state obtained by flipping every alternate elementary plaquette
in the reference state which we denote as $|fu\rangle_j$. From the
equivalence of the wire to the 1D PXP chain
shown in Sec.~\ref{treelargedrums}, it
is clear that while  local operators starting from the state
$|r\rangle_w$ will thermalize quickly, since the initial state maps to the
Rydberg vacuum state of the PXP chain, this will not be the case
from the initial states $|fu\rangle_w$ and $|fuu\rangle_w$ which map to
the period-$2$ $|\mathbb{Z}_2\rangle$ and the
period-$3$ $|\mathbb{Z}_3\rangle$
Fock states of the PXP chain, respectively.

\begin{figure}[!tbh]
  \begin{center}
    \includegraphics[width=0.46\linewidth]{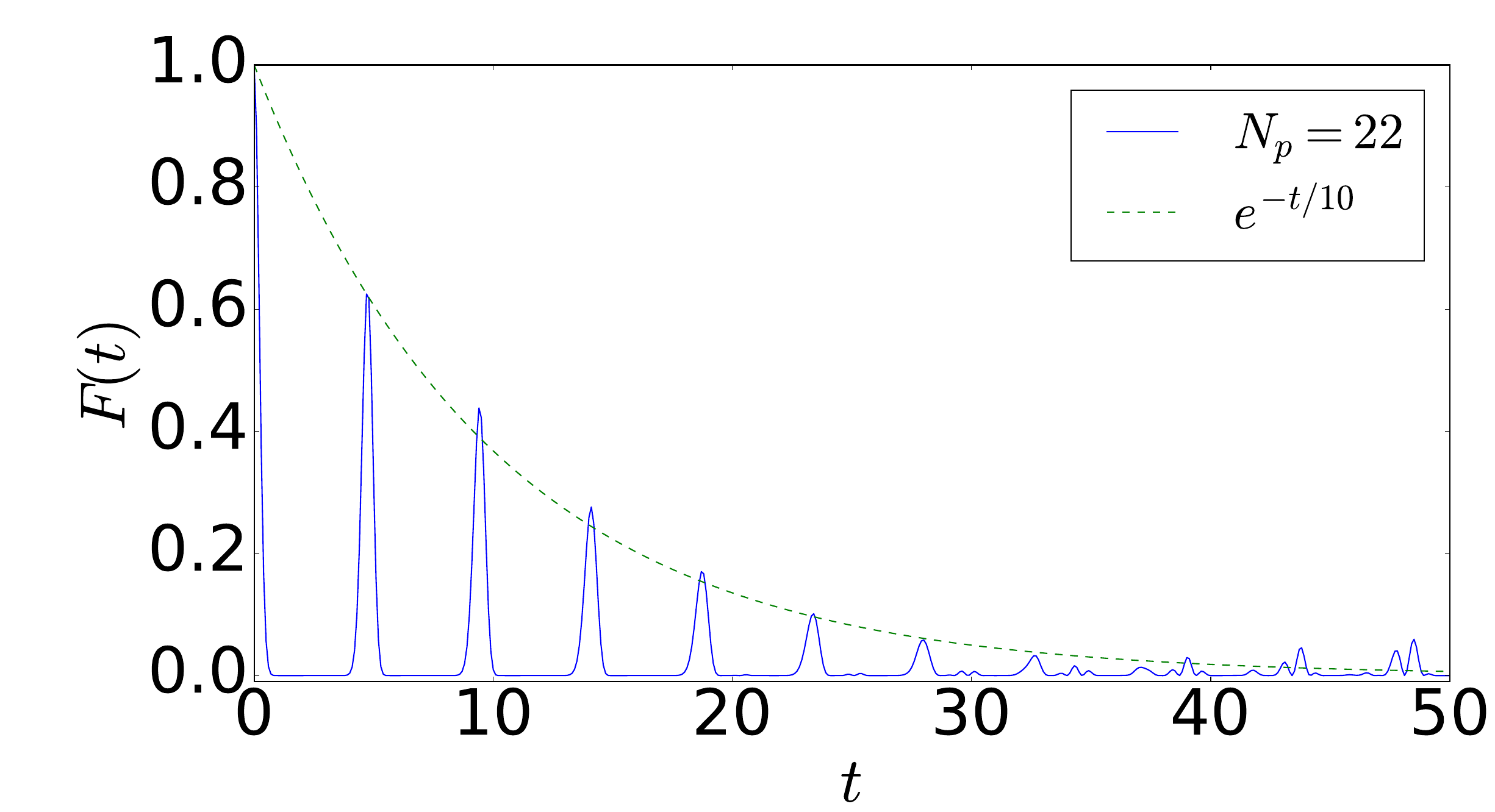}%
    \includegraphics[width=0.46\linewidth]{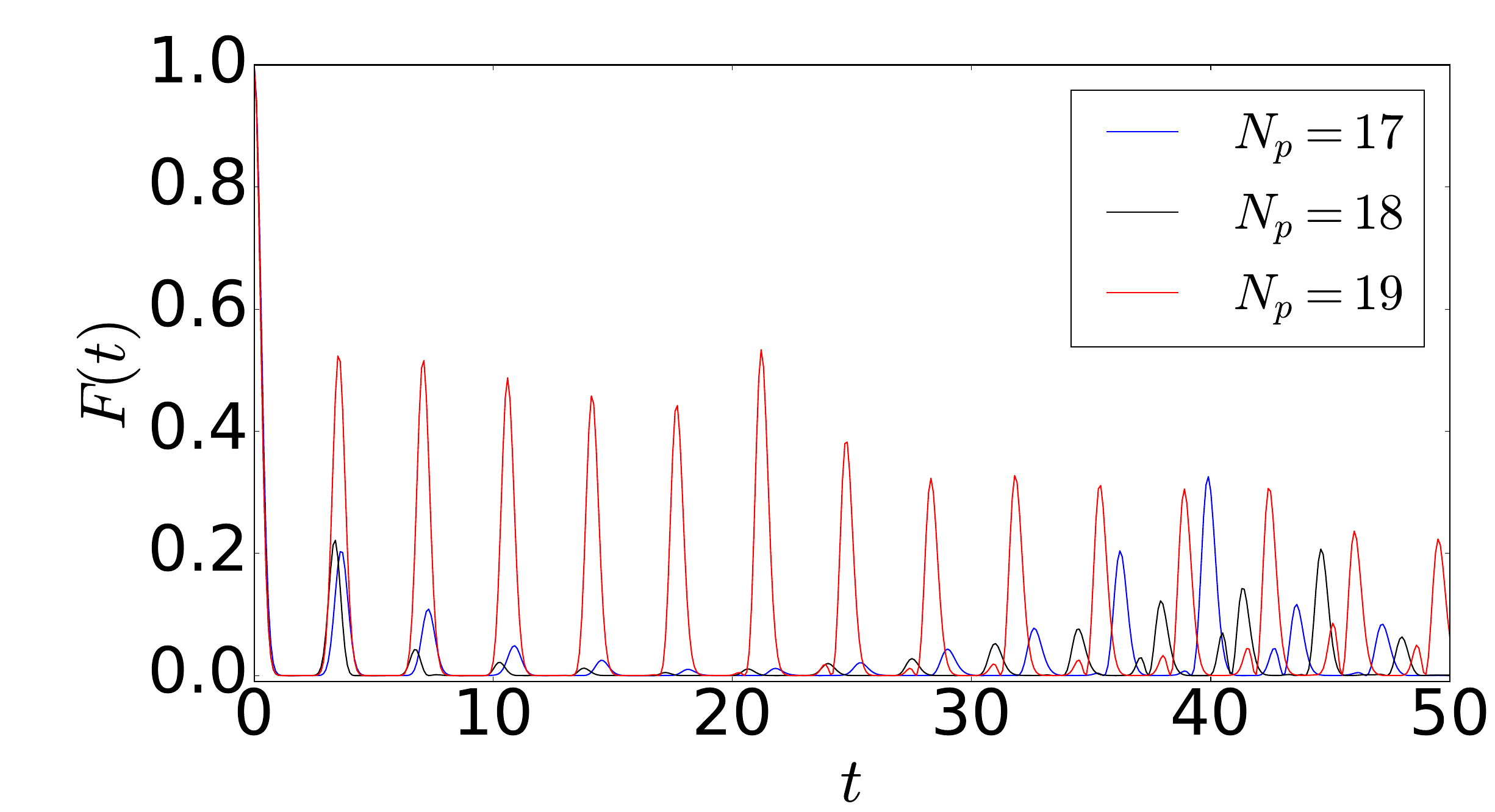}\\
    \includegraphics[width=0.46\linewidth]{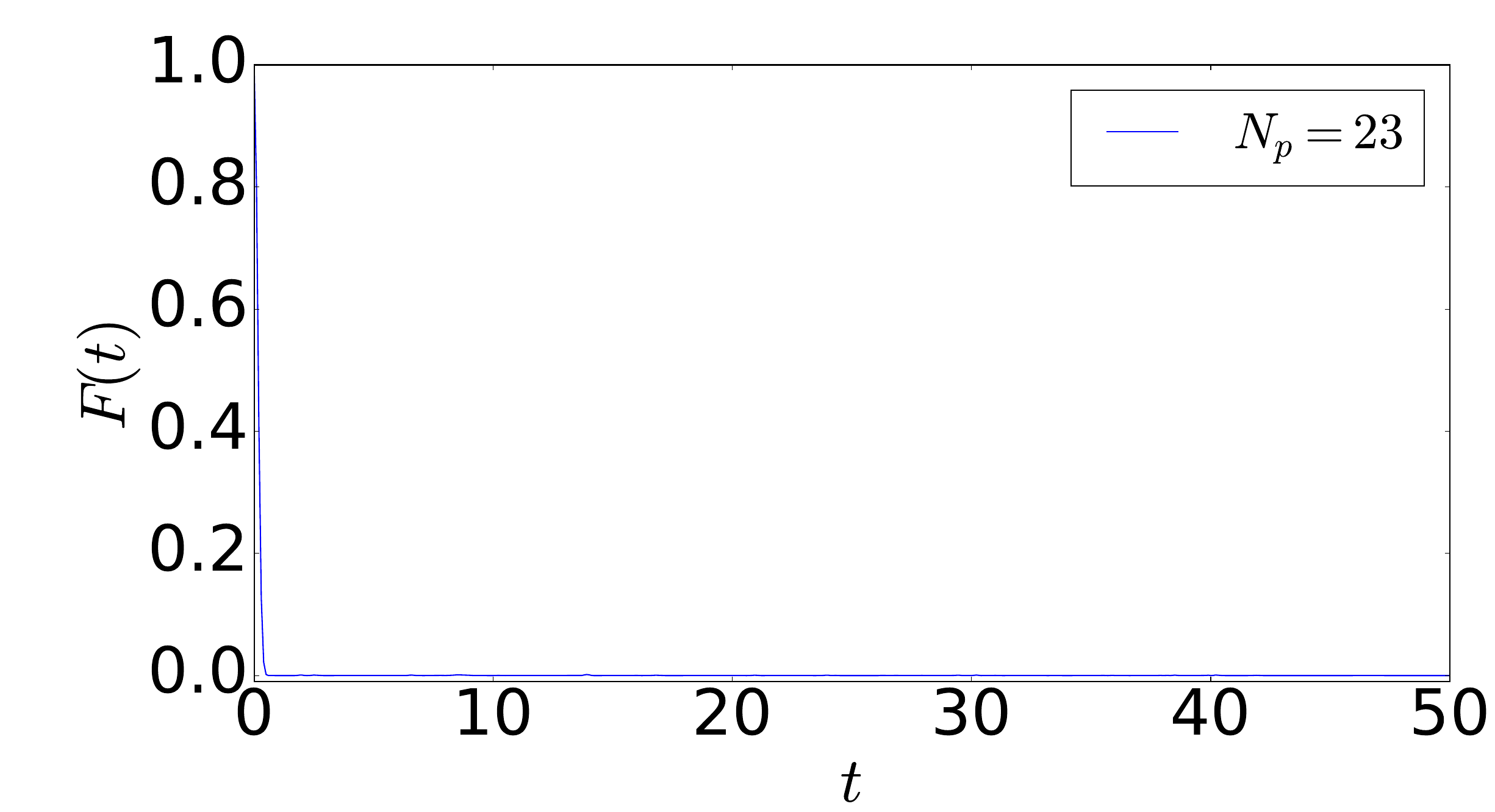}%
    \includegraphics[width=0.46\linewidth]{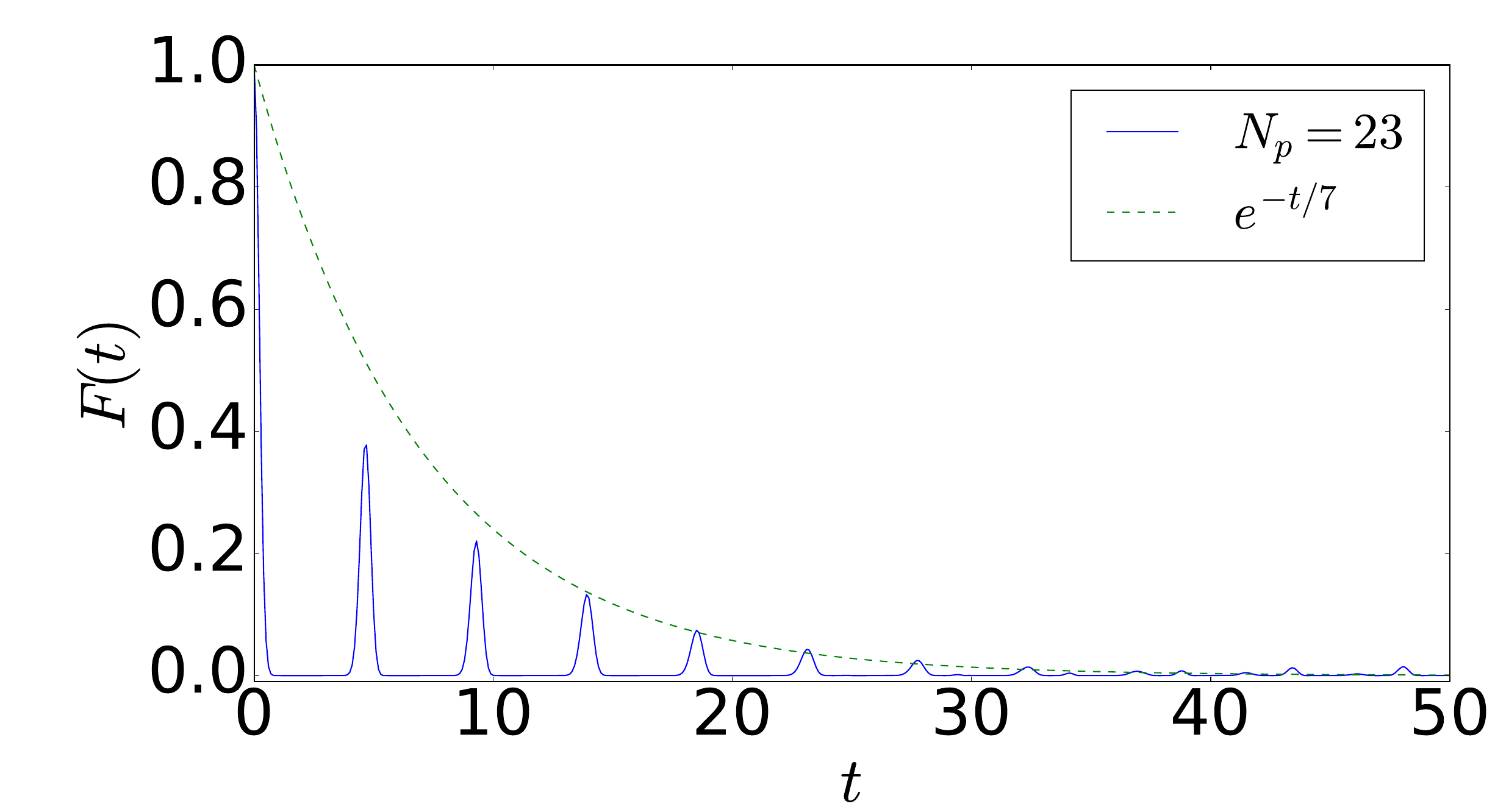}
    \caption{The behavior of fidelity $F(t)$ shown for the wire with two
      different initial Fock states with the top-left panel for $|fu\rangle_w$,
      and the the top-right panel for $|fuu\rangle_w$. The bottom-left (bottom-right)
      panel shows the fidelity as a function
      of time with the initial state being $|r\rangle_j$
      ($|fu\rangle_j$) for a junction of
    two wires.} \label{FIG17}
 \end{center}
\end{figure}

This can indeed be checked by monitoring the fidelity $F(t)$ $=$
$|\langle s |\exp(-iHt)|s\rangle|^2$ using ED for these
representative Fock states (denoted by $|s\rangle$) in
Fig.~\ref{FIG17}. Most initial states show a rapid drop in $F(t)$
within $t \sim O(1)$ which is expected for a high-energy initial
state in an interacting system. However, for the wire, the behaviour
of $F(t)$ for $|fu\rangle_w$ and $|fuu\rangle_w$ are markedly
different, with both showing periodic revivals with an emergent
time-scale $T^* \sim 5$ for $|fu\rangle_w$ (Fig.~\ref{FIG17},
top-left panel) and $T^* \sim 4$ for $|fuu\rangle_w$
(Fig.~\ref{FIG17}, top-right panel). The periodic revivals of $F(t)$
starting from $|fu\rangle_w$ show a decaying envelope in time that
can be reasonably described by the envelope function
$\exp(-t/\tau_w)$ with $\tau_w \approx 10$ (Fig.~\ref{FIG17},
top-left panel). This decaying envelope to the periodic revivals
distinguish this phenomenon from the persistent oscillations
starting from initial Fock states discussed in Sec.~\ref{intstates}.
In fact, the fidelity revivals for the $|fuu \rangle_w$ state shows
a very interesting finite-size effect with such revivals being
strongest when $N_p=3n+1$ where $n$ is an integer. For example, the
peak value of the first fidelity revival in time equals $0.52$ for
$N_p=19$ while it is much smaller for $N_p=18$ and $N_p=20$ ($0.22$
and $0.18$ respectively). Furthermore, the $N_p=19$ data for $F(t)$
starting from the initial state $|fuu\rangle_w$ shows no sign of a
decaying exponential envelope till $t=50$ (see Fig.~\ref{FIG17},
top-right panel). For the junction of two wires with $N_p=23$, we
again see rapid decay of $F(t)$ starting from $|r\rangle_j$
(Fig.~\ref{FIG17}, bottom-left panel) while $F(t)$ shows non-trivial
periodic revivals from $|fu\rangle_j$ with the same $T^*\sim 5$ as
in the wire case. The periodic revivals are weaker for the junction
of two wires compared to the single wire and again have a decaying
exponential envelope described by $\exp(-t/\tau_j)$ with a smaller
$\tau_j \approx 7$, but these fidelity revivals are nontheless
clearly visible up to $t \sim 20$.

\begin{figure}[!tbh]
  \begin{center}
    \includegraphics[width=\linewidth]{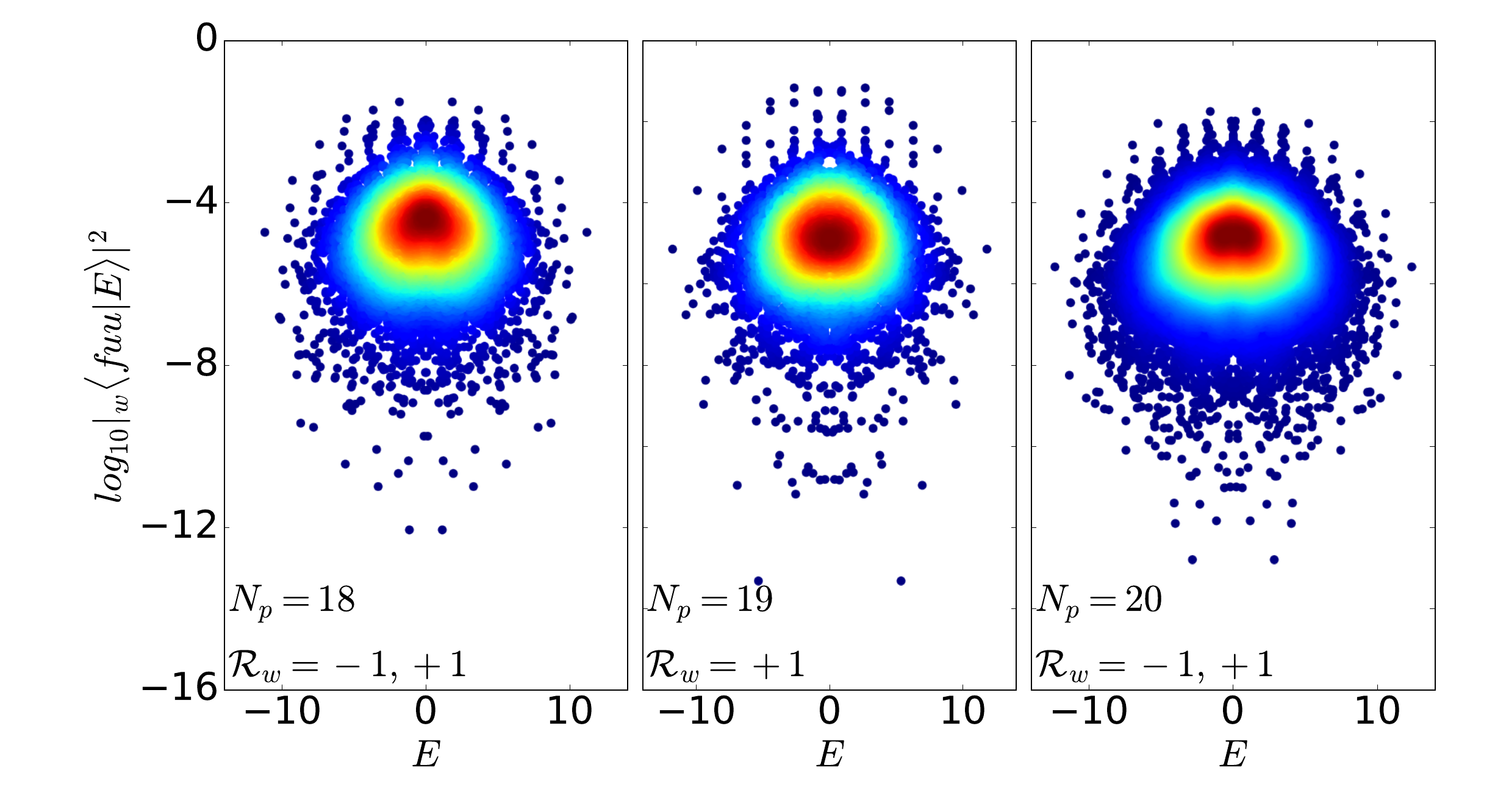}
    \caption{Density plots showing the overlap of the $|fuu\rangle_w$ state
      with energy eigenstates
      of the wire with $N_p=18$ (left panel), $N_p=19$ (middle panel)
      and $N_p=20$ (right panel) respectively.  In all the panels,
      the density of states is indicated by the
      same color map where warmer color
      corresponds to higher density of states.} \label{FIG18}
 \end{center}
\end{figure}

It is useful to point out that the enhanced revivals observed for
$|fuu\rangle_w$ for $N_p=3n+1$ for the wire, which is equivalent to
a $|\mathbb{Z}_3\rangle$ initial state in a
1D PXP chain with $N_p$ sites and OBC, was not pointed out in the
literature previously
and additional terms were added to the PXP Hamiltonian to cause
enhancement of fidelity revivals from the $|\mathbb{Z}_3\rangle$ state
~\cite{Bull2020}. As is well-known from the 1D PXP
chain~\cite{Turner2018a, Turner2018b}, these
fidelity revivals from certain special initial states is due to a large
overlap with {\it approximate}
towers of QMBS that are equally spaced in energy.
These towers are most clearly seen by plotting the overlaps of the initial
Fock state $|fuu\rangle_w$ with the many-body eigenstates
$|E\rangle$ as a function of energy (see Fig.~\ref{FIG18}). We see
that at $N_p=3n+1$ (Fig.~\ref{FIG18}, middle panel), these towers are
much more clearly formed compared to $N_p=3n$ (Fig.~\ref{FIG18}, left panel)
and to $N_p=3n+2$ (Fig.~\ref{FIG18}, right panel).
We also note that at the system sizes, $N_p=3n+1$,
the $|fuu\rangle_w$ Fock state becomes orthogonal
to the zero mode subspace of the system (up to machine precision)
even though the
initial state has zero average energy.
A deeper understanding of all these
striking finite-size effects at $N_p=3n+1$ for the wire/open PXP chain
would be highly desirable.

\begin{figure}[!tbh]
  \begin{center}
    \includegraphics[width=0.5\linewidth]{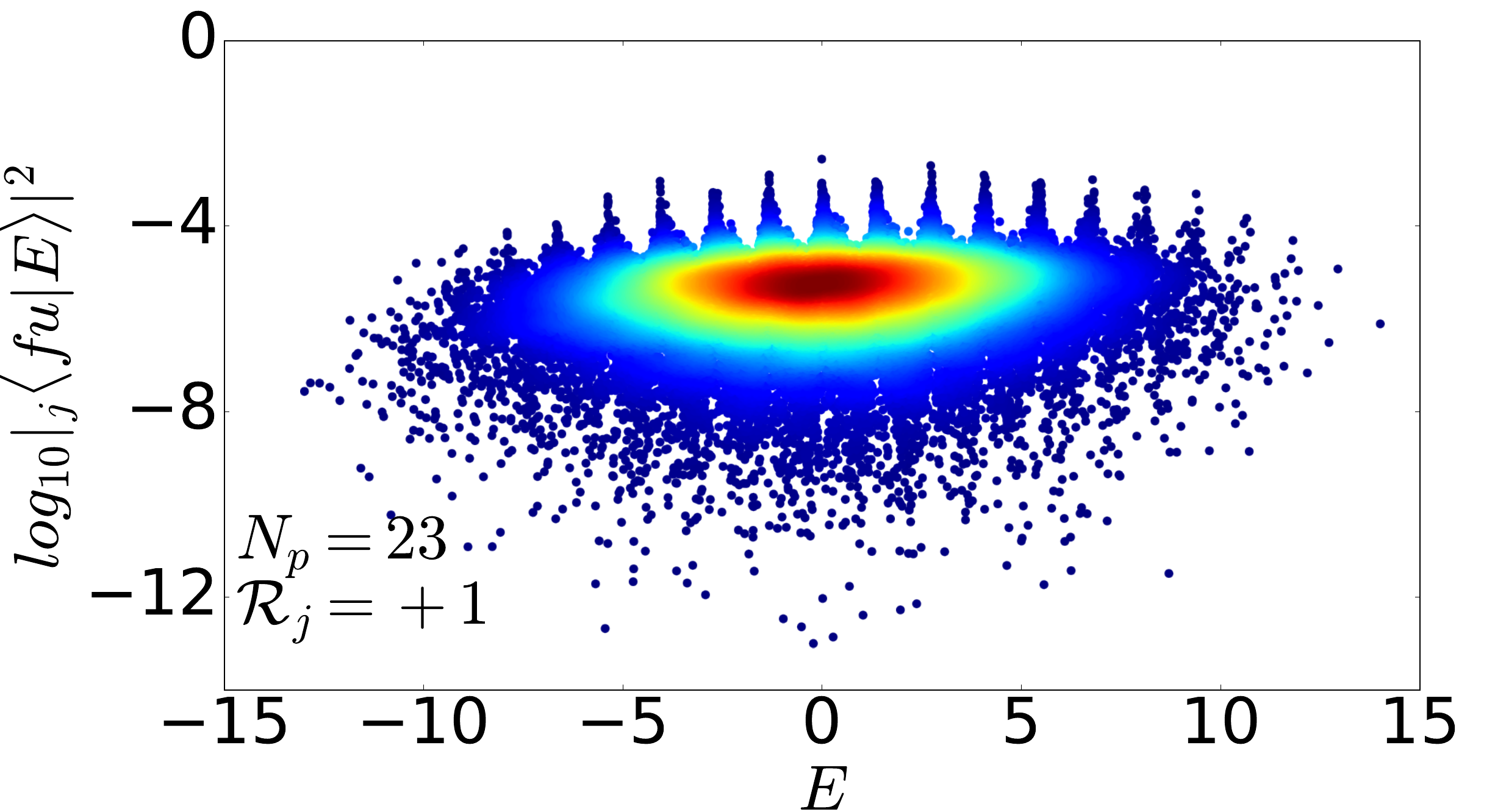}%
    \includegraphics[width=0.5\linewidth]{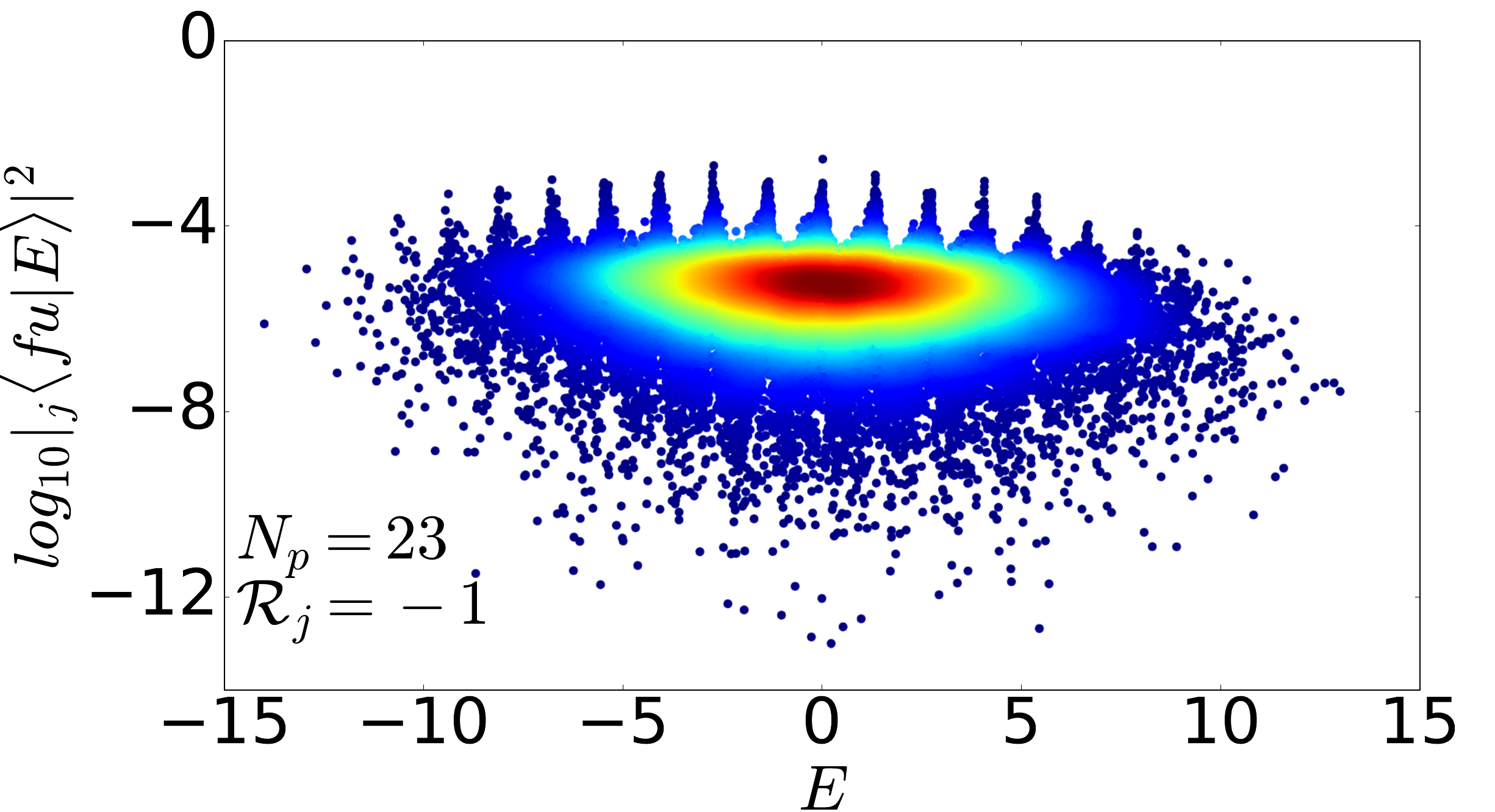}
    \caption{Density plots that show the
      overlap of the $|fu\rangle_j$ state for a junction of two wires
      with energy eigenstates of the
      junction with two wires with $N_p=23$ for the $\mathcal{R}_j=+1$
      (left panel) and the $\mathcal{R}_j=-1$ (right panel) sectors
      respectively. In both panels, the density of states is indicated by the
      same color map where warmer color
corresponds to higher density of states. } \label{FIGjunc}
 \end{center}
\end{figure}

Even though the junction of two wires cannot be reduced to the 1D
PXP chain, this model also admit {\it approximate} towers of QMBS
that are equidistant in energy. In Fig.~\ref{FIGjunc} (two panels),
the overlap behavior of the $|fu\rangle_j$ Fock state
(Fig.~\ref{FIG16}, right panel) with the eigenstates of the junction
of two wires with $N_p=23$ is shown for the $\mathcal{R}_j=+1$ and
the $\mathcal{R}_j=-1$ sectors respectively. In this case, the
towers of states with higher overlap to the Fock state are somewhat
less clearly separated from the bulk of the spectrum as compared to
the 1D PXP model, explaining the weaker fidelity revivals in the
junction of two wires as compared to the single wire case
(Fig.~\ref{FIG17}, top left and bottom right panels). Since
$[\mathcal{R}_j, \mathcal{C}_j] \neq 0$, the overlaps are not
symmetric with respect to zero energy; rather, the overlap behavior
for $\mathcal{R}_j=+1$ sector is a {\it mirror image} (with the
mirror axis being $E=0$) of the $\mathcal{R}_j=-1$ sector.


\begin{figure}[!tbh]
  \begin{center}
    \includegraphics[width=0.46\linewidth]{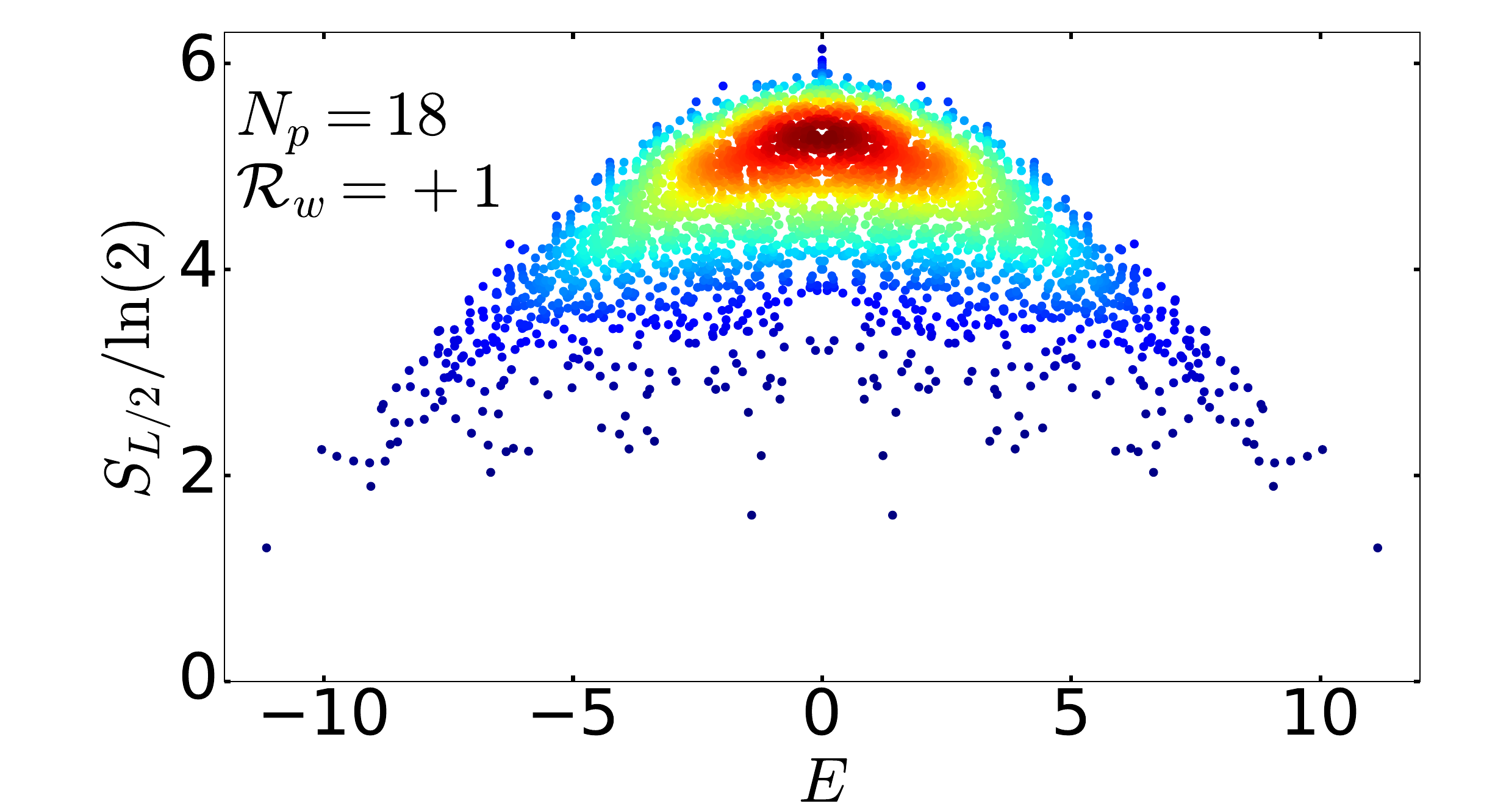}%
    \includegraphics[width=0.46\linewidth]{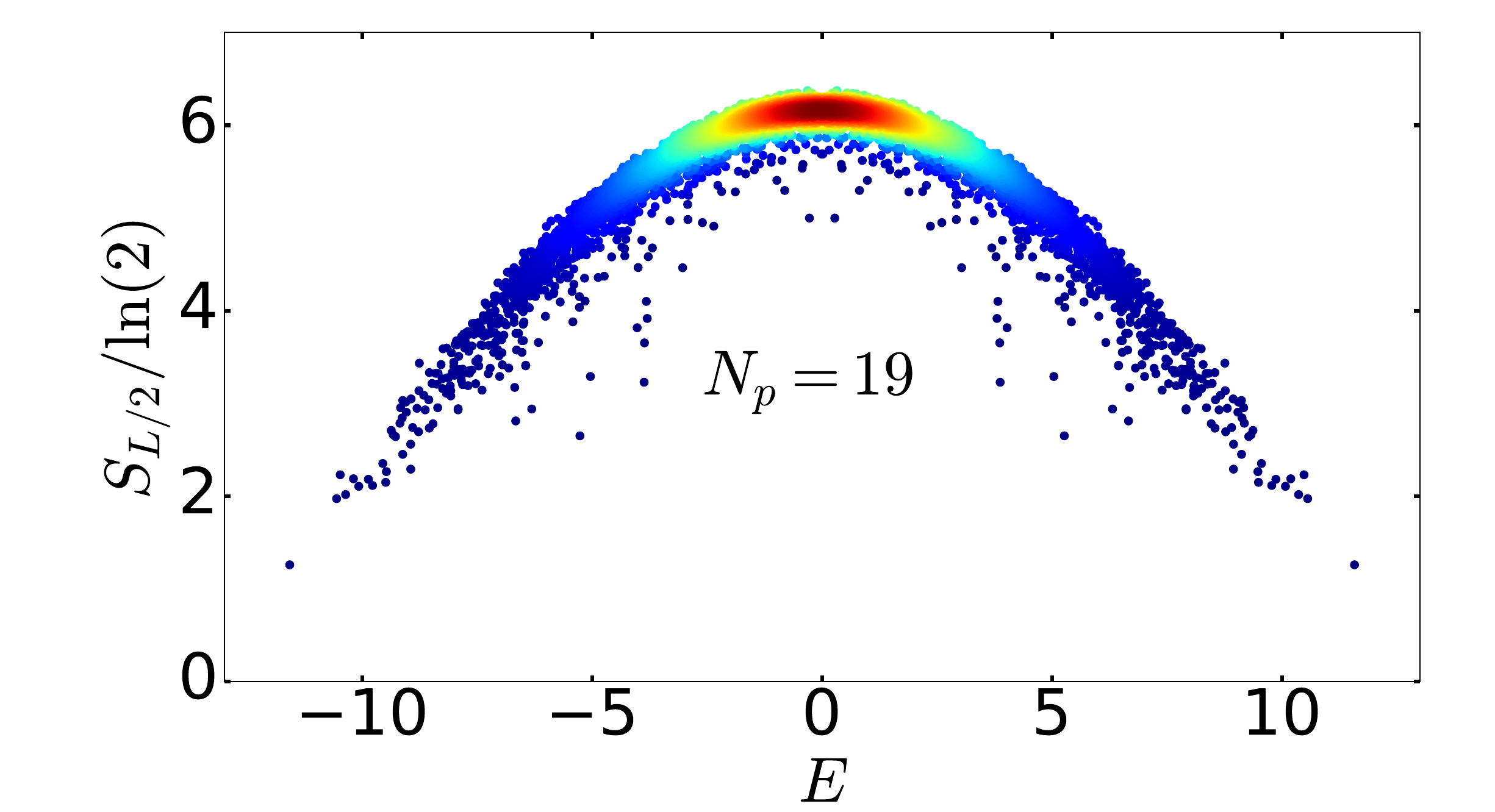}\\
     \includegraphics[width=0.46\linewidth]{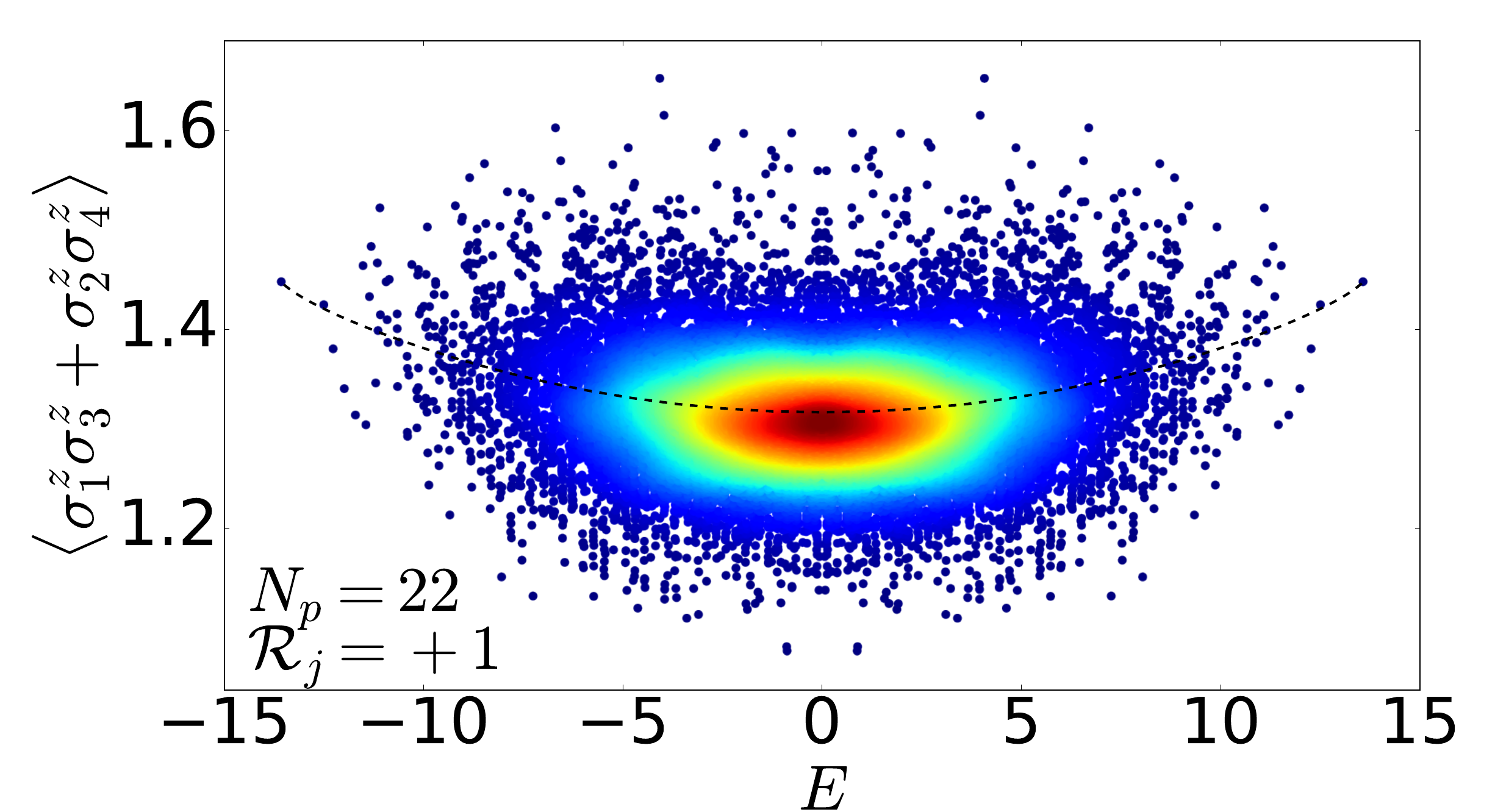}%
    \includegraphics[width=0.46\linewidth]{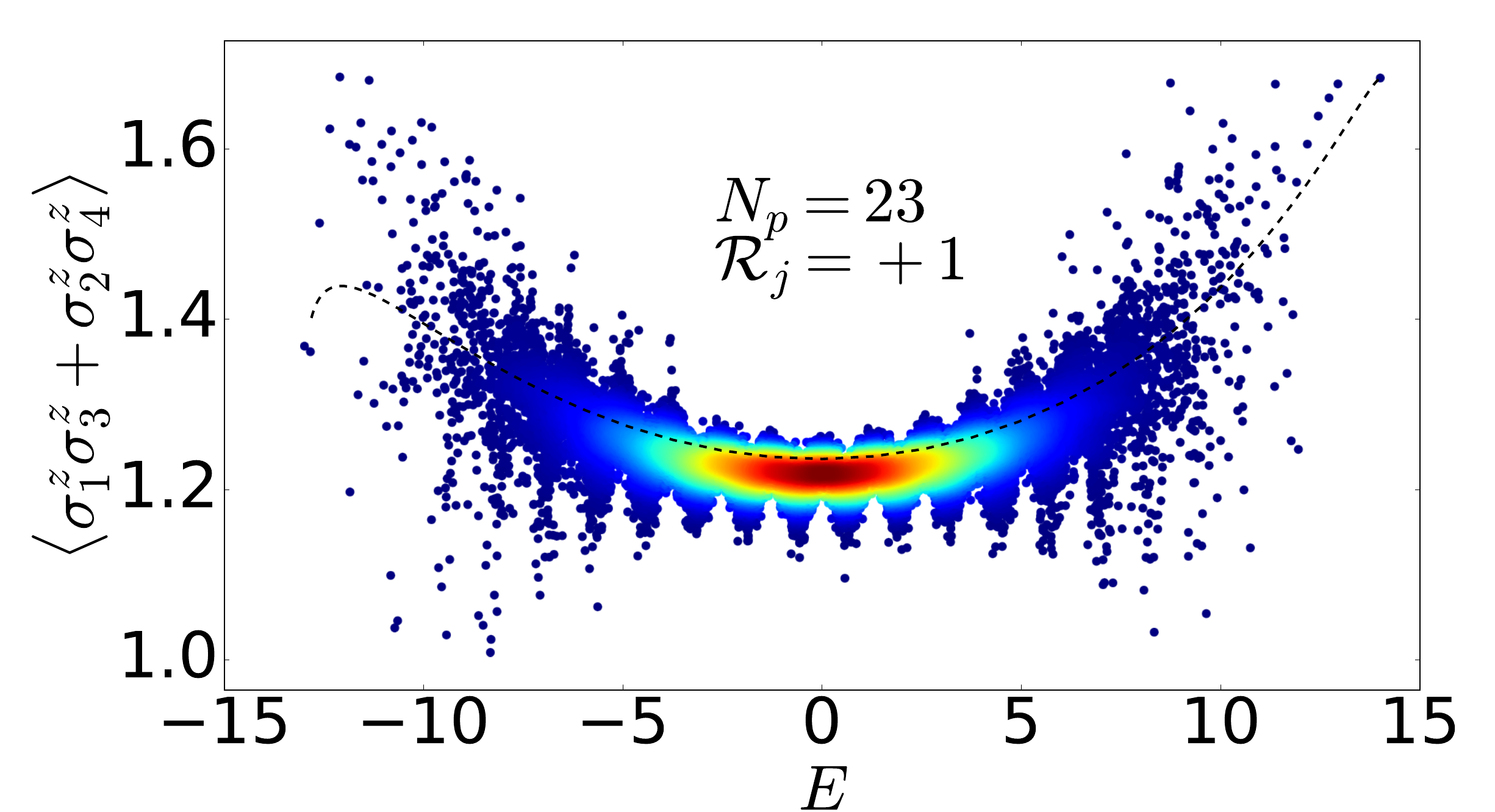}
    \caption{The behavior of the bipartite entanglement entropy for each
      eigenstate shown for (top left panel) a wire with $N_p=18$ in the
      $\mathcal{R}_w=+1$ sector and for (top right panel) a junction of two
      wires with $N_p=19$ for both $\mathcal{R}_j = \pm 1$ sectors
    together. The expectation value of a
      local diagonal operator defined on an elementary plaquette for each
      energy eigenstate shown for the wire with $N_p=22$ (bottom left panel)
      and the junction of two wires with $N_p=23$ (bottom right panel)
      respectively, both in the symmetry sector $\mathcal{R}_{w/j}=+1$. The
      dotted lines in both the lower panels indicate the thermal
      values as a function of the energy $E$. In all the panels,
      the density of states is indicated by the
      same color map where warmer color
      corresponds to higher density of states.} \label{FIG19}
 \end{center}
\end{figure}

Another tell-tale signature for the presence of QMBS is that such states
have anomalously low bipartite
entanglement entropy compared to neighboring eigenstates
with similar energies. The bipartite entanglement entropy is given by
\begin{eqnarray}
  S(A) = -\mathrm{Tr}[\rho_A \ln \rho_A]
\end{eqnarray}
for each eigenstate $|\Psi\rangle$ where $\rho_A =
\mathrm{Tr}_{\overline{A}}|\Psi \rangle \langle \Psi|$ where
$\rho_A$ represents the reduced density matrix obtained by
partitioning the system in to two spatial regions, $A$ and its
complement $\overline{A}$. We find it convenient to compute the
bipartite entanglement entropy by adopting the one-to-one mapping of
Fock states in a wire or a junction of two wires to pseudospins with
values $0, \pm 1$ in a 1D open chain with $N_p$ sites, with the
mapping explained in Sec.~\ref{treelargedrums}. We then take
$\overline{A}$ to be the first $N_p/2$ sites of the 1D chain for the
wire (as shown in Fig.~\ref{FIG15}) and the first $(N_p/2)-1$ sites
of the 1D chain for the junction of two wires (as shown in
Fig.~\ref{FIG16}). The results of such a computation from ED are
shown in Fig.~\ref{FIG19} for the wire (top-left panel) and the
junction of two wires (top-right panel) respectively. While both the
panels show a presence of several anomalous eigenstates with lower
bipartite entanglement entropy than the bulk of the spectrum, the
wire shows a broader distribution of values especially in the
neighborhood of $E=0$ compared to the junction of two wires.

The expectation value of any local operator in a high-energy eigenstate
is supposed to approach the thermal result, with the inverse temperature
being fixed by the energy density of the eigenstate, for a system that
satisfies ETH. In Fig.~\ref{FIG19} (bottom panels), we consider the expectation
value $\langle \Psi|\mathcal{O}|\Psi \rangle$ as a function of energy,
where $|\Psi\rangle$ denotes an
eigenstate of the wire (bottom left panel) or the
junction of two wires (bottom right panel) and $\mathcal{O}=
\sigma_1^z\sigma_3^z+\sigma_2^z\sigma_4^z$ where sites $1,2,3,4$ represent
the four sites (in a clockwise manner) of the $N_p/2$-th elementary
plaquette from the top-left for a wire and the central junction plaquette
for the junction of two wires. Since this local operator is located away
from the edges of the system, it represents a bulk operator in both the
cases. The thermal result as a function of energy is represented by dotted
curves on both the lower panels of Fig.~\ref{FIG19}. While the expectation
value of the local operator for the bulk of
the spectrum indeed approaches the thermal result, several eigenstates do
show an expectation value that is quite far from the corresponding thermal
result. The wire again shows a much larger variation in the range of
expectation values compared to the junction of two wires, especially in the
vicinity of $E=0$. Interestingly, the
latter case shows tower-like structures that are equidistant in energy
(Fig.~\ref{FIG19} (bottom right panel)) with a similar spacing between them
as the tower of scar states visible in the overlap plots (Fig.~\ref{FIG18},
two panels).


\section{Discussion}
\label{diss}
In conclusion, we have considered a spin-$1/2$ model on the two-dimensional
square lattice in a constrained Hilbert space where no two nearest-neighbor
sites can have up-spins simultaneously.
The interaction Hamiltonian is composed of
ring-exchange terms on elementary plaquettes that not only conserve the
total magnetization but also the magnetization along each column and row of
the square lattice. These additional
subsystem symmetries imply conservation of a
global dipole moment that leads to the phenomenon of Hilbert
space fragmentation. While microscopic models of both weak and strong
fragmentation are known in one dimension, we show that this particular
interacting model with
both hard-core constraints and subsystem symmetries presents a rich
structure of emergent quantum drums as well as
a rare example of strong Hilbert space fragmentation in two dimensions.

All the many-body eigenstates of this model can be expressed in terms of
the tensor product of modes of appropriate quantum drums and
any left-over inert spins. Given an initial unentangled product state in
the computational basis, the
associated quantum drums get fixed and come in a variety of shapes
and sizes starting from one-plaquette drums to truly extensive structures
made of plaquettes that share edges and/or vertices with each other.
Specifying the plaquettes that belong to a drum uniquely fixes its
spectrum. Crucially, these drums can be
``shielded'' from each other by shielding regions that only
grow as the perimeter and not the area of such drums.

Large quantum drums and their associated fragment dimensions can be most
easily estimated by using a ``wire'' decomposition of such drums and
then counting the number of ways in which such wires can fluctuate
simultaneously without violating the kinematic constraints. This allows us
to identify the appropriate drums that dominate statistically for a given
density of up-spins (bosons). The largest Hilbert space
  fragment is generated by the ``checkerboard drum'' at a density of
  $n=1/4$ for the up-spins (bosons). It is shown
that initial states that belong to such fragments
evade ETH-predicted thermalization (in the
full Hilbert space) due
to the presence of either an extensive number of inert spins or an extensive
number of next-nearest neighbor spin correlations that 
  retain the memory of the initial state. In particular, initial
  states that belong to the checkerboard drum fragment
  contain zero density of inert spins but a finite density of
  next-nearest neighbor correlations that are pinned to athermal
values under time evolution with $H$.

We consider the spectrum of some small drums analytically to show the
emergence of interesting zero, non-zero integer and irrational modes.
Close packing an extensive number of the
elementary one-plaquette drums already generate many-body eigenstates with
integer energies (including zero) and strict area-law scaling of entanglement
entropy. Large quasi-one-dimensional and two-dimensional quantum drums can be
viewed as interesting interacting systems with constrained Hilbert spaces.
A class of these drums harbor a large number of exact zero modes. The
simplest quasi-one-dimensional drum, which we dub as a wire, is shown to
be exactly equivalent to the well-known PXP chain with open boundary
conditions. However, a particular junction of two wires is also studied
which cannot be mapped in to the PXP chain and represents a different
constrained model. Both these quasi-one-dimensional drums support
distinct families of quantum many-body scars that cause
periodic revivals from certain simple initial states. Our numerics for the
wire also shows that the period-$3$ state with Rydberg excitations on
every third site shows strong revivals for open chains of length $3n+1$
without the necessity of adding further perturbations to the PXP chain.
This result can have possible implications for experiments with Rydberg
atoms.

Several possible open directions emerge from our study. Other
junctions of wires, like junctions of three wires and four wires, as
well as some of the two-dimensional drums introduced here should
have interesting high-energy properties. It is further possible to
add diagonal interactions in the computational basis which preserve
the fragmented structure of the model.
Using such additional
interactions, one can possibly access different phases and phase
transitions at zero temperature in both quasi-one-dimensional and
two dimensional theories in the presence of subsystem symmetries. Whether
many-body localized phases can emerge
in quasi one-dimensional and two-dimensional
drums in the presence of subsystem symmetries
on adding diagonal interactions with random couplings presents
another interesting research direction. \\

{\it{Note added}}: While preparing this manuscript, we came to know
of a related work by Lehmann {\it et al.}~\cite{Lehmann} which
discusses strong Hilbert space fragmentation in higher dimensions
using a different Hamiltonian (correlated hopping model).\\

\section*{Acknowledgements}
A.C. thanks Madhumita Sarkar for help with cluster facilities at IACS, Kolkata
and DST, India
for support through SERB project PDF/2021/001134.
B.M. has been funded by the European Research Council (ERC) under the
European Union's Horizon 2020 research and
innovation programme (Grant No. 853368).
K.S. thanks DST, India for support through
SERB project JCB/2021/000030. 




\bibliography{2dmodel_v4}

\nolinenumbers

\end{document}